\documentclass[useAMS,usenatbib]{mn2e}
\usepackage{graphicx}
\usepackage[draft]{hyperref}
\usepackage{xcolor}

\hypersetup{
    colorlinks,
    linkcolor={red!50!black},
    citecolor={blue!50!black},
    urlcolor={blue!80!black}
}

\usepackage{caption}
\usepackage{amsmath}
\usepackage{natbib}
\usepackage{color}
\usepackage[english]{babel}
\usepackage{array}

\title[Age-metallicity relation in simulated galaxies]{The history of stellar metallicity in a simulated disc galaxy}
\author[Snaith et al.]{O. N. Snaith$^1$, J. Bailin$^{1,2}$, B. K. Gibson$^{3}$, E. F. Bell$^4$,  G.  Stinson$^5$,\newauthor M. Valluri$^4$, J.  Wadsley$^6$, H. Couchman$^6$\\
$^1$ Department of Physics and Astronomy, University of Alabama, Box 870324, Tuscaloosa, AL 35487-0324, USA\\
$^2$ National Radio Astronomy Observatory, P.O. Box 2, Green Bank, WV, 24944, USA \\
$^3$ E.A. Milne Centre for Astrophysics, Dept of Physics \& Mathematics, University of Hull, Hull, HU6 7RX, United Kingdom \\
$^4$ Department of Astronomy, University of Michigan, 500 Church Street, Ann Arbor, MI 48109, USA \\
$^5$ Max-Planck-Institut fur Astronomie, Konigstuhl 17, D-69117, Heidelberg, Germany \\
$^6$ Department of Physics and Astronomy, McMaster University, Hamilton, Ontario, L8S 4M1, Canada \\
}
\date{\today}
\begin{document}

\maketitle

\begin{abstract}
We explore the chemical distribution of stars in a simulated galaxy. Using simulations of the same initial conditions but with two different feedback schemes (MUGS and MaGICC), we examine the features of the age-metallicity relation (AMR), and the three-dimensional age-[Fe/H]-[O/Fe] distribution, both for the galaxy as a whole and decomposed into disc, bulge, halo, and satellites. The MUGS simulation, which uses traditional supernova feedback, is replete with chemical substructure. This substructure is absent from the MaGICC simulation, which includes early feedback from stellar winds, a modified IMF and more efficient feedback. The reduced amount of substructure is due to the  almost complete lack of satellites in MaGICC. We identify a significant separation between the bulge and disc AMRs, where the bulge is considerably more metal-rich with a smaller spread in metallicity at any given time than the disc. Our results suggest, however, that identifying the substructure in observations will require exquisite age resolution, on the order of 0.25 Gyr. Certain satellites show exotic features in the AMR, even forming a `sawtooth' shape of increasing metallicity followed by sharp declines which correspond to pericentric passages. This fact, along with the large spread in stellar age at a given metallicity, compromises the use of metallicity as an age indicator, although alpha abundance provides a more robust clock at early times. This may also impact algorithms that are used to reconstruct star formation histories from resolved stellar populations, which frequently assume a monotonically-increasing AMR.
\end{abstract}

\section{Introduction}
Gaia, APOGEE and extragalactic surveys such as CALIFA and MaNGA will provide ever more detailed data on the chemical evolution of galaxies. It is important to understand the fine structure of the Milky Way in order to interpret these observations. However, simulations contain detailed `sub-grid' physics that can have strong effects on the end result \citep{Scannapieco2012} and remain uncertain. One avenue to understanding the chemical evolution of galaxies is to compare the chemistry in simulated galaxies using the same initial conditions but different sub-grid physics, which we address in this paper. 

The metallicity of the gas in a galaxy is controlled by the rate of star formation, the distribution of stars and the flow of infalling and outflowing material \citep[e.g.][]{Tinsley1972,Pagel1981}. These processes play off against each other, and the evolution of the ISM is encoded  in the properties of stars which form at a given time. In essence, the formation of stars `freezes out' the ISM, and provides a historical record of how the chemical properties of the galaxy have evolved. 

Stellar metallicity data provides one of the only windows through which we can view the history of star formation in a galaxy.  This is because observations of other properties, such as galaxy morphology and kinematics, provide only a single snapshot in the lifetime of a galaxy. Such structures evolve, and break up, due to radial migrations \citep{Sellwood2002}, interactions such as the scattering of disc stars by satellites which heats stars overall, mergers, and other stochastic effects such as the influence of the galactic bar, spiral heating and external tidal effects etc. Once a star has formed, however, its surface metallicity does not change (on the whole).  However, although the metallicity of individual stars is constant with time, various metrics (such as integrated metallicities or the {\it local} age-[Fe/H] relation) are influenced by radial motions. These properties are, however, more robust {\it globally} than other galactic properties over time,  and require detailed modelling to help in interpreting data from current and future surveys, such as Gaia, APOGEE etc.

It has become possible to reconstruct the age metallicity relation (AMR) of local galaxies \citep{Skillman2003, Cole2007, Williams2009}. These authors have concluded that local dwarf galaxies have varying metallicity histories: IC 1613 shows a rising mean metallicity with time \citep{Skillman2003}, Leo A \citep{Cole2007} and the outer disc of M81 \citep{Williams2009} show a flat AMR, and M32 \citep{Monachesi2012} shows an AMR which rose early and flattened. \citet{Holmberg2009} found the Milky Way has a flat AMR, while \citet{Haywood2013} find that the Galaxy has a shallowly rising AMR after a steep initial increase.  PHAT \citep{Dalcanton2012} will make similar measurements for M31. HST ACS colour-magnitude diagrams have been used \citep{Weisz2011} to explore the star formation in a sample of Local Group dwarf galaxies, while \citet{Kirby2011} have used Keck DEIMOS spectra to calculate star formation histories using abundance ratios. \citet{Snaith2014} used high signal to noise stellar abundances from \citet{Adibekyan2012} and ages from \citet{Haywood2013} to reconstruct the SFH of the Milky Way. All these different approaches show the strengths of chemical data in reconstructing the past history of galaxies. However, in each case, various assumptions have to be made which can strongly affect the outcome of the reconstruction.  Other observers have decomposed galaxies into radial bins \citep{Gogarten2010} and measured the AMR in each bin. The AMR of dwarf galaxies in simulations has been explored by \citet{Pilkington2012b}, who also analyse the observed galaxy IC 1613, and considerable differences between observations and theory were identified. However, as distant objects cannot be studied in the same detail as the Milky Way, those authors did not attempt a direct comparison between their simulations and our own Galaxy.

The specific elemental abundance of different components of a galaxy (bulge, disc, halo etc.), along with metallicity, provide detailed information about its assembly history in a form that can be reconstructed from detailed observations. Alpha elements, usually traced using oxygen, are overwhelmingly produced by core collapse supernovae (CCSNe), with a time delay of the order of several Myrs. Iron, however, is produced mainly by  SNIae, which contribute over 8 times as much iron as CCSNe \citep{Iwamoto1999}. SNIae take  a longer time to release metals back into the ISM, beginning after 50~Myr with a time delay distribution that peaks at 100~Myr to 1~Gyr depending on the SFH \citep{Gibson1997}. As a result of the difference between the timescales of the two types of supernovae, the ratio of oxygen to iron encodes information of the star formation rate, providing a further avenue of investigation.

We will demonstrate the key features in the chemical evolution of a simulated galaxy in detail. We will also compare this to a simulation carried out using the same initial conditions but with a different implementation of stellar feedback. This will allow us to contrast the predictions of the two models. 

Our goal in this paper is to study the ages, metallicities, and chemical abundances of stars in a simulated Milky Way-like disk galaxy. In particular, we will explore the different signatures of evolution in the bulge, disc and halo, while comparing the results of both the MUGS \citep{Stinson2010} and MaGICC \citep{Stinson2013} galaxy simulations. Compared to previous work on the chemical evolution of these simulated galaxies \citep[e.g.][]{Pilkington2012b,Pilkington2013,Calura2012,Gibson2013} we will explore the fine structure of the chemical evolution of the MUGS and MaGICC simulations in detail in terms of age, [Fe/H] and [O/Fe]. While \citet{Calura2012} examined the MDF of these galaxies, and \citet{Pilkington2012} examined the gradients, we explore the detailed fine structure and the origin of the different features by decomposing the full AMR into different galactic components.  This is particularly important at the present time because we are seeing a growing interest in novel feedback implementations \citep[e.g.][]{Hopkins2014,Bird2013}.

 We will first outline the simulations used (Section~\ref{sec:simulations}), our methods (Section~\ref{sec:defns}), and present the age-metallicity, age-[O/Fe] and metallicity-[O/Fe] distributions of a simulated galaxy using two  distinct implementations of the supernova feedback but the same initial conditions (Section~\ref{sec:results}). We will dissect the simulated galaxy, and examine the variation in the chemistry of stars of the different components (bulge, disc, halo), and the properties of current and former satellites. Further discussion and conclusions are presented in Section~\ref{sec:conclusions}.

\section{Simulations}\label{sec:simulations}

In this paper we use the McMaster Unbiased Galaxy Simulations \citep[MUGS,][]{Stinson2010} sample and the Making Galaxies in a Cosmological Context \citep[MaGICC,][]{Stinson2013} sample. We selected the disky galaxy known as g15784 which is common to both samples, and which has been analysed in a number of other papers \citep[e.g.][]{Nickerson2011,Brook2012,Brook2014,Calura2012,Valluri2013,Obreja2014, Gibson2013,Pilkington2012,Woods2014}. 

Previous work on the chemistry of the MUGS  galaxies has explored the radial and vertical metallicity gradients \citep{Pilkington2012} and the MDF of the solar vicinity and bulge \citep{Calura2012}. \citet{Calura2012} find notable differences between the simulated galaxy and the Milky Way. These authors found that the median metallicities in MUGS are 0.2 to 0.3 dex lower than in the Milky Way disc and bulge,  with larger dispersions.

The initial conditions assume a $\Lambda$CDM, WMAP3 cosmology $H_0=73~\mathrm{km~s^{-1}~Mpc^{-1}}$, $\Omega_{m} = 0.24$, $\Omega_{\Lambda} = 0.76$, $\Omega_{b} = 0.04$ and $\sigma_8=0.79$ \citep{Spergel2007a}. The galaxy sample was chosen at random from a catalogue with halo masses between $\sim5\times 10^{11}$ to $\sim2\times 10^{12} M_\odot$. Further selection criteria required that there was no structure within 2.7 Mpc with a mass greater than $\sim5\times 10^{11}M_\odot$. The simulation volume was large enough to ensure a realistic angular momentum distribution and merger history. 

In order to achieve sufficient mass and spatial resolution the simulations employ the commonly adopted zoom technique. This method adds high resolution particles in the region of interest, while following other regions with much lower resolution particles. In the highest resolution region of each simulation the dark matter, gas and star particles have masses of $1.1\times 10^6$, $2.2\times 10^5$ and $<6.3\times 10^4$ $M_\odot$  respectively, and a gravitational softening length of 310 pc. The simulation was advanced through time using the SPH code GASOLINE \citep{Wadsley2004} and includes low-temperature metal cooling  \citep{Shen2010} based on CLOUDY \citep{Ferland1998}, a Schmidt-Kennicutt star formation law \citep{Kennicutt1998} and UV background radiation. For further detail on MUGS and MaGICC, see \citet{Stinson2010} and \citet{Stinson2013} respectively.

MUGS and MaGICC use the same initial conditions and cosmology but have a different implementation of the 
stellar feedback. They both employ the `blast wave' model of SN feedback, where gas cooling is locally suspended in order to mimic the thermal heating of gas from supernovae \citep{Stinson2006}. MaGICC also includes early energy input into the ISM, from massive stars. This early feedback heats the gas from the moment a star forms, rather than waiting until the first  CCSNe which are triggered after $\sim$4 Myr \citep{Stinson2013}. Since MUGS galaxies lack this early feedback, they suffer from overcooling \citep{Pilkington2012}. For example, in \citet[][Figure 13]{Stinson2010} the r-band magnitude of galaxies in MUGS are systematically too bright for their halo mass, compared to observations.  This is the principal difference between MUGS and MaGICC and it is expected to have the dominant effect  on the resulting galaxies. However, there are a further series of differences between the simulations which we expect to have a less significant effect than the differences in feedback:
\begin{itemize}
\item  The diffusion prescription for metals was changed. The original diffusion prescription for MUGS was first discussed in \citet{Wadsley2008}. In this model, the amount of mixing depends on the local velocity of the shear field and the  spatial resolution of the simulation. In MaGICC this was modified, so that diffusion did not occur between particles which had cooling shut off by feedback processes. This was implemented because the method tended to unphysically reduce the efficiency of outflows. For a longer discussion see \citet{Stinson2013}. This {\it may} have second order effects on the metallicity distribution.
\item The metallicity, Z, is underestimated, in MUGS, by a factor of 1.8  \citep{Pilkington2012}. This is because Z was calculated on the basis of O+Fe in MUGS, while in MaGICC the metals of other species were accounted for. This difference will not directly affect the chemistry, but influences processes such as cooling which are metal dependent.
\item The minimum SPH smoothing length is set to 0.25 times the gravitational softening length ($r_{softening}$) in MaGICC, while it is 0.01$r_{softening}$ in MUGS \citep{Pilkington2012b}. This is expected to have only a minor impact on the simulation and was done to improve the computation time in high density regions.
\end{itemize}

Although the most important difference in the simulations was the implementation of the early feedback in MaGICC, the stellar feedback in MaGICC was additionally altered in three further ways in order to increase the energy fed back into the ISM:

\begin{itemize}

\item MUGS uses the Kroupa IMF \citep{Kroupa1993}, while the MaGICC sample uses the Chebrier IMF \citep{Chabrier2003}. This change of IMF means that 4$\times$ as many  CCSNe explode per generation of stars in MaGICC than in MUGS. 
\item In MUGS, feedback was immediately radiated away if the cooling shut off was shorter than 1~Myr, and so never coupled to the ISM. This was corrected in MaGICC.
\item The feedback efficiency was increased 2.5$\times$ per SNe. 
\end{itemize}

\citet{Stinson2013}, however, showed that the total amount of energy dumped back into the ISM by stellar feedback was less important than the addition of the {\it early} feedback. This means that if the energy put into early feedback was instead added to traditional supernovae the effect on galaxy morphology, which is one of the principle successes of MaGICC, is not as pronounced. However, the changing IMF, will have an effect on the chemistry.

An in depth analysis of these issues is beyond the scope of this paper but are mentioned as potential sources of difference beyond the change in feedback. \\

We can use MUGS vs. MaGICC as a proxy for the range of plausible possibilities for feedback in the real universe. MaGICC represents a step forwards in attempts to simulate realistic galaxies, matching numerous scaling laws \citep{Brook2012} which former simulations, such as MUGS, could not reproduce \citep[e.g.][]{Brook2014, Stinson2013, Gibson2013}.  

One difference between MUGS and MaGICC is the  decrease in the number of luminous satellites orbiting the main galaxy.  \citet{Nickerson2013} showed that MUGS effectively reproduced the number of luminous satellites expected around Milky Way sized galaxies.  The MUGS galaxy has $\sim$ 20 luminous satellites,  although the most massive ones tend to be overly massive.   MaGICC has only 4 such satellites. Since real galaxies have numerous luminous satellites, we must use MUGS to understand their effects even though we expect MUGS to overestimate their impact. MUGS satellites have higher stellar masses dark matter mass ratios than observations \citep{Stinson2010}. 

The `true' feedback situation is assumed to be similar to, but not quite as extreme, as used in MaGICC. If we see similar patterns in both MUGS and MaGICC that are compatible with the differences in their SFHs, then we can feel confident that we are drawing realistic conclusions. 

We identify the halos and subhalos using AHF \citep{Gill2004, Knollmann2009}\footnote{AHF can be downloaded from http://popia.ft.uam.es/AHF.}, which uses adaptive mesh refinement to locate halos in a smoothed density field. For each density peak, the potential of the surrounding particles is identified, and those particles bound to the density peak are classed as (sub)halo members. AHF assumes that particles within the virial radius, that are bound to the halo, are members of that halo. In this way the code always returns spherical halos. Subhalos, however, are not assumed to expand to their virial radius, but to the saddle point of the density profile in the host potential. This is one of a number of ways to define dark matter subhalos, none of which has been shown to be substantially superior to any other \citep[e.g.][]{Knebe2011}.

\section{Definitions}\label{sec:defns}
We decomposed the galaxy into various components (halo, bulge, disc), and subdivide the disc component  by radius. We also mark stars according to whether they formed in-situ or in satellites.  \\

\subsection{Dynamical Decomposition} 
\label{Sec:DecompDef}
We decompose the  galaxy into a disc, bulge and halo using the dynamical decomposition approach  presented in \citet{Stinson2010}, which is based on the method of \citet{Abadi2003}. Our algorithm is based on the one supplied with  PYNBODY \citep{pynbody}\footnote{We made use of PYNBODY (https://github.com/pynbody/pynbody) in our analysis for this paper. } We decompose the galaxy in both MUGS and MaGICC in the same way, and examine the detailed chemical evolution for the first time.

In order to calculate the distribution of $J_z/J_{circ}$ for stars in the galaxy we follow the method of \citet{Stinson2010}. While \citet{Abadi2003} used the value of the total binding energy of the particles, and thus a careful accounting of the shape of the potential, to calculate $J_{circ}$, the approach of \citet{Stinson2010} assumes spherical symmetry.  Therefore, $|J_z/J_{circ}| \le 1$ in the \citet{Abadi2003} method, but can extend beyond these bounds when using the approach of \citet{Stinson2010}. We have adopted the simpler \citet{Stinson2010} method, which clearly produces a good separation between the stellar populations in the bulge, disc, and halo; the $J_z/J_{circ}$ distribution for MUGS g15784 using the \citet{Abadi2003} method can be found in \citet{Calura2012}. Populations of stars which show features of more than one component according to the decomposition are discarded to reduce interlopers in our samples. 

\begin{figure}
    \centering
     \includegraphics[scale=.4]{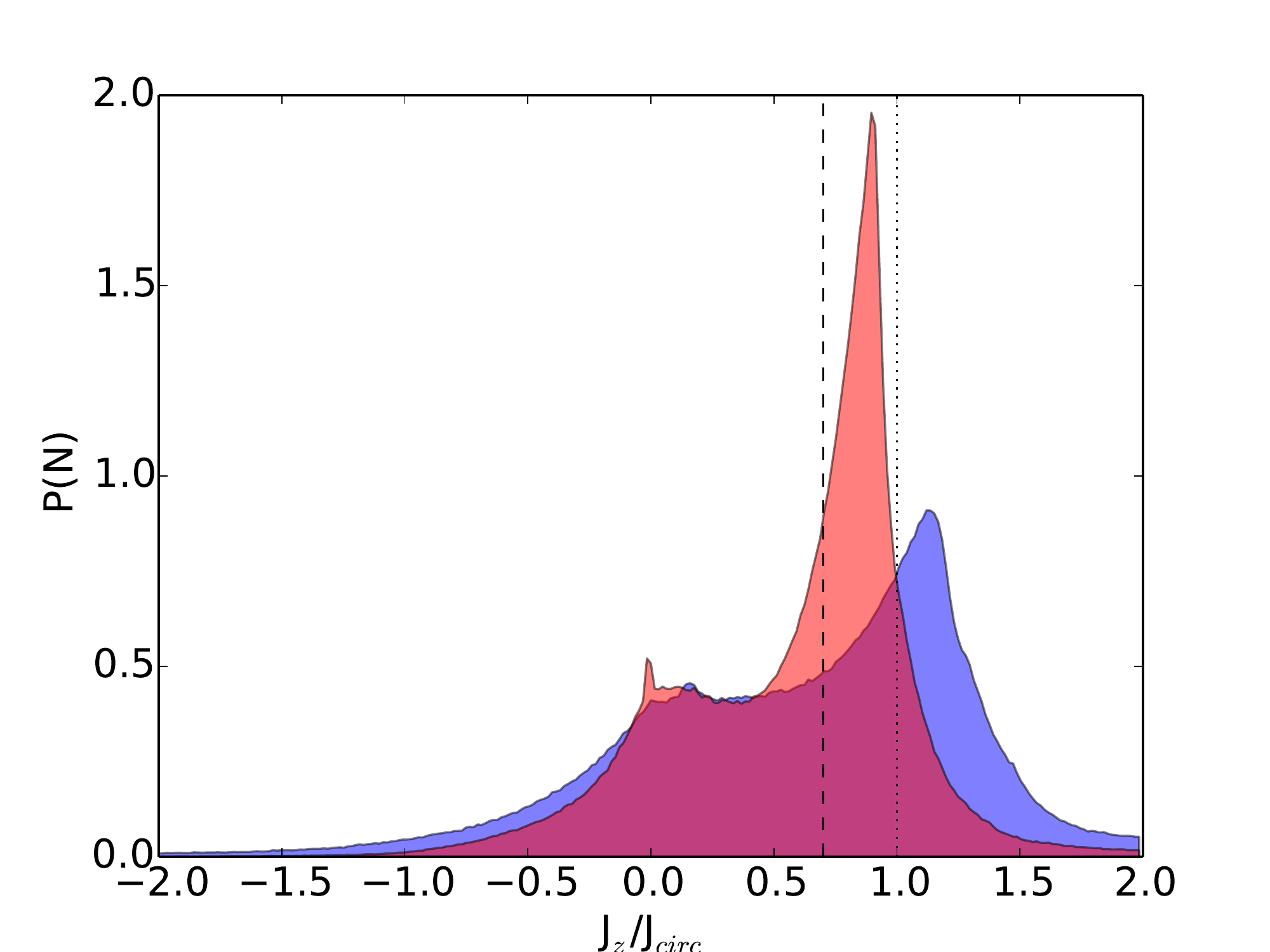} 
     \caption{The probability distribution of $J_z/J_{circ}$ for MUGS (blue) and MaGICC (red) stars. The dotted line shows the expected value if all the stars are on circular orbits, and the dashed line shows the lower limit for selecting disc stars. In this plot we discard all stars with energy less than a given value, as these are deep in the potential well and assigned to the bulge.}
 \label{Fig:jzjcirc}    
\end{figure}

The probability distribution of the  $J_z/J_{circ}$ distribution for MUGS and MaGICC are shown in Fig. \ref{Fig:jzjcirc}. The disc is defined as those stars with $0.7<J_z/J_{circ}<5$ (called disc 2 in Table \ref{Tab:parts} ). Unless described otherwise we constrain the disc to also lie inside of tight positional bounds. The disc is defined as those stars  that satisfy the dynamical definition with radii less than 20 kpc, heights above the plane of less than 5 kpc (disc 1). We chose an inner radius cut off of R$>$2 kpc to avoid contamination by bulge stars. Even though the bulge extends out to $\sim$5 kpc, for MUGS, and $\sim$2.4 kpc, for MaGICC, the dynamical decomposition becomes more effective  at splitting up the bulge and disc outside this inner region. We also remove all satellite stars to remove interlopers which contaminate the disc.

The bulge members are defined, as in PYNBODY, as those stars with $J_z/J_{circ}<J_{crit}$ and a binding energy less than the median energy of the galaxy (for bound particles the binding energy is negative, meaning that a lower energy means the particle is {\it more} bound). The calculation of the $J_{crit}$ criterion is an iterative process, but is ultimately where the total angular momentum of the bulge is equal to zero. This defines a classical bulge where the bulge is entirely pressure supported. 

Halo stars are those stars not in the disc but with binding energies greater than the median. Any star which does not fit these criteria are neglected; the algorithm in  PYNBODY also includes definitions of the pseudobulge and kinematical thick disc, but the resolution of MUGS and MaGICC is considered insufficient to resolve these components. Thus, these  leftover stars are of ambiguous origin, and have properties which overlap the various other components. They appear to form `transition' populations in terms of their chemical properties. As can be seen in Table \ref{Tab:parts}, this is 10\% of stars in MUGS and 20\% of stars in MaGICC.

An important caveat is that the dynamical decomposition is imperfect.  We choose to define the disc in terms of the value of $J_z/J_{circ}$. However, this separation between the spheroidal component and the disc is somewhat arbitrary. We expect that the distribution of $J_z/J_{circ}$ in the disc and halo to be more correctly modelled by two overlapping Gaussians, one centered at 0 with a large  width, and other narrower and centered at 1. However, it must be noted that for the bulge this is an approximation only. It assumes that the bulge has no circular velocity, and is only pressure supported, which is not true. Indeed, the bulge of the Milky Way  has $V_{circ}/\sigma \sim 0.5$ (70/140 km/s) \citep[e.g.][]{Howard2008}.

A simple cut in $J_z/J_{circ}$ will result in some cross-contamination between components.
However, as the halo is diffuse, and the bulge is centrally concentrated, we can reduce the contamination  with the `strict' definition of the disc given above.

\subsection{Where stars are formed.}

We define four types of star:
\begin{enumerate}
\item {\bf in-situ}: stars which form within the dark matter halo of the host galaxy, and not in one of the subhalos.
\item {\bf accreted}: stars which form in another halo, separate from the host but are now members of the host.
\item {\bf commuter}: stars which formed in subhalos of the host but now lie in the host. Commuter stars have also been called `endodebris' by \citet{Tissera2013}, and `ex-situ' stars by \citet{Pillepich2014}. As a satellite falls into the host halo (thus becoming a satellite) the newly forming stars will become classified as `commuter stars' whereas if they were formed before the satellite entered the halo of the host they are `accreted stars'. This is different from the ST ACC stars defined in \citet{Brook2014} which include both accreted and commuter stars.
\item {\bf satellite}: stars which lie within  the subhalos of the host at the current time.
\end{enumerate}

Membership of types (i), (ii) and (iii) are identified by looking back in time at the membership of stars during the first output in which they can be identified.

\subsection{Galaxy properties}
The properties of g15784 in MUGS and MaGICC can be found in a number of papers, but are summerized here. 

\begin{table}
\centering
\begin{tabular}{cccc}
\hline
	& MUGS & MaGICC & MW  \\
	\hline
M$_{vir}$ & 1.5$\times 10^{12}$ M$_\odot$ & 1.5$\times 10^{12}$ M$_\odot$ & 1.3$\times 10^{12}$ M$_\odot^3$  \\ 
M$_{cold gas}$ & 3$\times 10^{9}$ M$_\odot$ & 4$\times 10^{10}$ M$_\odot$ & $\sim 1\times 10^9$ M$_\odot^4$\\
M$_{*}$ &1.1$\times 10^{11}$ & 8.3$\times 10^{10}$ & 6.4$\times 10^{10}$  M$_\odot^3$ \\
R$_s$& 3.38$^2$ kpc  & 2.7$^1$ kpc &  2.6-3.6 kpc$^3$\\
R$_z$& 0.6$^2$ kpc& 0.7$^1$ kpc & 0.3-0.9 kpc$^3$ \\
B/T & 0.6$^2$ & 0.21$^1$ &  0.14$^3$ \\
\hline
\end{tabular}
\caption{Bulk properties of the simulated galaxies and the Milky Way. $^1$ from \citet{Brook2012}, $^2$ from \citet{Stinson2010} and $^3$ from \citet{McMillan2011}, $^4$ \citet{Putman2012}. For the Milky Way the two scale lengths are for the thin and thick discs respectively.} 
\label{Tab:MWcomp}
\end{table}

Table \ref{Tab:MWcomp} shows that the simulated galaxy has a mass comparable to the Milky Way. It has a fairly quiescent merger history since z=1. Both galaxies have similar scale lengths and scale heights to the Milky Way. MaGICC has a B/T which is closer to the Milky Way. Thus, we can consider g15784 a Milky Way type galaxy, which should share global properties with the Milky Way even if it differs in the details.  However, we must be careful in comparing this galaxy to the Milky Way in detail, because no attempt was made to ensure that the assembly history of g15784 bore any similarity to our own Galaxy, except in terms of halo mass.

\section{Results}\label{sec:results}

\subsection{Overview}
\label{overview}
\begin{figure*}
\centering
     
\begin{tabular}{cccc}

      \parbox[c]{.35\textwidth}{\includegraphics[scale=.34, trim={0cm 0cm 0cm 0cm},clip]{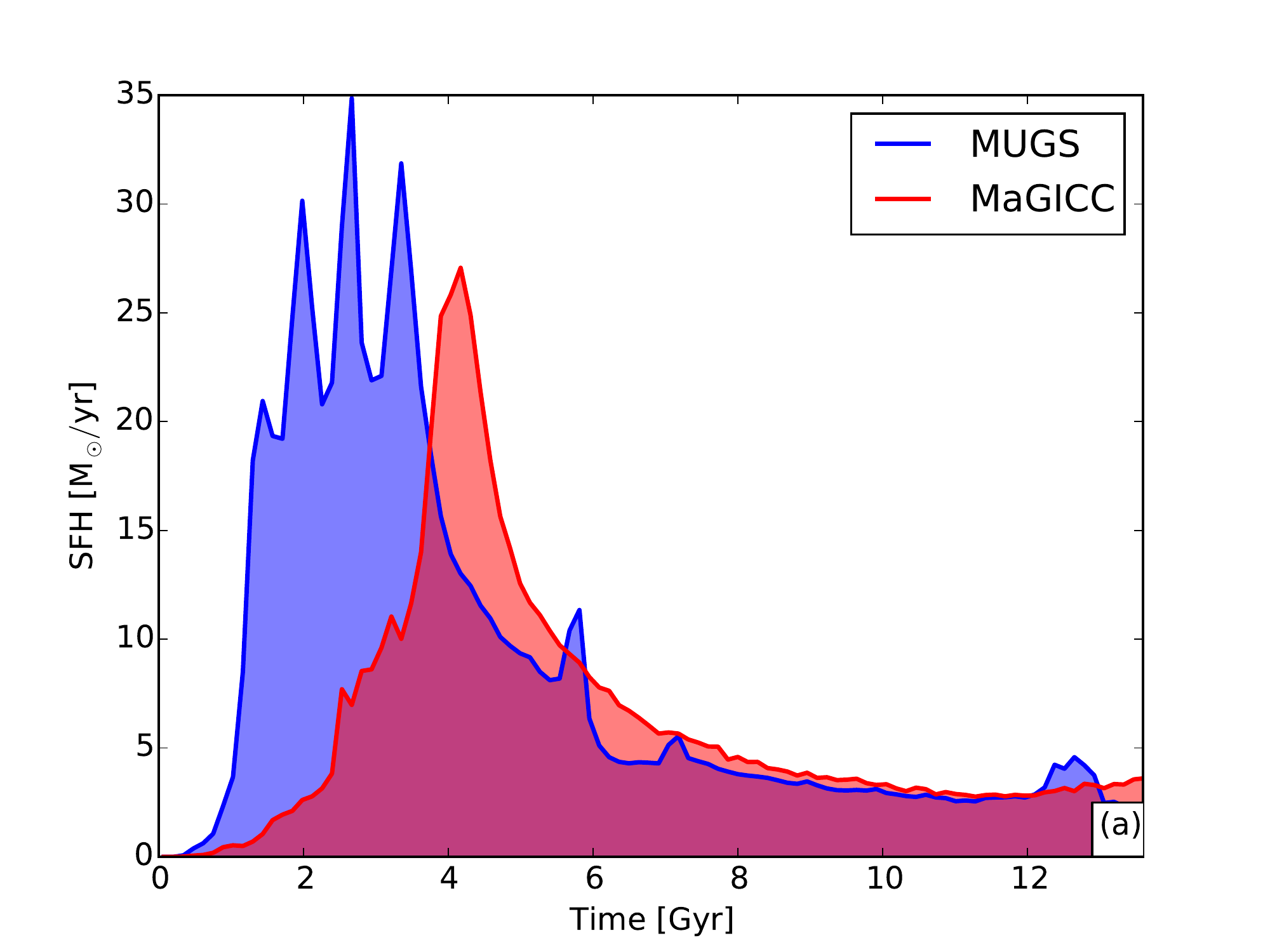}} & 
      \parbox[c]{.3\textwidth}{\includegraphics[scale=.34, trim={0.5cm 0 1.5cm 0cm},clip]{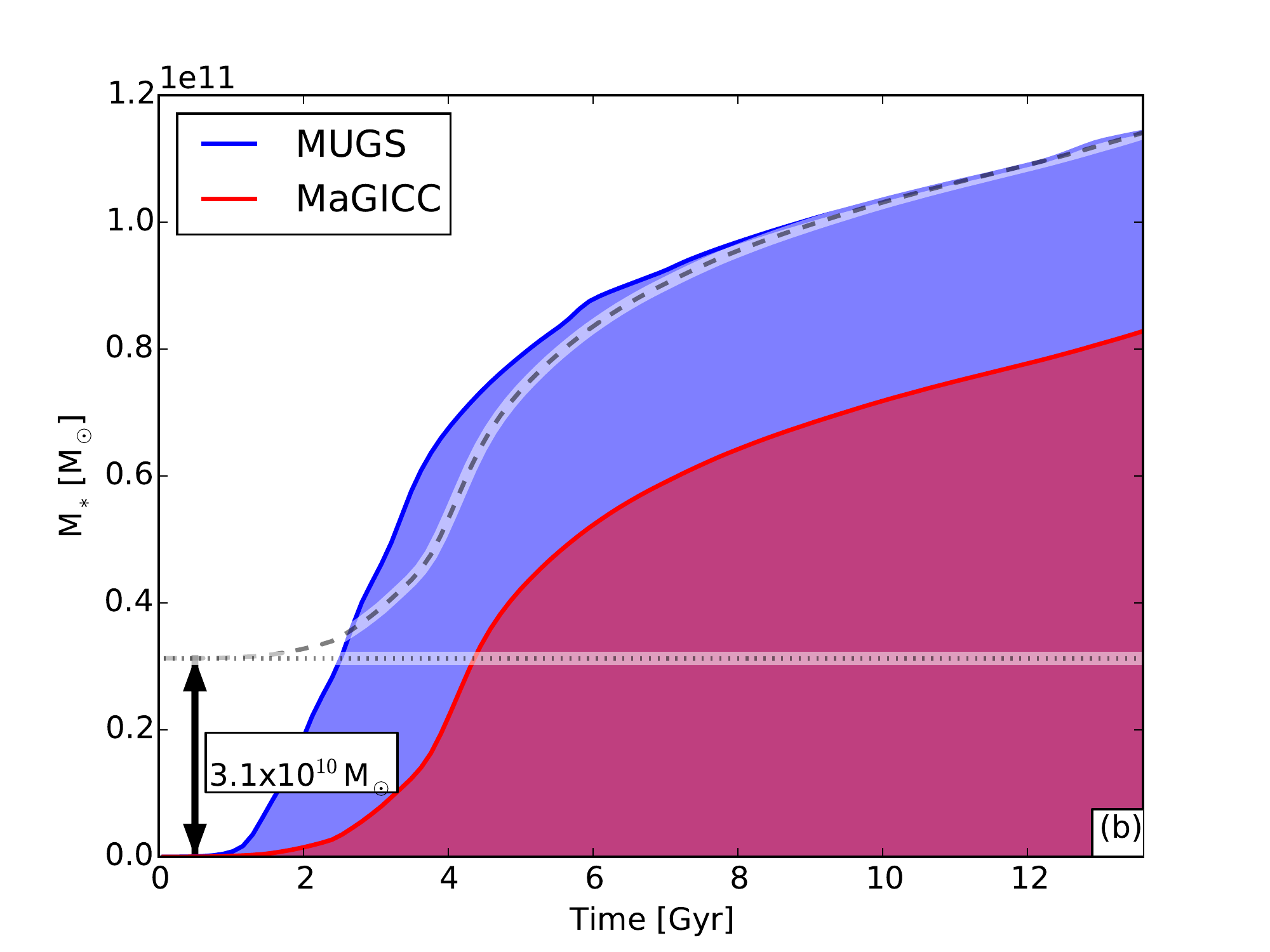}}  \\
     
      \hline   
  MUGS & MaGICC  \\   
      \hline 
      
    \parbox[c]{.35\textwidth}{\includegraphics[scale=.34, trim={0.cm 0 1.5cm 0cm},clip]{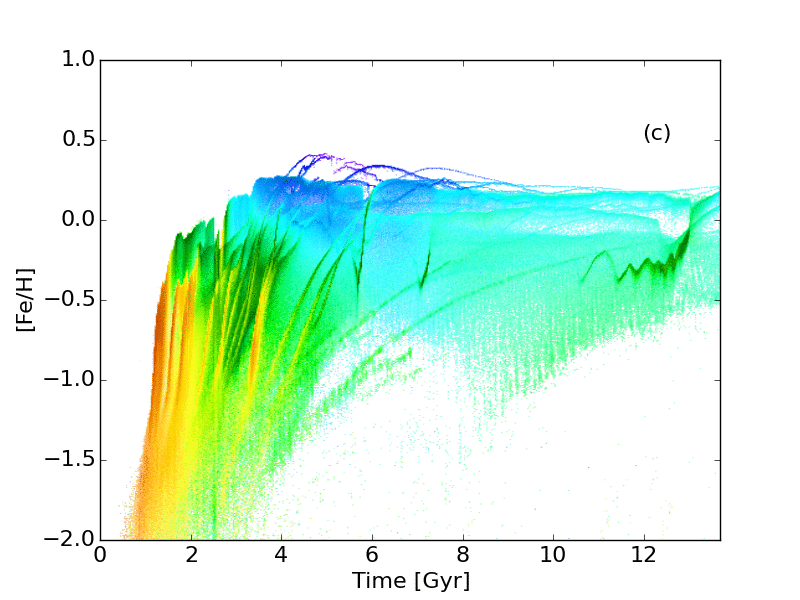}} & 
    \parbox[c]{.35\textwidth}{\includegraphics[scale=.34,trim={0.cm 0 1.5cm 0cm},clip]{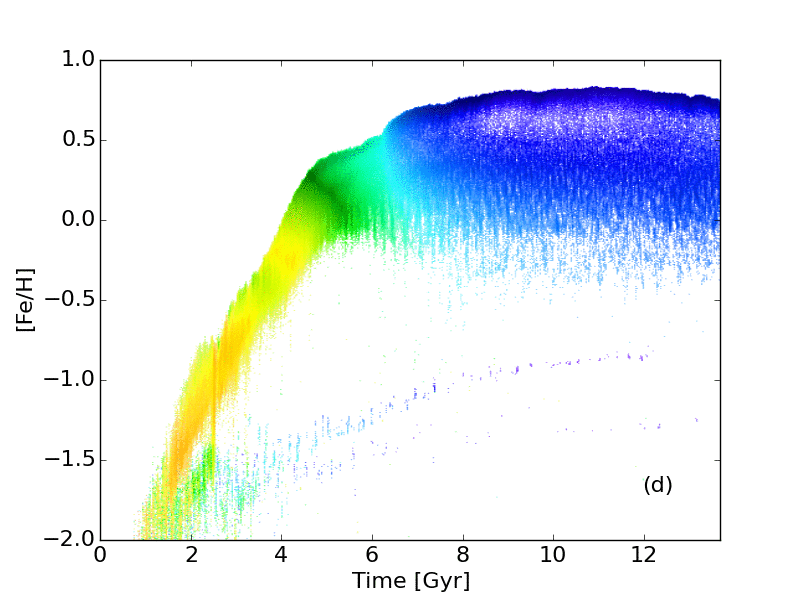}} & 
    \parbox[c]{.3\textwidth}{\includegraphics[scale=.34,trim={7cm 0 8.5cm 1cm},clip]{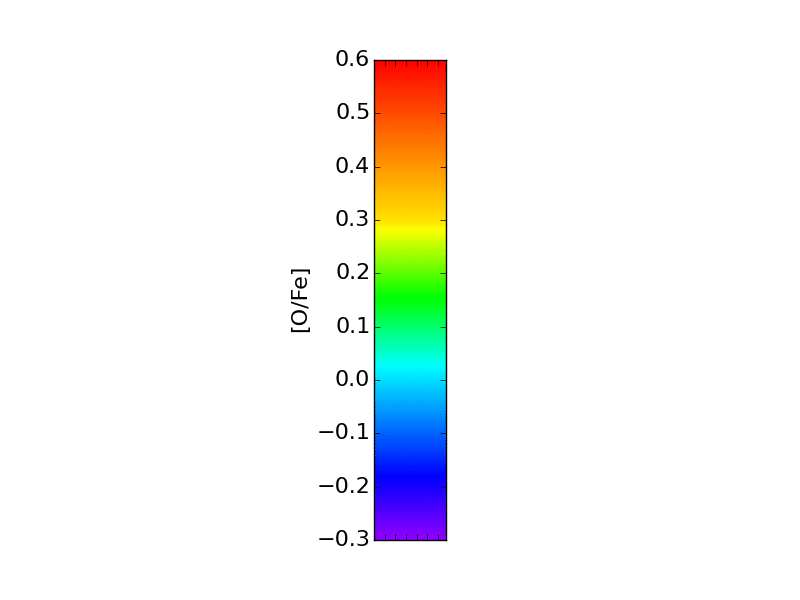}}\\
\\    
          
   \parbox[c]{.35\textwidth}{ \includegraphics[scale=.34, trim={0.cm 0 1.5cm 0cm},clip]{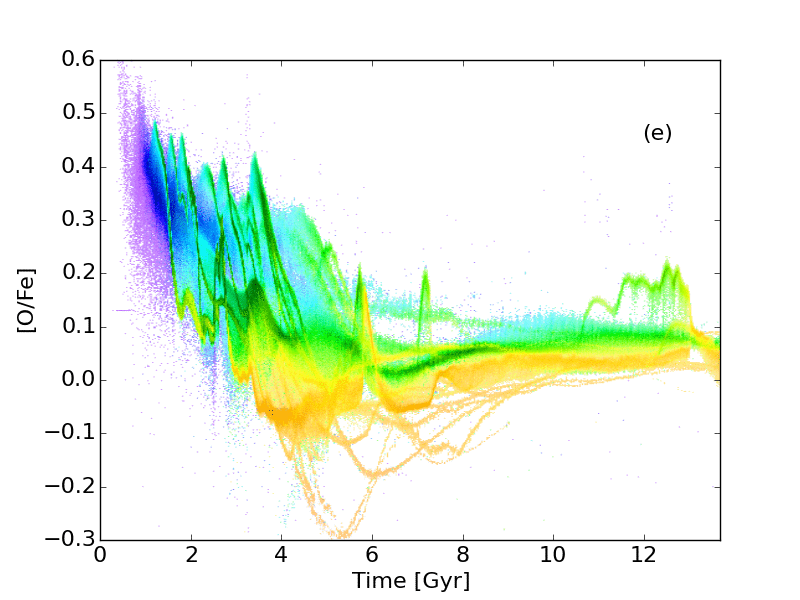}} & 
    \parbox[c]{.35\textwidth}{\includegraphics[scale=.34,trim={0.cm 0 1.5cm 0cm},clip]{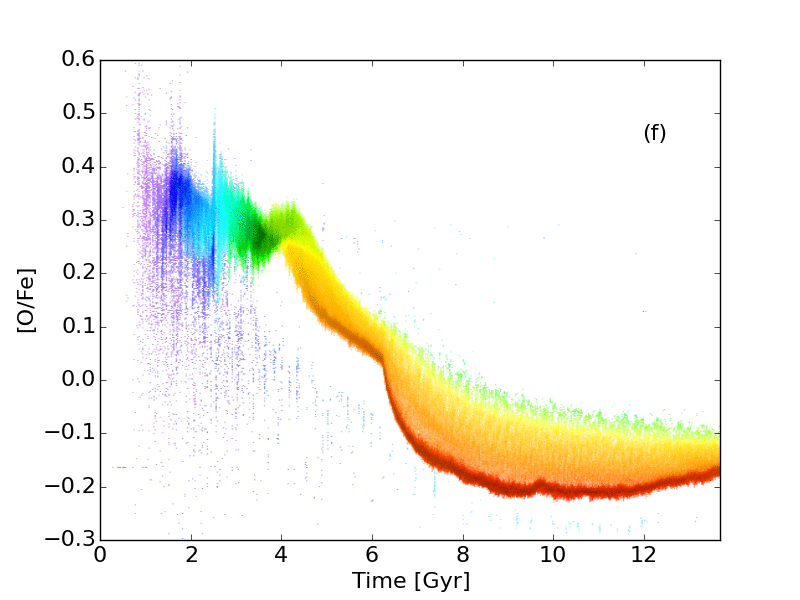}} & 
    \parbox[c]{.3\textwidth}{\includegraphics[scale=.34,trim={7cm 0 8.5cm 1cm},clip]{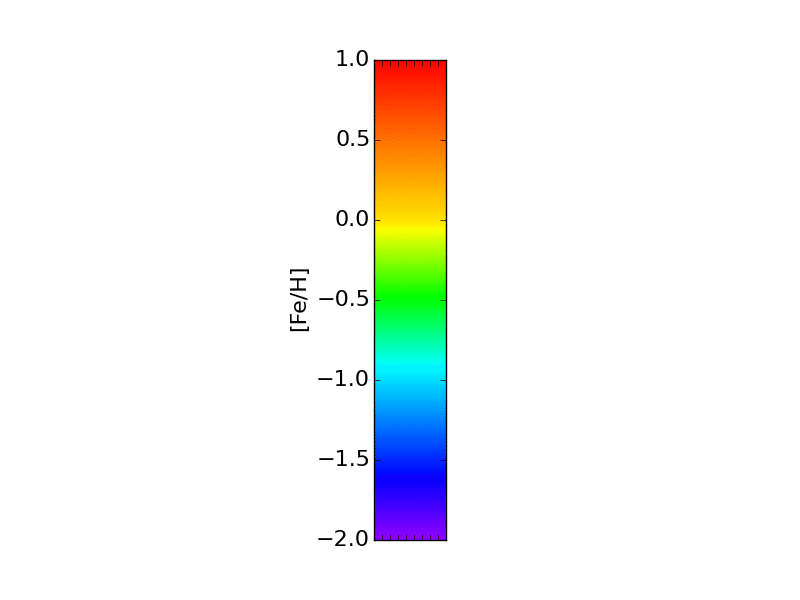}}
\\
    
   \parbox[c]{.35\textwidth}{ \includegraphics[scale=.34, trim={0.cm 0 1.5cm 0cm},clip]{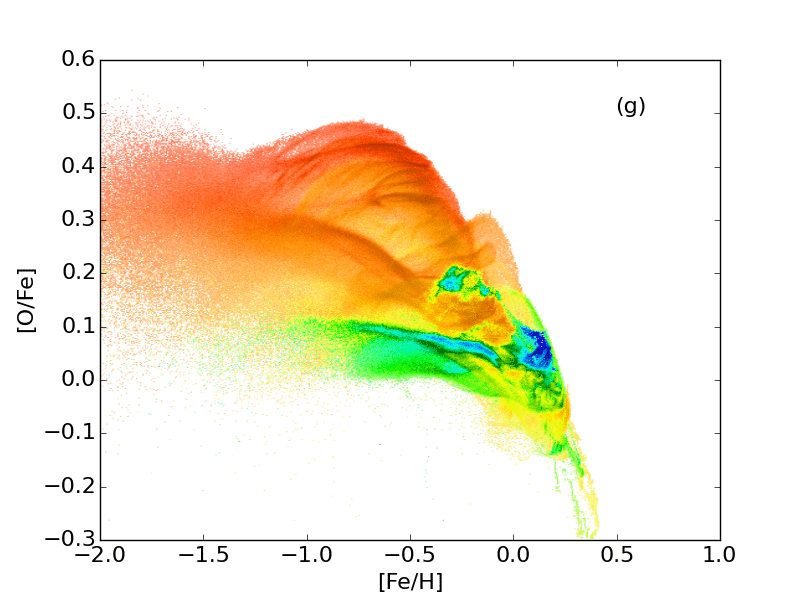}} & 
    \parbox[c]{.35\textwidth}{\includegraphics[scale=.34,trim={0.cm 0 1.5cm 0cm},clip]{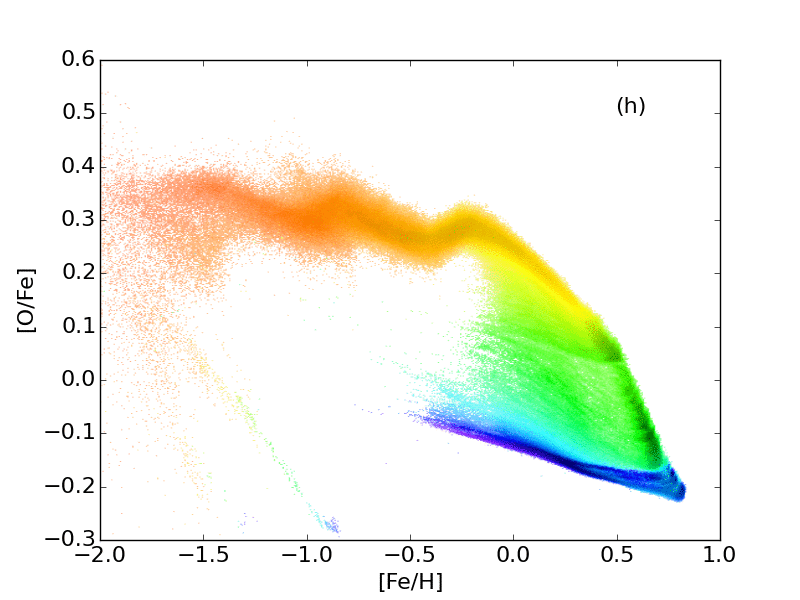}} & 
    \parbox[c]{.3\textwidth}{\includegraphics[scale=.34,trim={7cm 0 8.5cm 1cm},clip]{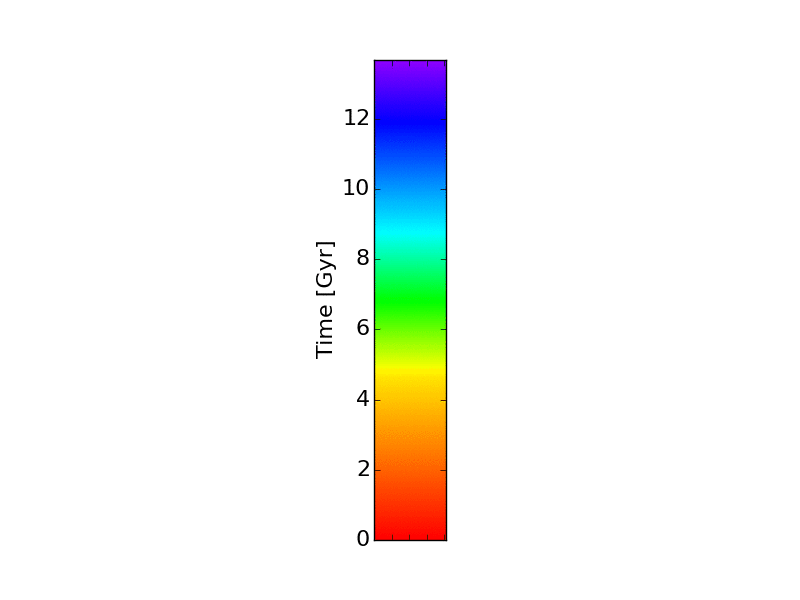}}\\

\end{tabular}
\caption{The evolution of galaxy g15784. The comparative star formation history and stellar mass growth of the two galaxies (top row), the AMR, the time-[O/Fe] evolution and the [Fe/H]-[O/Fe] distribution (next three rows). The bottom three rows are coloured by [O/Fe], [Fe/H] and time respectively, and the darkness of the   colour is the histogram weighted density of stars. We have used histogram equalization  in order to emphasize structure in both high-density and low-density parts of the parameter spaces. All plots were produced for the galaxy at z=0 ($t=13.7$~Gyr). Panel (a) is the comparison between the MUGS and MaGICC star formation histories. Panel (b) shows the stellar mass growth of MUGS and MaGICC. The dashed line follows the mass growth of the MaGICC galaxy but displaced by the difference between the final stellar mass of the two galaxies. The other panels are described in detail in the text.}
\label{Fig:allgalallplots}
\end{figure*}

Figure \ref{Fig:allgalallplots} shows the star formation history and chemical properties (time-[Fe/H], time-[O/Fe] and [Fe/H]-[O/Fe]) of g15784 in both MUGS and MaGICC. We use [Fe/H] as a proxy for metallicity in order to mimic observations, as it is very often the iron abundance which is used to trace the metallicity, rather than any other element \citep[e.g.][]{Haywood2013}. 

In order to generate the chemical evolution distributions,  we produced  two-dimensional histograms of the chemical evolution, (time, [Fe/H], [O/Fe]) and coloured the distribution according to the parameter not given in the x and y axes (for example, the time-[Fe/H]  plot is coloured according to the [O/Fe] value). The darkness of the colour is a function of the number of particles in each bin. In order to ensure the maximum contrast and to bring out the substructure, we  have used histogram normalisation on each individual figure. This has the advantage of picking out the detailed structure of the galaxy but at the cost of a consistent intensity scale across the different figures. 

As a first order approximation, the chemical evolution of a galaxy is a play-off between star formation, which enriches the ISM, infall, which dilutes it, and outflows which eject gas from the galaxy. However, these processes are not independent, as gas is the fuel which drives star formation and the rate of star formation is related to the gas surface density by the well known Schmidt-Kennicutt relation, 

\begin{equation}
\Sigma_{SFR} \propto \Sigma_{gas}^{k},
\end{equation}
\noindent
where $\Sigma_{SFR}$ is the star formation rate,  $\Sigma_{gas}$ is the gas surface density and $k$ is a constant (\citet{Schmidt1959}, 1.4 after \citet{Kennicutt1998}). Thus, the amount of cold gas present and the star formation rate are closely linked, as gas and star formation play off against one another to mold the chemical evolution of galaxies. Stars also generate various feedback processes which affect the properties of the gas, a considerable amount of which, rather than being cold, is in the warm circumgalactic medium or in the hot halo \citep{Sommer2006}. Further, as the galaxy is a diffuse object comprised of various components (disc, bulge, halo, satellites, etc.), it is unsurprising that the chemical distribution of stars is rich and complex. 

	The most obvious point to take from Fig. \ref{Fig:allgalallplots} is that the MUGS galaxy is replete with substructure, while the MaGICC galaxy is not. Observations, such as APOGEE, \citep{Hayden2015} do not show such fine structure, but these are limited by observational errors (see Section \ref{Sec:Obs}) which may hide considerable details. 
	
	The galaxy in both MUGS and MaGICC has the same initial conditions but a different star formation history (top row) and chemical evolution (other rows) due to the influence of feedback.    All the left hand panels in Fig. \ref{Fig:allgalallplots} show significant amounts of substructure. This can only arise if star formation is occurring in relatively isolated regions. In the rest of the paper we will dissect the galaxy and identify the origin of substructure. However, much of the filamentary sub-structure comes from satellites which have merged hierarchically with the host, and from others that have not yet merged. Further, we can expect differences due to distance from the centre of the galaxy, and bulge/disc/halo identification. 

\subsubsection{Star formation Rate}
Despite having the same initial conditions the MUGS and MaGICC feedback implementations produce very different star formation histories (see Fig. \ref{Fig:allgalallplots}, panels (a) and (b)). The early radiative feedback delays the beginning of the peak in star formation in g15784 for  around 2-3 Gyrs in MaGICC, with the star formation strongly suppressed for the first 3-4 Gyr. The peak in star formation in MaGICC takes place at z$\sim$1.5, which is 1 Gyr later than the peak in the cosmic star formation history \citep[e.g][]{Madau2014}, and is  due to the early feedback. This lack of early star formation means that the stellar mass in MaGICC is 72\% the stellar mass in MUGS. The MaGICC galaxy takes 1.5 Gyr longer to assemble half its final stellar mass, making the galaxy `younger'. The lack of early star formation results in a thinner galaxy disc, and smaller spheroidal component \citep{Stinson2013}.  Enhanced feedback essentially means that only dark matter halos of considerable mass can efficiently form stars. The higher feedback in smaller halos inhibits the formation of stars.  
 
The MUGS galaxy shows a number of peaks in star formation before 4 Gyr, which are the result of interactions between the host and its satellites causing starbursts. The absence of star formation in low mass objects in MaGICC means there are fewer dense objects to interact with the host (see \S \ref{Sec:Sat}). Although dark matter subhalos are present in MaGICC, the dense inner regions caused by star formation are absent. This means that the very low mass subhalos are missing from MaGICC. The mass distribution of subhalos in MUGS and MaGICC is not greatly dissimilar, particularly at higher masses. The principal difference is the baryons rather than the dark matter.  There are, however, numerous dark matter subhalos, which contain dark matter and gas, but no stars. 

In the MUGS version of g15784 there is another local maximum in the star formation rate at 5.5 Gyr, which is due to an interaction. This peak is followed by a brief fall in SFR in the disc because the interaction causes the gas to redistribute in the galaxy disc (discussed in more detail in Section \ref{Sec:DecompRad}). 

As the stars produce metals to enrich the gas and subsequent generations of stars, the difference in early star formation has a considerable impact on the early time enrichment. A more gradual star formation rate will result in slower enrichment (when diluting in the same amount of gas). This has a direct consequence on the age-metallicity distribution. In MUGS, the ISM enriched very rapidly (2 dex in less than 1 Gyr for the outer envelope of the distribution). This leaves us with the characteristic `handgun' form of the MUGS AMR. The more gradual rise in the SFR in MaGICC leads to a more slower increase in the metallicity of the ISM (the upper envelope takes approximately 3.5 Gyr to rise from -2 to 0.5 dex). The rising arm flattens only at 5 Gyr, almost 3 Gyr later than in MUGS. This delayed star formation can also be seen in the evolution of [O/Fe], which takes longer to reach its minimum value. The low rate of star formation at early times is due to the more energetic and earlier feedback that inhibits star formation in low mass objects. Clearly, while the potential of the MaGICC galaxy is shallow, the feedback is strong enough to considerably reduce the SFR at early times (1-4 Gyr) compared with MUGS. As the dark matter halo grows,  the galaxy in MaGICC becomes able to more efficiently form stars. Even so, even at later times the star formation rate efficiency is five times lower in MaGICC than in MUGS. At early times (before 4 Gyr) MaGICC is 25 times less efficient at forming stars. At later times the two SFHs are very similar, with very similar star formation rates for a given time. 

\subsubsection{time-[Fe/H]}
The strong suppression of star formation at early times and in low mass objects has a considerable impact on the metallicity evolution of MaGICC compared to MUGS. 

The stars are considerably more metal rich in the MaGICC run than in MUGS. The total mass of oxygen formed in the entire simulation volume (accounting for both gas and stars) at z=0 is over two times higher in MaGICC. The ratio in oxygen mass per unit of stars formed exceeds 2.6. This implies that MaGICC stars produce significantly more metals than MUGS stars, i.e., that it is not a matter of the distribution of metals in the galaxy, but a greater net production per unit of stellar mass formed. This is a result of the use of the Chebrier IMF \citep{Chabrier2003} in MaGICC, which produces more high mass stars than the Kroupa \citep{Kroupa1993} IMF used in MUGS. For example, the IMF used in MUGS generates 4 times fewer stars with masses greater than 8~M$_\odot$. 

Although the stars are twice as metal rich over all in MaGICC, the gas is {\it eight} times more metal rich than the stars. This suggests that ejection of metals into the warm/hot gas component is more efficient, and metals are not locked up in stars to the same degree as in MUGS. The rate of metal-dependent cooling will therefore differ in the two simulations, but the effect of this on the star formation rate is expected to be dwarfed by the  dynamical influence of early radiative feedback. 

The MUGS galaxy shows a trend from low $Z$ for the very oldest stars to high $Z$ for the youngest stars. However,  the metallicity saturates fairly quickly in the history of the galaxy to between 1.5 and 1.2 times solar metallicity. The metallicity increases by 3 dex in the first 3 - 4 Gyr and then the upper envelope of the distribution is essentially flat, or even shows slight dilution at later times with a peak metallicity at 4 Gyr.  The initial rise in MaGICC is considerably slower, enriching from -2 to 0 dex over the first 4 Gyr of the simulation while in MUGS it takes just over 1 Gyr. In MaGICC the peak metallicity in the bulge (the most metal rich component) is at 11 Gyr. 

Observational data from \citet{Haywood2013} for stars in the solar vicinity shows a more gradual slope for old stars than MUGS, but faster than in MaGICC. This implies that `reality' is somewhere between MUGS and MaGICC, with a few caveats. g15784 is not the Milky Way, and can be expected to diverge significantly in the details of its history. Further, the Haywood data is local data and Fig. \ref{Fig:allgalallplots} shows all stars within the virial radius of g15784.

There is spread of at least 1~dex  (this can rise to as much as 2 dex) in the metallicity of stars at any given time in MUGS. Even though we see a wide spread in metallicity between 1 and 3 Gyr much of this apparent spread is due to the histogram normalisation procedure (see \S \ref{overview}). The standard deviation of the metallicities of stars in the different age bins varies between 0.4 dex at 3 Gyr to 0.3 dex at 12 Gyr. At a given metallicity the age range of stars is also large, with standard deviations ranging from 1.8 Gyr at -2 dex to 3.5 Gyr at -0.3 dex and 2.8 Gyr at 0.14 dex. 
MaGICC, however, demonstrates narrower scatter in metallicity with age at early times (0.18 dex at 3 Gyr but 0.3 dex at 12 Gyr) and a rapidly increasing scatter in age with increasing metallicity (0.8 Gyr at -2 dex,  3 Gyr at 0.14 dex)

Our visualization approach is designed to emphasize substructure and so may exaggerate apparent differences, at first glance. There are, however, notable difference between the two simulations and our plots demonstrate this difference well.   \citet{Gibson2013} notes that although the AMRs in MUGS and MaGICC appear different the metallicity distribution functions (MDF) are not dissimilar.

Metallicity is sometimes considered as a rough proxy for the age of stars, and so any scatter in the age-metallicity relation must be understood and taken into consideration.  The spread in metallicity is the smallest for young stars, while the spread in age is smallest at  low metallicity. It is evident that any hope using $Z$ to recover stellar age would introduce immense errors using all stars in the galaxy.

The MUGS AMR contains many streamers and rich substructure, but the only evidence of substructure in MaGICC is a bifurcation between the upper limit of the AMR and the skirt beyond 6 Gyr, with a large gap (this is a gradual 'u' shaped feature with a FWHM $\sim$0.3 dex at 10 Gyr) between the two sequences. A similar gap exists in MUGS, but it is much smaller (0.15 dex). The existence of the two sequences is a result of the contributions of two different galactic components (the bulge and disc) and will be discussed in greater detail in \S  \ref{Sec:Decomp}.

The uniformity of the plot is also evident in the colour table in MaGICC, which changes gradually from alpha overabundance to lower alpha with time, without the peaks and undulations seen in the MUGS galaxy. The age-metallicity distribution is extremely tight for the whole evolution, particularly at early times.  The same behaviour was seen in the dwarf late-type disks shown in \citet[][Fig 2; upper 2 panels]{aPilkington2012}, also, \citet{Gibson2013}. \citet{aPilkington2012} showed that the AMR scatter is very dependent on the degree of metal diffusion. In both MUGS and MaGICC, there is a sharp upper limit on [Fe/H] at a given age, which should be kept in mind when comparing the AMR to observations.

A significant deviation from monotonicity can be observed in the metallicity evolution in MUGS (MaGICC is more monotonic). In various tracks the metallicity of some of the substructures can move from higher to lower metallicity. This implies that star forming regions are acquiring new low-metallicity gas, and/or that the locus of star formation is moving into less-enriched regions.

\subsubsection{time-[O/Fe]}

The [O/Fe]-age distribution is more tightly correlated than the metallicity in both MUGS and MaGICC  (around 0.1 and 0.2 dex at 12 Gyr for MUGS and MaGICC respectively), although non-monotonic features remain. This distribution was also discussed in \citet{Miranda2015b}. Early star formation in MUGS shows a wide spread in [O/Fe] of  around 0.4 dex at 4 Gyr, compared to 0.07 dex in MaGICC.  \citet{Stinson2013b} showed that mono-abundance populations show less than 1 Gyr spread in their ages. This is consistent with recent measurements in the Milky Way for several $\alpha$ elements \citep{Haywood2013}.  The general trend is one of decreasing [O/Fe] with time for t$<$ 5 Gyr, and an almost flat relation thereafter. This implies that [O/Fe] is only a good timer during the early phase of galaxy evolution, which corresponds to the rapid star formation phase (top left panel). The transition between the fast evolution and flat phases is reasonably sudden, leading to a kinked [O/Fe] evolution, with a knee at around 5 Gyr.  \citet{Haywood2013} shows this same feature in the Milky Way, and \citet{Snaith2014} identify this as the location of a sudden transition from rapid star formation to lower rates of star formation. In MUGS, this change from high SFR to low SFR is more gradual than found for the Milky Way in \citet{Snaith2014,Snaith2014b}, but the shallow time-[O/Fe] evolution does correspond to the low SFR phase. This property, however, will also be dependent on the SNIa formalism which starts to dominate the IMF on a similar timescale. 

In MaGICC, the age-[O/Fe] distribution as a more `sickle' shape, where the [O/Fe] value  continues to fall even after the peak of the SFR. The tight fit in the chemical evolution is also evident in age-[O/Fe] evolution, and the kink in the age-[O/Fe] co-coincides with the beginning of the bifurcation in the AMR discussed above (\S \ref{Sec:DecompDef}). This takes place approximately 1 Gyr after the peak in the SFR, which is the typical SNIa time delay. This makes the onset of SNIa very clear from the star formation history. 

Due to the importance of substructure in the early history of the MUGS galaxy, the spread in [O/Fe] is greater at early times.  The opposite is true in MaGICC, because of the absence of substructure. {\it The [O/Fe] evolution shows events in the assembly history of MUGS much more clearly than the AMR. This same effect can be seen in local Milky Way data, \citep[e.g.][]{Haywood2013, Snaith2014, Snaith2014b}.}  The Milky Way also shows a tighter correlation between age-[O/Fe] at early times \citep{Haywood2013}. See \citet{Haywood2015} for a detailed discussion of the early time SFH of the Milky Way.

Some of the galactic chemical enrichment tracks in MUGS are almost vertical (such as at 6 Gyr where the metallicity jumps over 0.5 dex in a few Myrs), indicating a very rapid enrichment. Over brief periods of time, the [O/Fe] value rises but soon falls back to the previous value (this can be seen at 6 Gyr, where the [O/Fe] value rises from around 0 dex to 0.2 dex and falls back to 0 dex in around 1 Gyr). This implies very rapid star formation, where the ISM is enriched by CCSNes. It is only after a delay do the SNIa add iron to the ISM, thus bringing the value down again. These [O/Fe] episodes coincide with peaks in the SFR, strengthening this idea. The feature at 6 Gyr is due to a small starburst which takes place just before it, and the SFR peak corresponds to the rising arm of the [O/Fe] peak, the falling arm is due to the the delayed SNIa. Because these peaks are due to interactions and starbursts, which do not occur in MaGICC, the time-[O/Fe] in MaGICC is more featureless.

An important caveat to this analysis, is, however, that Gasoline \citep{Wadsley2004} does not use metallicity dependent yields, meaning that some behaviour in the [O/Fe] evolution is lost \citep{Haywood2013,Snaith2014}. Gasoline uses the  $Z/Z_{\odot}=1$ yields from \citet{WW95} for stars of all metalicities. 

\subsubsection{metallicity-[O/Fe]}

The lower left hand panel shows how [O/Fe] evolves with metallicity in MUGS. The [O/Fe]-[Fe/H] distribution is the easiest to compare with observations. Calculating ages of stars from observational data is difficult, and some of the best age related data shows uncertainties on the order of 1 Gyr  \citep[e.g.][]{Chaplin2014,Epstein2014,Haywood2013,Ramirez2013}, even for the Milky Way.  This plot, however, is not as easy to dissect as the other projections.  We do see three large and distinct evolution paths, with one oxygen-rich and one intermediate path, both of which are old, along with a young oxygen-poor path. These apparently separate evolutions are due to the different components of the galaxy, and will be discussed in \S \ref{Sec:Decomp}. Interestingly, the youngest stars are not the most metal rich. We also see a distribution of young stars with -1.0$<$[Fe/H]$<$0.0 dex and around [O/Fe]=0.1 dex. This corresponds to the metallicity distribution of the gas disc at the current time, while the cluster of very young stars at [Fe/H]=0 and [O/Fe]=0.1 corresponds to the bulge. The gas sequence is very narrow in [O/Fe] and the spread in [Fe/H] is directly correlated with radius as expected.

The [Fe/H]-[O/Fe] distribution in MaGICC is  tight, with two narrow, distinct paths. One is at [O/Fe]=0.3, the other at [O/Fe]=-0.2, suggesting two separate regions of star formation. The spine of the upper path is due to the bulge, while the diffuse distribution and the lower sequence are due to stars in the disc of the galaxy. This tighter correlation is due to the higher feedback, which disturbs the gas and keeps the ISM well mixed. As with MUGS, the gas distribution overlaps the youngest stars.

\subsection{Decomposed AMR}
\label{Sec:Decomp}
The above analysis was based on all stars within the virial radius of the galaxy. In this section we subdivide the stars according to their component (halo, disc, satellite) and origin (in-situ, accreted, commuter). The relative sizes of these populations can be seen in Table \ref{Tab:parts}. It is worth noting that in each simulation we see considerable intermediate age star formation in the bulge, which is not found in the Milky Way, where the bulge tends to be older.  In MUGS between 50\% and 10\% of the star formation at any given time is in the bulge. This ratio is highest at early and late times. In MaGICC the bulge fraction of the total star formation rate is around 25\% at all times, falling off at later times. In the Milky Way the  total stellar mass fraction of the bulge is 10\%. 

\begin{figure*}
\centering
     \begin{tabular}{ccc}
        \hline        
         \multicolumn{3}{c}{MUGS} \\
        \hline
         \includegraphics[scale=.3,trim={0.cm 0 1.5cm 1cm},clip]{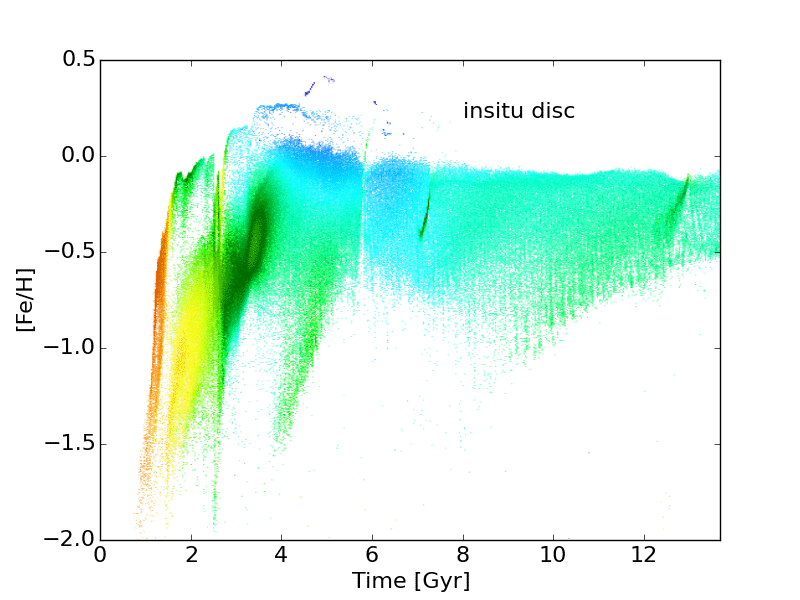} &
         \includegraphics[scale=.3,trim={0.cm 0 1.5cm 1cm},clip]{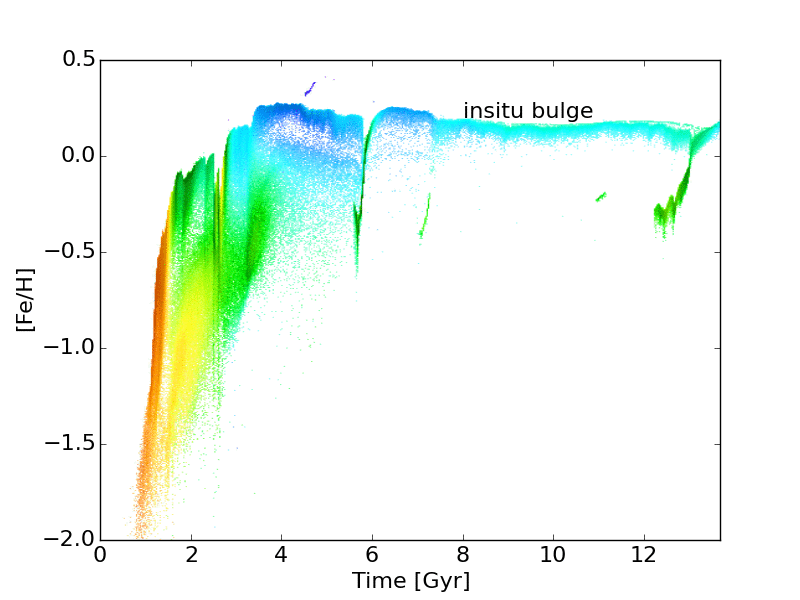} &
         \includegraphics[scale=.3,trim={0.cm 0 1.5cm 1cm},clip]{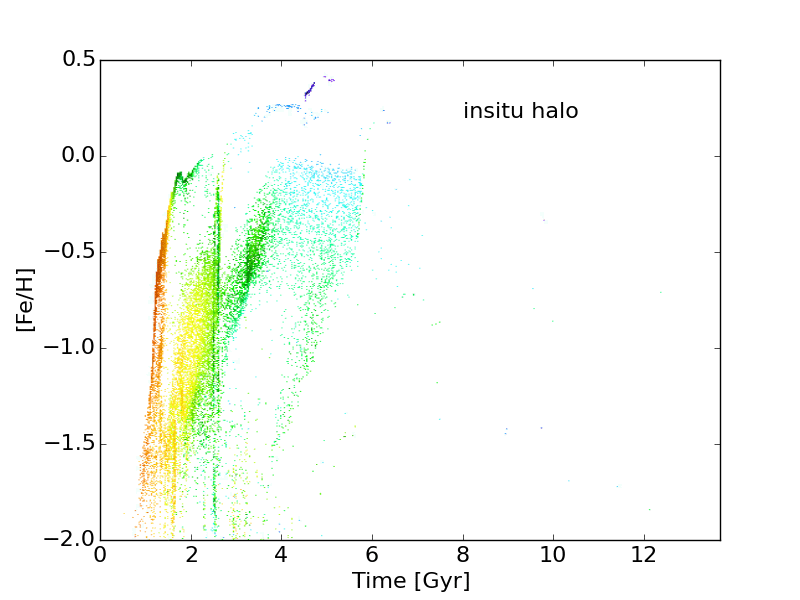} \\
         (a) & (b) & (c) \\
         \includegraphics[scale=.3,trim={0.cm 0 1.5cm 1cm},clip]{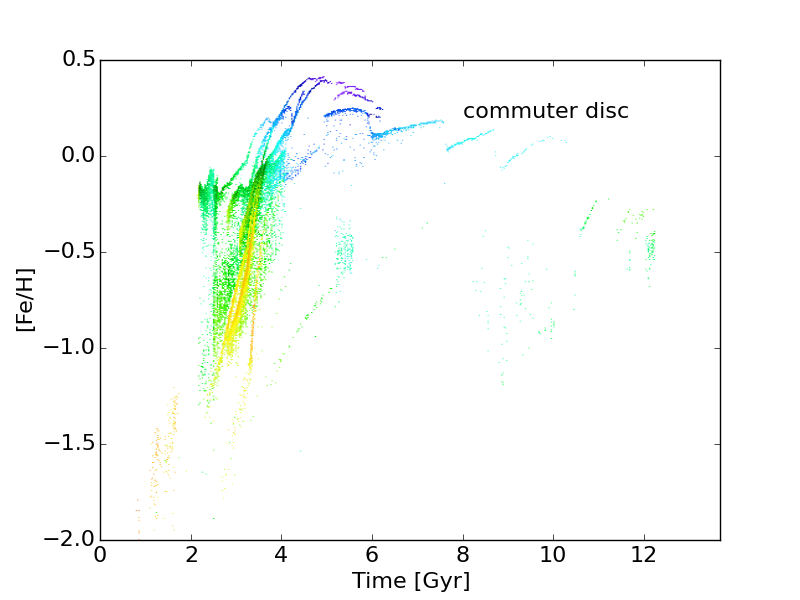}&
         \includegraphics[scale=.3,trim={0.cm 0 1.5cm 1cm},clip]{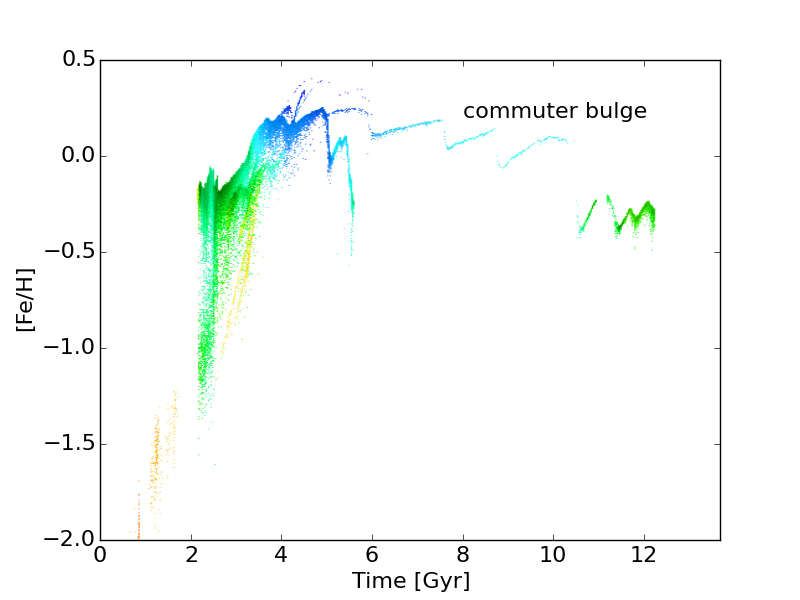}&
         \includegraphics[scale=.3,trim={0.cm 0 1.5cm 1cm},clip]{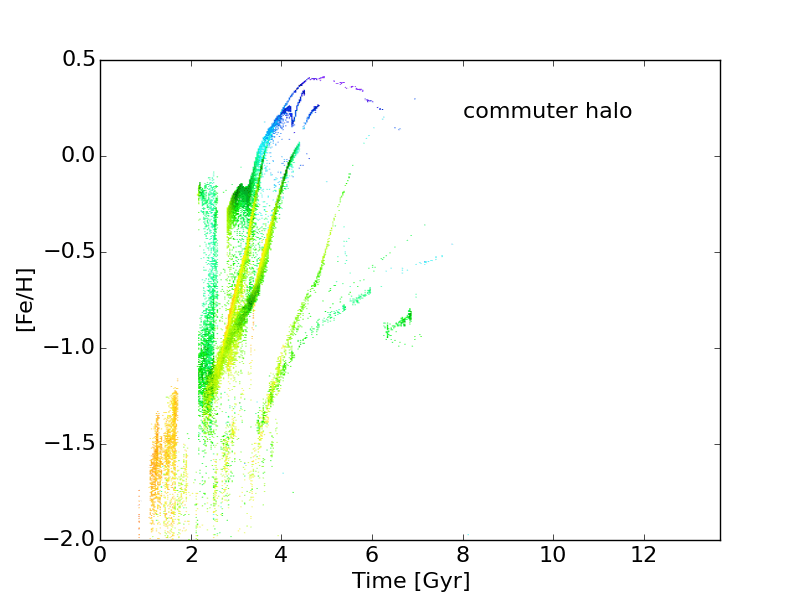}\\
         (d) & (e) & (f) \\              
         \includegraphics[scale=.3,trim={0.cm 0 1.5cm 1cm},clip]{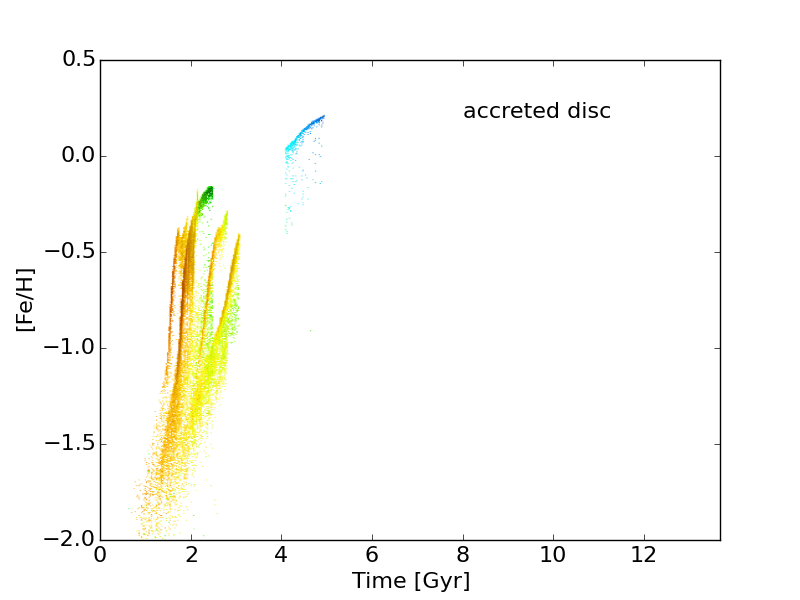}&
         \includegraphics[scale=.3,trim={0.cm 0 1.5cm 1cm},clip]{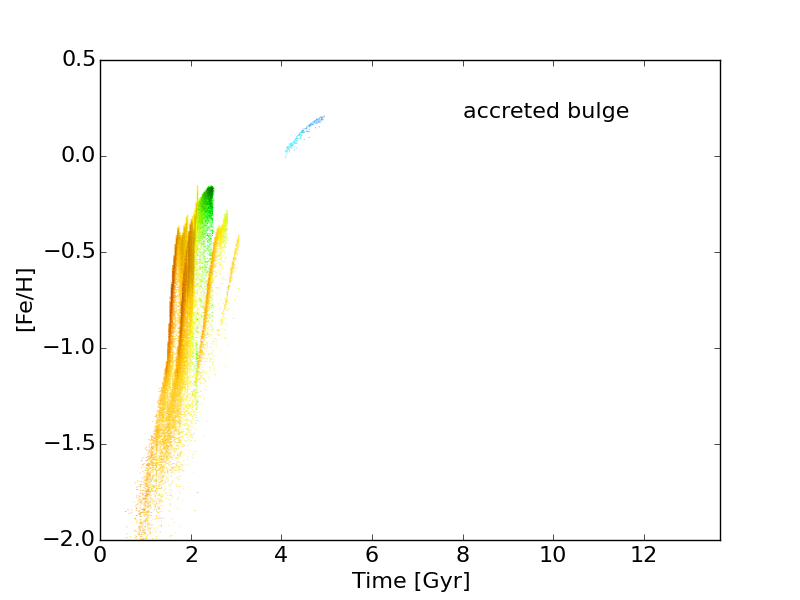}&
         \includegraphics[scale=.3,trim={0.cm 0 1.5cm 1cm},clip]{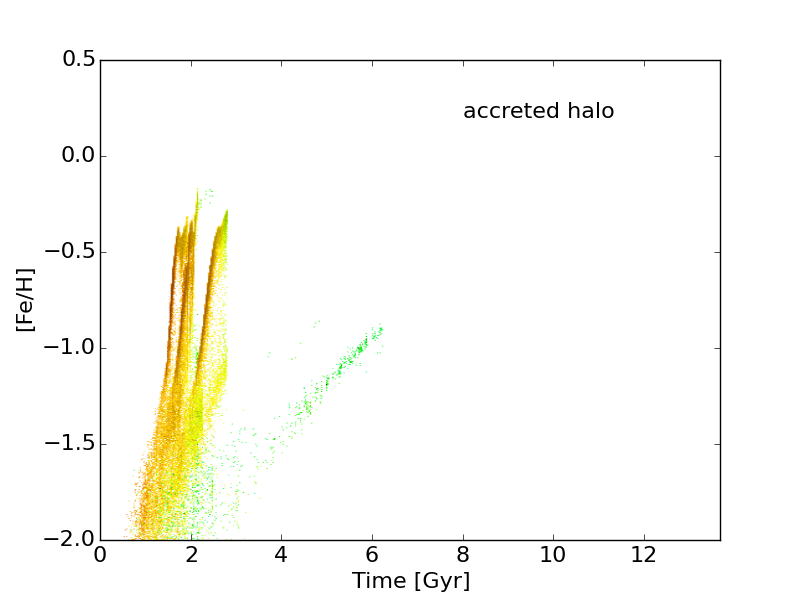}\\
         (g) & (h) & (i) \\  
         \hline
         \multicolumn{3}{c}{MaGICC} \\
         \hline
         \includegraphics[scale=.3,trim={0.cm 0 1.5cm 1cm},clip]{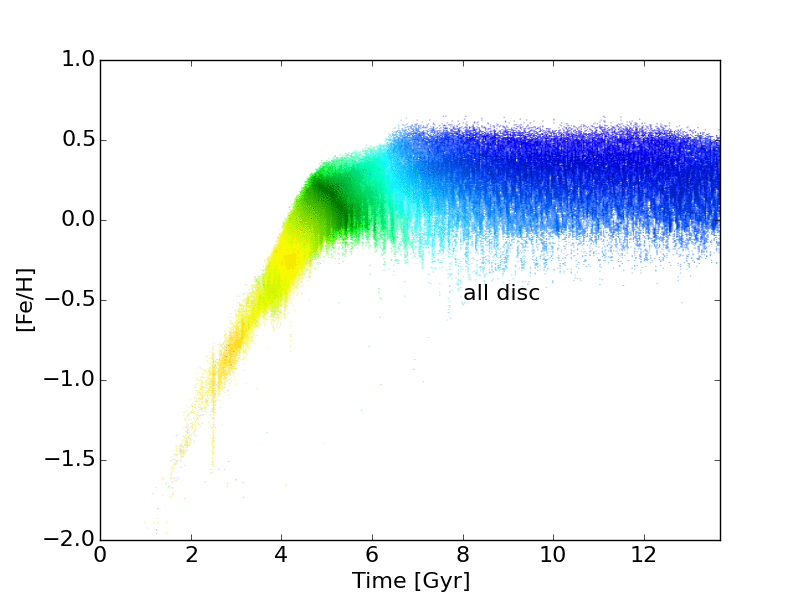} &
         \includegraphics[scale=.3,trim={0.cm 0 1.5cm 1cm},clip]{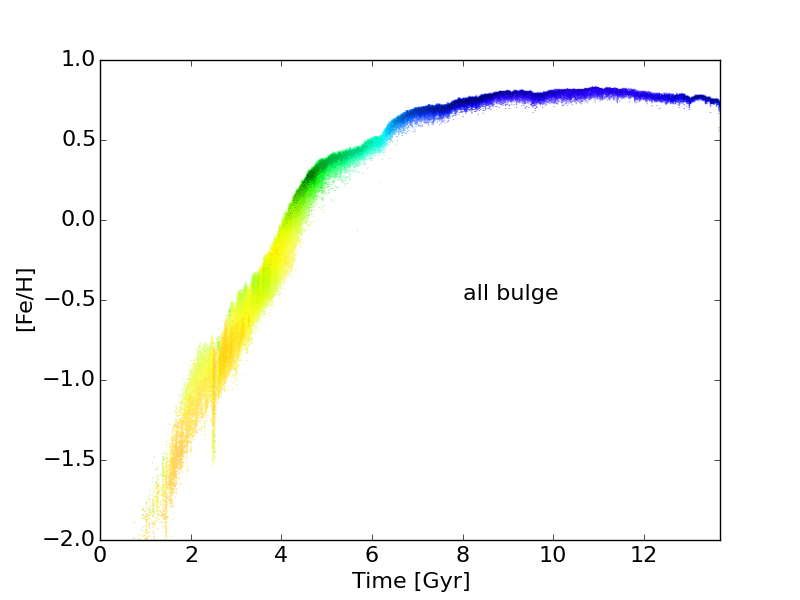} &
         \includegraphics[scale=.3,trim={0.cm 0 1.5cm 1cm},clip]{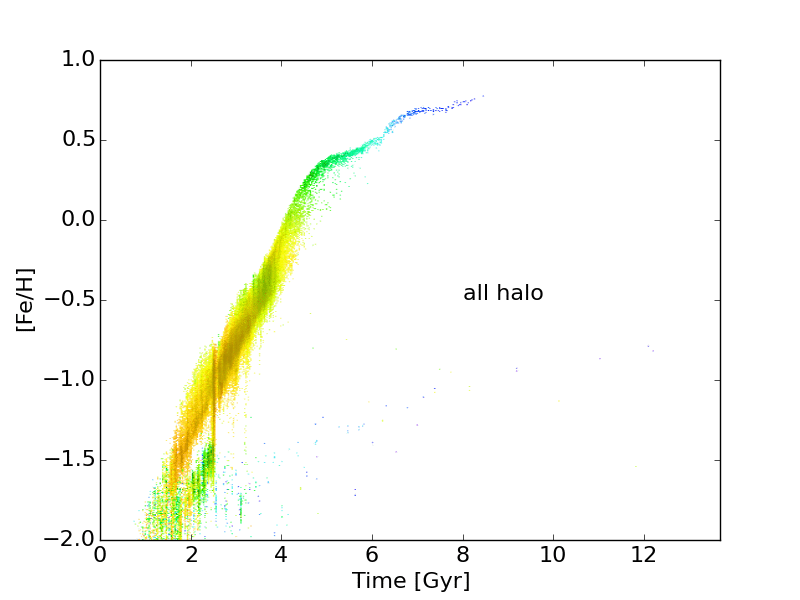} \\                                                    
         (j) & (k) & (l) \\ 
    \end{tabular}
\caption{The AMRs of the MUGS (top three rows) and MaGICC (bottom row) galaxy after decomposition into components. The left column (panels a,d,g) shows the insitu, commuter and accreted stars for the disc, while the centre (b,e,h) and right hand columns (c,f,i) show the same for the bulge and halo respectively. The colours are scaled the same as Fig. \ref{Fig:allgalallplots}, intensity is scaled by the histogram equalization approach. }
\label{Fig:decompose}
\end{figure*}

Fig. \ref{Fig:decompose} takes the stars in each of these populations and shows how the stars classified into each group evolves as a function of time and metallicity. The panels in this figure use histogram equalization (described in \S \ref{overview}) to make the substructure more apparent. In MaGICC we do not decompose the galaxy into in-situ, accreted or commuter stars because of the overwhelming dominance of in-situ stars (see Table \ref{Tab:parts}).

\begin{table}
\centering
\begin{tabular}{ccccccc}
 \hline
                 & all & disc$^1$ & disc$^2$  & bulge  & halo  & other   \\
\hline
\hline
                 \multicolumn{7}{c}{MUGS (N$_*$=2594942)}\\

 \hline
all             &  100.0 & 23.9 & 49.8 & 26.1 & 13.6 & 10.5 \\
\hline
\hline
no sats     & 80.4 & 23.8 & 39.7 & 26.1 & 7.4 & 7.3 \\
sats          & 19.6 & 0.05 & 10.1 & 0.0 & 6.2 & 3.3 \\
\hline
in-situ       & 57.3  &  19.6  &  31.2  &  19.1  &  1.9  &  5.1 \\
commuter & 16.9  &  2.1  &  8.7  &  2.9  &  3.5  &  1.8\\
accreted   & 25.8 & 2.2  & 9.95    & 4.1   & 8.1   & 3.7\\ 
\hline 
\hline
                 \multicolumn{7}{c}{MaGICC (N$_*$=2167946)}\\

\hline             
all & 100.0 & 31.0 & 52.6 & 19.3  &  8.8  &  19.3
 \\
\hline
\hline
no sats & 99.8 & 31.0 & 52.6 & 19.30 &  8.6  &  19.3 \\
sats & 0.2 & 0.0 & 0.001 & 0.0 & 0.2 & 0.0 \\
\hline
insitu & 98.4 & 30.9 & 52.3 & 19.2 & 8.6 & 19.0 \\
commuter & 1.50 &  0.1  &  0.27  &  0.06  &  0.9  &  0.2 \\
accreted & 0.05  &  0.0  &  0.0  &  0.0  &  0.1  &  0.00 \\
\hline                 
 \end{tabular}
 \caption{The percentage of star particles in MUGS and MaGICC in each different subset at z=0. ``All'' includes all stars in the galaxy or subset, ``no sats'' refers to removing all stars that lie in satellites, ``other'' accounts for all stars not contained in the rigorous definitions of the disc, bulge and halo defined in \S 3, ``disc 1'' is the disc defined in \S 3 with height and radial cut, ($|z|<$5 kpc and r$<$20 kpc. In order to avoid overlap with the bulge we also include an inner radial cut r$>$2 kpc), and ``disc 2'' is defined by angular momentum ratios alone.}
\label{Tab:parts}
\end{table}

Over half (57\%) of the stars within the virial radius of the MUGS galaxy are formed insitu, while in MaGICC this is over 98\% because of the strong suppression of star formation in low mass objects.  

Most of the accreted stars in MUGS were formed at early times (time $<$ 4 Gyr),  which is around the same time the stars in the halo formed. Commuter stars, by definition, formed after the satellite was accreted, and before it was disrupted. The presence of commuter stars at a range of different times implies that most of the satellites which have ever fallen into the host were accreted at early times, and also demonstrates the length of time it takes for a satellite to merge with a galaxy.

In the MUGS (MaGICC) galaxy 49\% (53\%) of the stars are in the `dynamical disc', 26\% (10\%) are in the bulge, 13\% (3\%) are in the halo, and the remaining 10\%  (19\%) are in the `other' category. `Other' stars are dynamically associated with the 'thick disc' and 'pseudo bulge' by the dynamical decomposition, but, at the resolution of MUGS and MaGICC, we do not trust the method to correctly distinguish these components. Chemically, they are transition regions between the bulge, halo and disc, with properties similar to each of the principle components. This implies that they are not distinct parts of the galaxy (in these simulations), but are a mixture of disc, bulge and halo stars falsely associated with other components.

In MUGS, commuter and accreted stars are present in similar amounts in the bulge and disc (between 15\% and 20\% of the stars in each component) but not the halo, which is dominated by accreted stars (60\% of halo stars are accreted stars and 26\% are commuter). It is worth noting that the population of stars which did not form in-situ, and which are no longer in satellites, is twice the size of the population currently in satellites. This implies that although the satellite galaxies we see are long lived,  the majority of satellites have been disrupted.  In MaGICC, there are over 7 times as many accreted or commuter stars as there are stars currently in satellites. Also relevant to this is the presence of two massive satellites which contain 12\% of the  total stellar mass at z=0 of the host galaxy between them. 

\begin{figure*}
\centering
     \begin{tabular}{ccc}
     \hline
     \multicolumn{3}{c}{MUGS} \\
     \hline
         \includegraphics[scale=.3,trim={0.cm 0 1.5cm 1cm},clip]{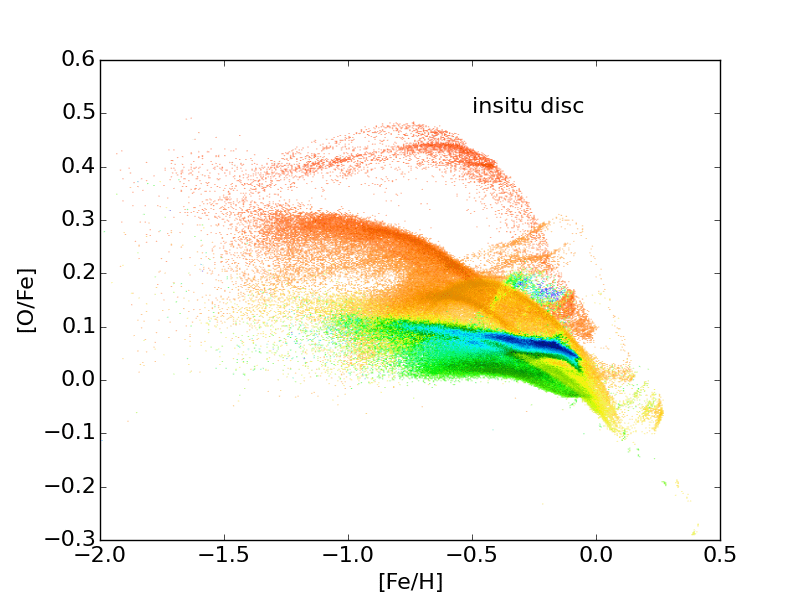} &
         \includegraphics[scale=.3,trim={0.cm 0 1.5cm 1cm},clip]{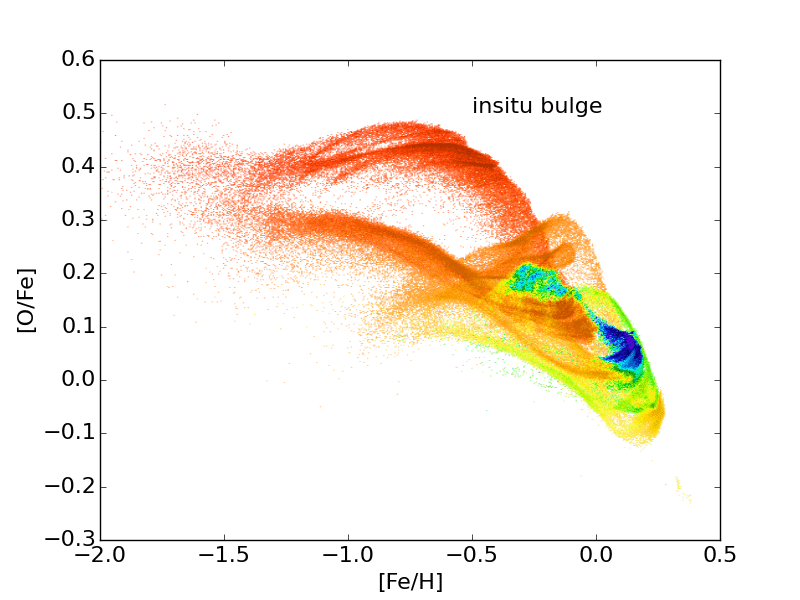} &
         \includegraphics[scale=.3,trim={0.cm 0 1.5cm 1cm},clip]{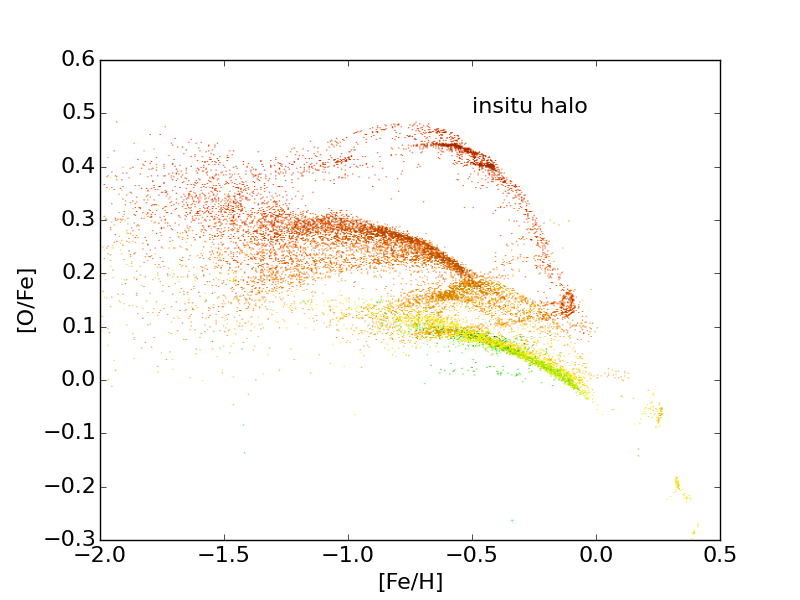} \\
         (a) & (b) & (c) \\
         \includegraphics[scale=.3,trim={0.cm 0 1.5cm 1cm},clip]{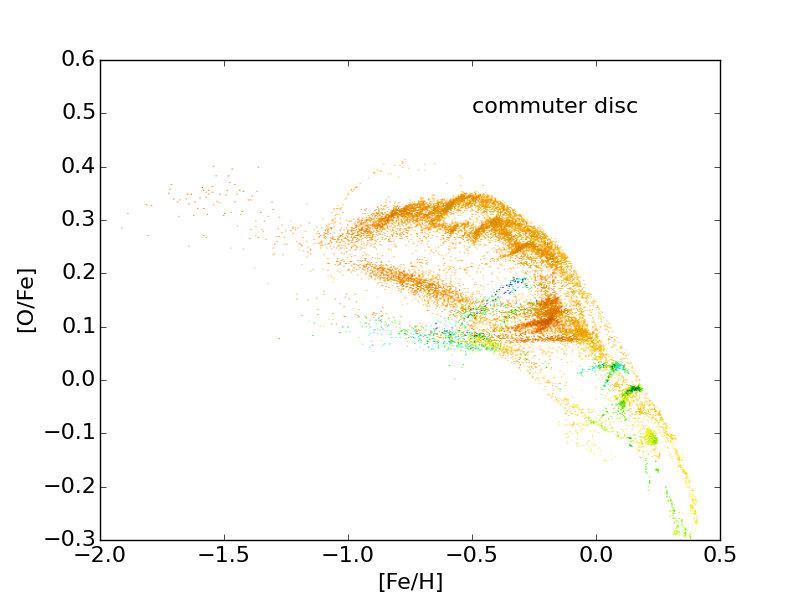}&
         \includegraphics[scale=.3,trim={0.cm 0 1.5cm 1cm},clip]{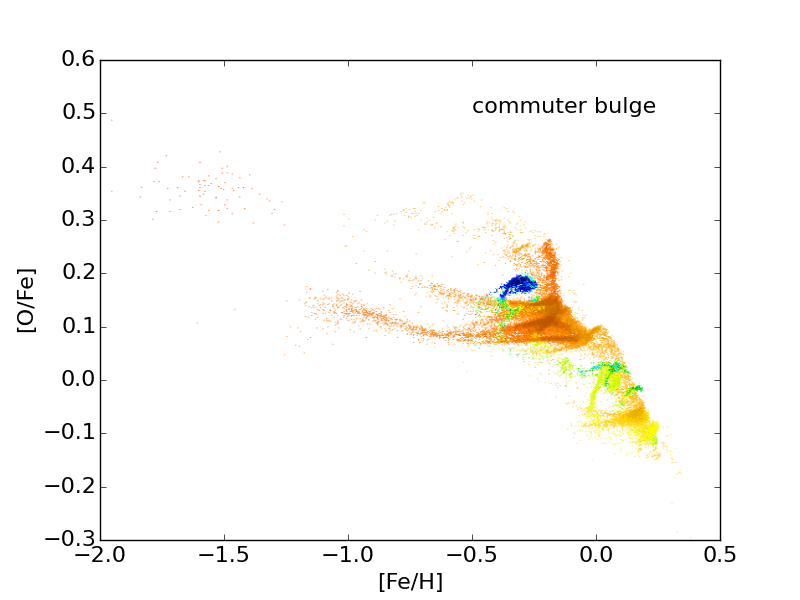}&
         \includegraphics[scale=.3,trim={0.cm 0 1.5cm 1cm},clip]{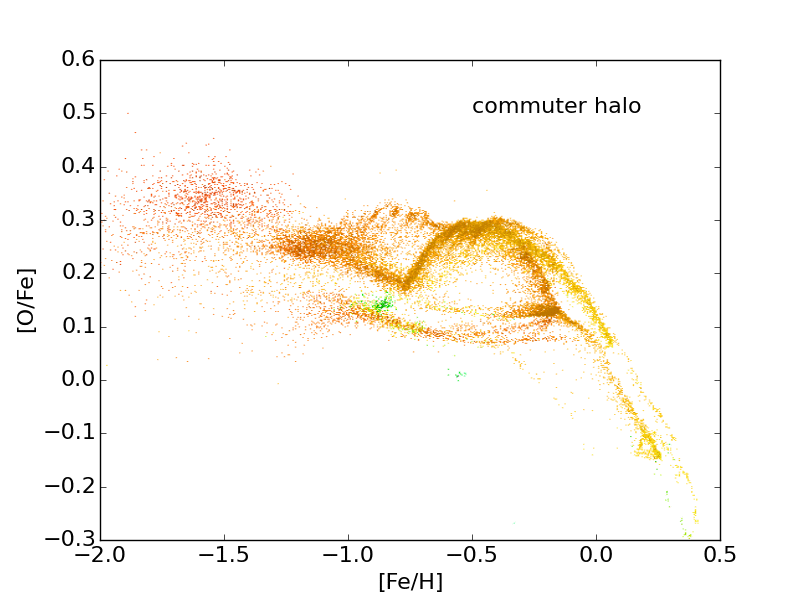}\\
         (d) & (e) & (f) \\              
         \includegraphics[scale=.3,trim={0.cm 0 1.5cm 1cm},clip]{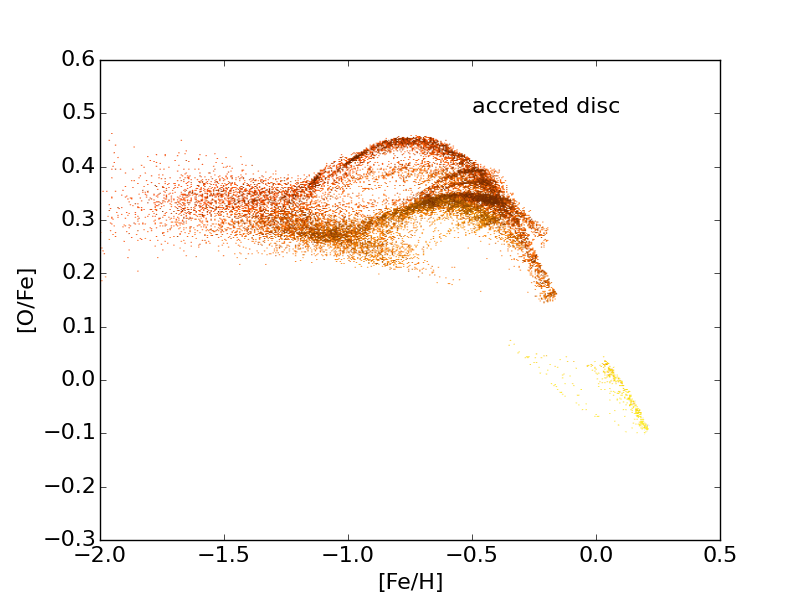}&
         \includegraphics[scale=.3,trim={0.cm 0 1.5cm 1cm},clip]{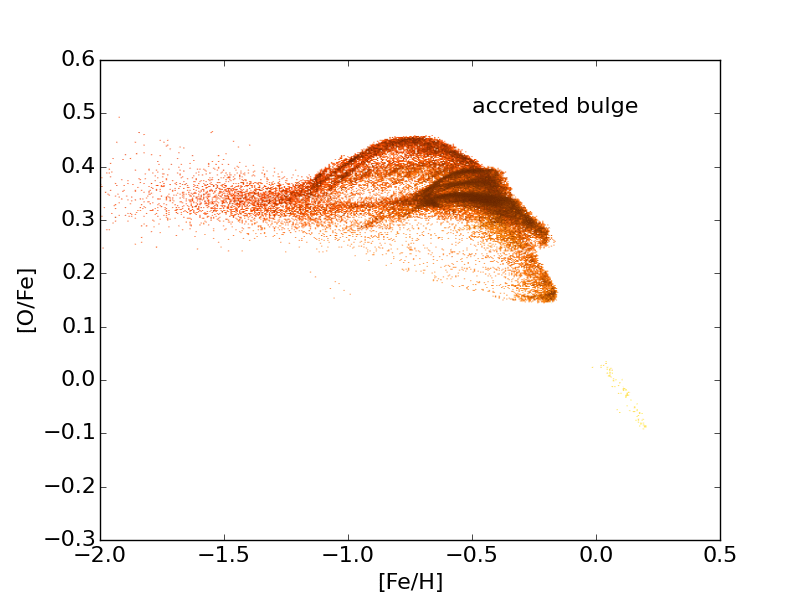}&
         \includegraphics[scale=.3,trim={0.cm 0 1.5cm 1cm},clip]{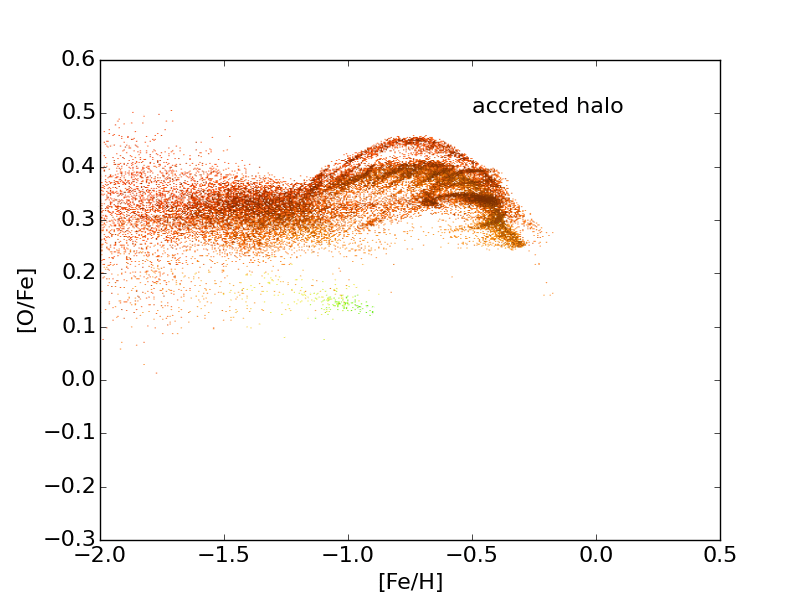}\\
         (g) & (h) & (i) \\  
         \hline
         \multicolumn{3}{c}{MaGICC} \\
         \hline
         \includegraphics[scale=.3,trim={0.cm 0 1.5cm 1cm},clip]{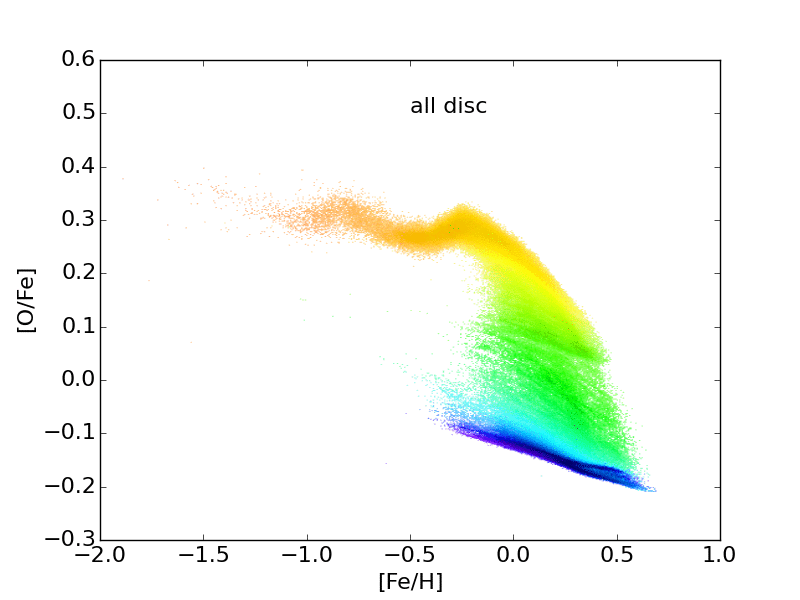} &
         \includegraphics[scale=.3,trim={0.cm 0 1.5cm 1cm},clip]{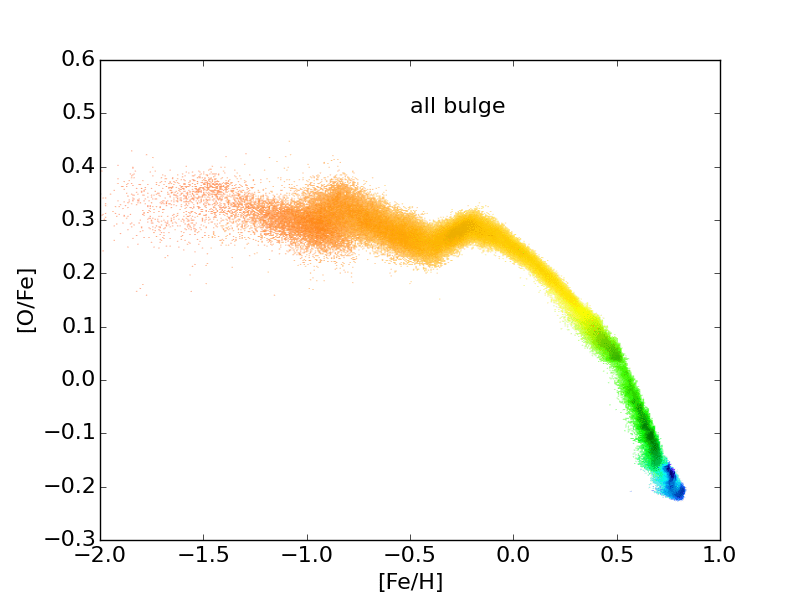} &
         \includegraphics[scale=.3,trim={0.cm 0 1.5cm 1cm},clip]{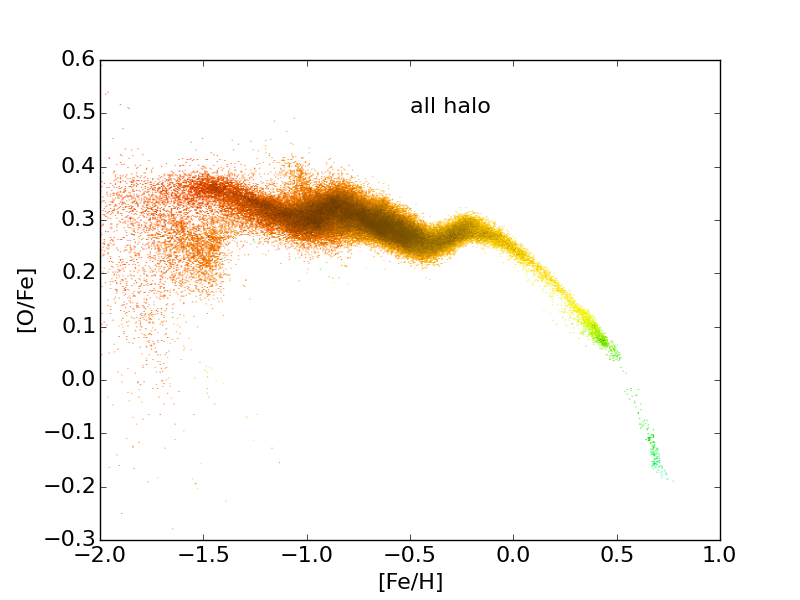} \\   
         (j) & (k) & (l) \\

    \end{tabular}
\caption{The metallicity-[O/Fe] distribution of the MUGS (top three rows) and MaGICC (bottom row) galaxy after decomposition into components. The left column (panels a,d,g) shows the insitu, commuter and accreted stars for the disc, while the centre (b,e,h) and right hand columns (c,f,i) show the same for the bulge and halo respectively. The colours are scaled the same as in Fig. \ref{Fig:allgalallplots}, intensity is scaled by the histogram equalization approach.  }

\label{Fig:decomposeObs}
\end{figure*}

The insitu stars in the disc of both simulations possess a wide range of metallicities. Their distribution is similar to the arrangement in Fig.\ref{Fig:allgalallplots}, but without much of the filamentary structure (in MUGS). The bulge, after the first 4 Gyrs, has a very tight, almost constant metallicity with time,  similar to the `cartoon' view of chemical evolution. These two populations result in the bifurcation in the AMR discussed in Section \ref{Sec:DecompDef}. The spread of metallicity in the disc ($\sim$0.5 dex), versus tight correlations in the bulge, is unsurprising. The disc is far more extended than the bulge and we would expect more variation throughout its structure. This spread is due to the well known metallicity gradient in galaxies \citep[e.g.][]{Gibson2013}

There is a very tight knot of gas in the very centre of the bulge in both MUGS and MaGICC, where the gas is drawn due to dissipation. This feature results in a high degree of enrichment, balanced by the inflow of gas. This dense  material in the inner 0.5 kpc contains 10\% of stars within 20 kpc of the halo centre in MUGS, and 16\% in MaGICC at z=0.  If we discard the stars in the inner 0.5 kpc from the panels in Fig. \ref{Fig:decompose}, then star formation in the bulge effectively halts after the starburst at z=1 in MUGS and at 10 Gyr,  after a decline starting sharply at 6 Gyr, in MaGICC. In MUGS (MaGICC) the SFR in the bulge versus the global star formation rate is around 12\% (21\%) but if we discard the inner 1 kpc this ratio falls to  0.8\% (1.4\%) at 8 Gyr. This suggests that the higher feedback in MaGICC keeps the inner regions more supplied with gas. Alternatively, the gas release from massive stars may maintain star formation for longer because of the more top heavy Chebrier IMF. The dense knot appears to be an intrinsic numerical effect of the code. If we ignore the inner region, the bulge appears older, more in keeping with observations of the Milky Way.

The disc in both MUGS and MaGICC is diffuse, with lower star formation rate densities, and is higher in the potential than the bulge. Thus, the range of metallicities is expected to have greater variation, and to vary with radius and height above the disc plane. The top envelope of the disc AMR is lower than in the bulge, suggesting either that there is a lower specific star formation  rate, or that more metals are lost from the disc environment than the bulge.  The high metallicity content in the bulge is a manifestation of the dependence of metallicity on the depth of the potential. If we interpret the mass-metallicity relation in terms of the potential depth, we expect that the very dense, deep, potential of the galaxy bulge to  be more metal rich than the disc which lies higher in the potential. 

In MaGICC, the stars in the halo are old, but follow the same chemical evolution `trajectory' as the bulge. Panels (a), (b) and (c) in Fig. \ref{Fig:decompose} show that the same is true in MUGS, except that the different populations overlap more noticeably. This implies that these two components have a common origin, or similar conditions. In all likelihood, many of the stars in the halo start to form in a 'bulge-like' environment, and are scattered up into the halo by secular or numerical processes.

The top of the envelope of the AMR is sensitive to the degree of metal diffusion in the code. In this respect the sharp cut off at the upper level of the envelope is possibly a numerical artifact.

 Figure \ref{Fig:decomposeObs} demonstrates the decomposition of the galaxy using the stellar [O/Fe]-[Fe/H] distribution. The distribution of stars in the different components overlap and cannot be easily separated. However, the bulge is very rich in substructure with distinct regions dominated by coeval clumps that evolve in a non-trivial way. This is due to satellites falling into the bulge via dynamical friction. Most of the bulge structure lies within the first gravitational softening length and is thus spatially unresolved. For insitu disc stars in MUGS, the majority of early star formation takes place in a narrow strip around 0.1 to 0 dex in [O/Fe], and 0.1 to 0 dex in [Fe/H]. Disc star formation shows a stronger evolution in [O/Fe] at intermediate times (-0.2 to 0.2 dex between 4 and 8 Gyrs). The bulge evolution is more complex in MUGS, even for insitu stars, indicative of a complex star formation history, while MaGICC produces a steady evolution from low [Fe/H] and high [O/Fe] to high [Fe/H] and low [O/Fe] over the course of the simulation.

\subsection{Disc Radial Trends}
\label{Sec:DecompRad}

\begin{figure*}
\centering
     \begin{tabular}{ccc}
         \includegraphics[scale=.3,trim={0.cm 0 1.5cm 1cm},clip]{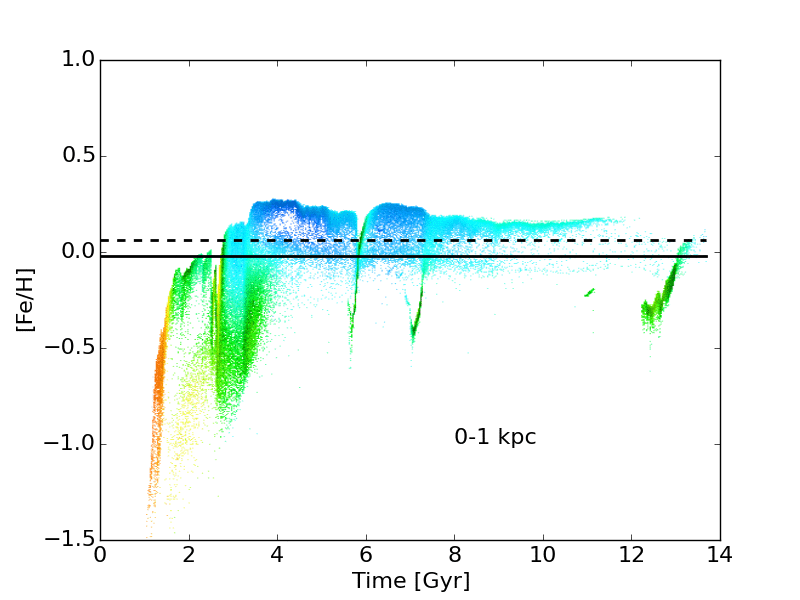} &
         \includegraphics[scale=.3,trim={0.cm 0 1.5cm 1cm},clip]{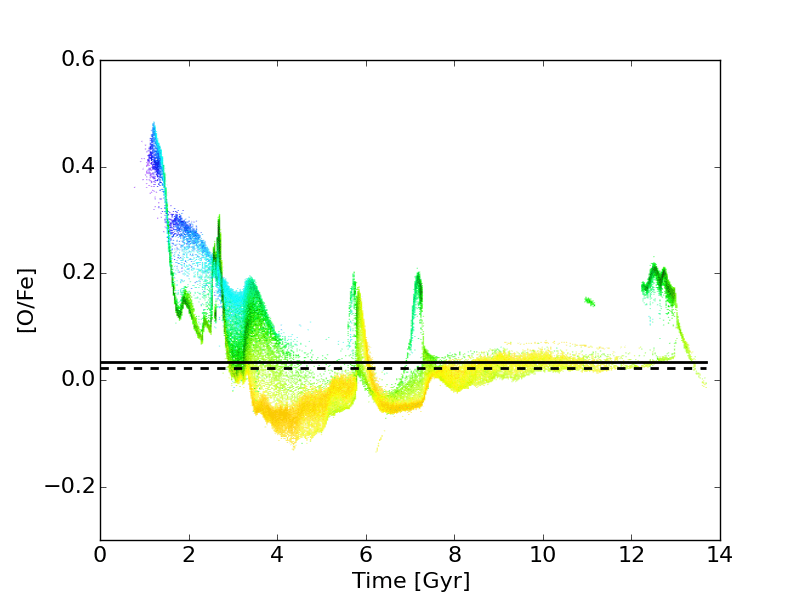}&
         \includegraphics[scale=.3,trim={0.cm 0 1.5cm 1cm},clip]{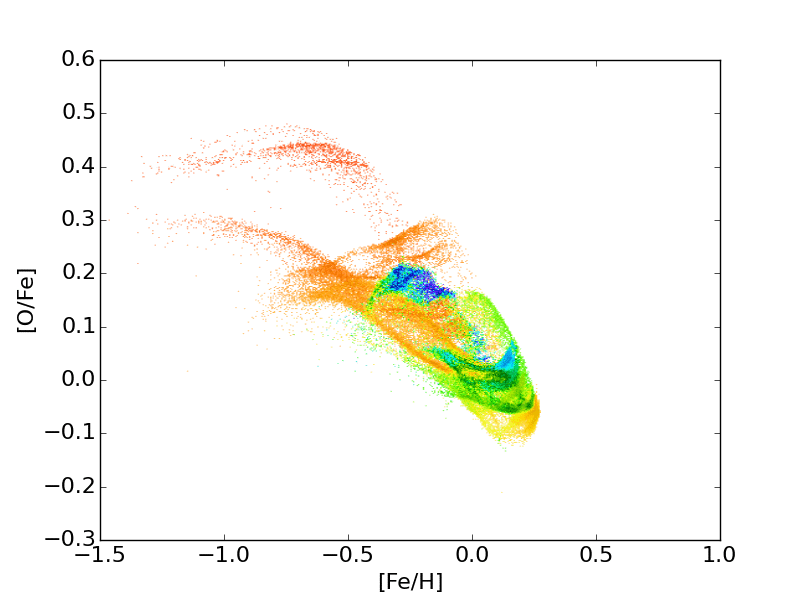} \\

         \includegraphics[scale=.3,trim={0.cm 0 1.5cm 1cm},clip]{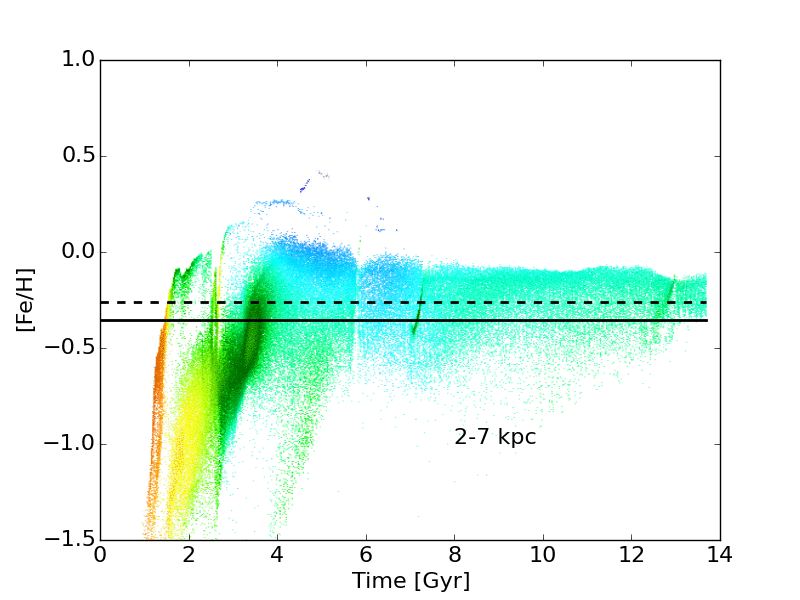} &
         \includegraphics[scale=.3,trim={0.cm 0 1.5cm 1cm},clip]{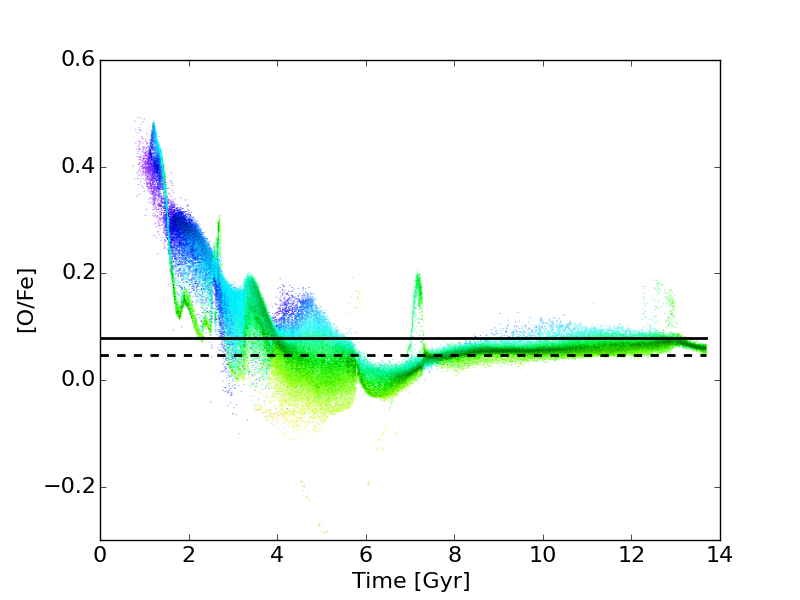}&
         \includegraphics[scale=.3,trim={0.cm 0 1.5cm 1cm},clip]{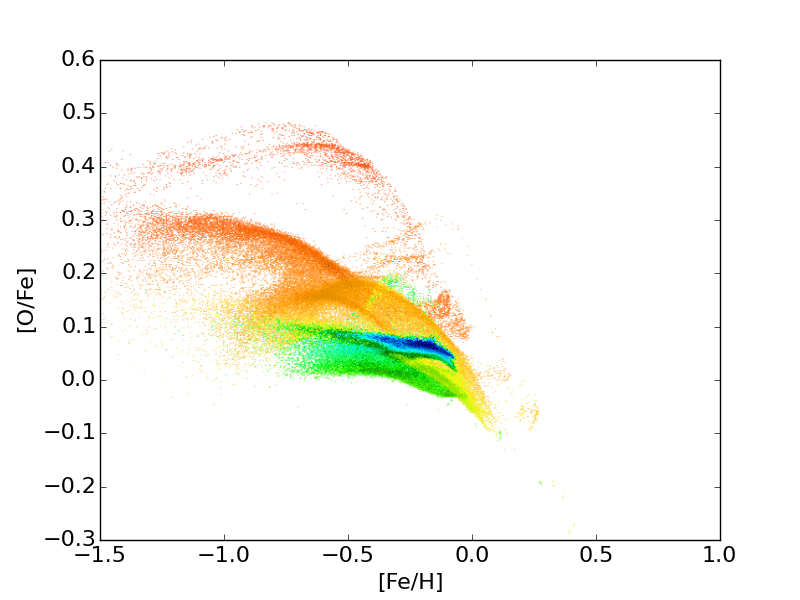} \\
         
         \includegraphics[scale=.3,trim={0.cm 0 1.5cm 1cm},clip]{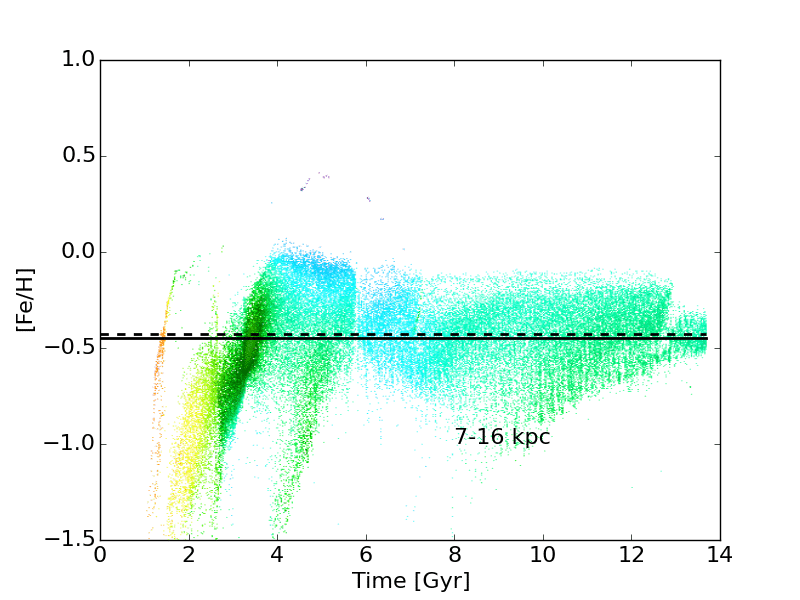} &
         \includegraphics[scale=.3,trim={0.cm 0 1.5cm 1cm},clip]{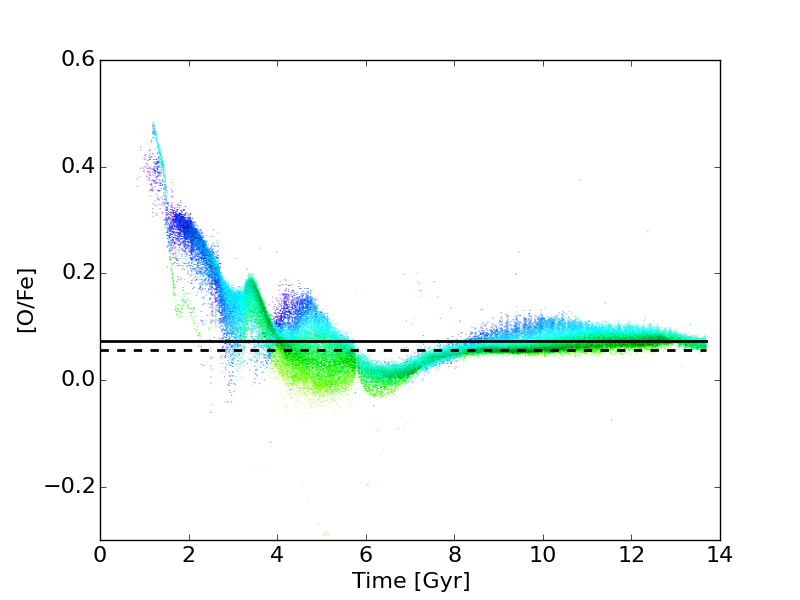}&
         \includegraphics[scale=.3,trim={0.cm 0 1.5cm 1cm},clip]{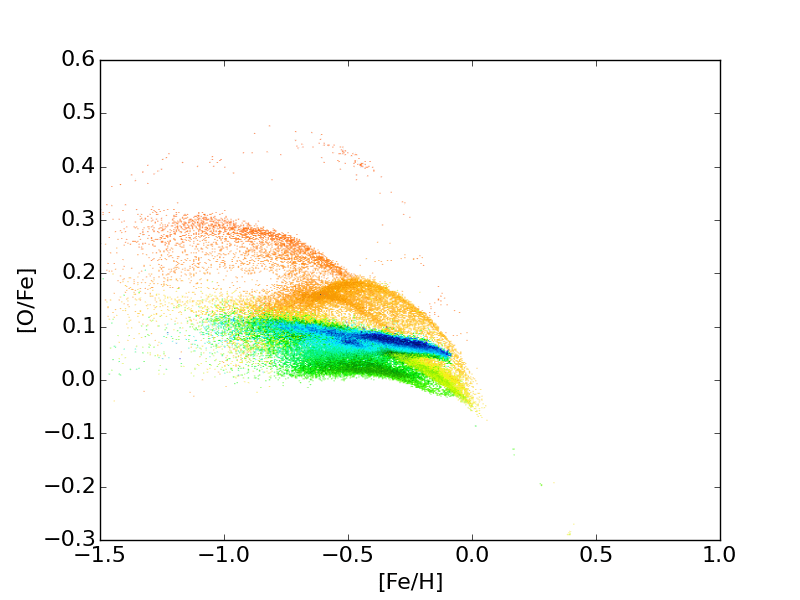} \\
         
         \includegraphics[scale=.3,trim={0.cm 5cm 1.5cm 5cm},clip]{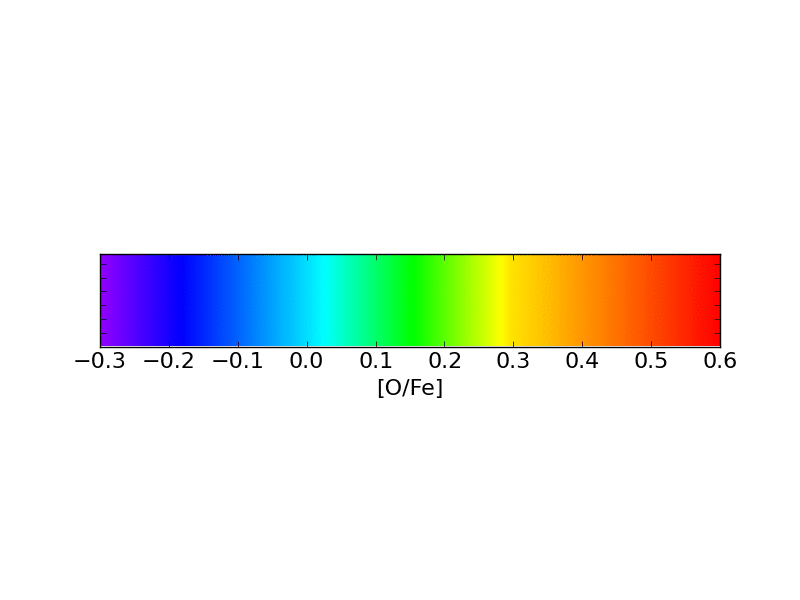} &
         \includegraphics[scale=.3,trim={0.cm 5cm 1.5cm 5cm},clip]{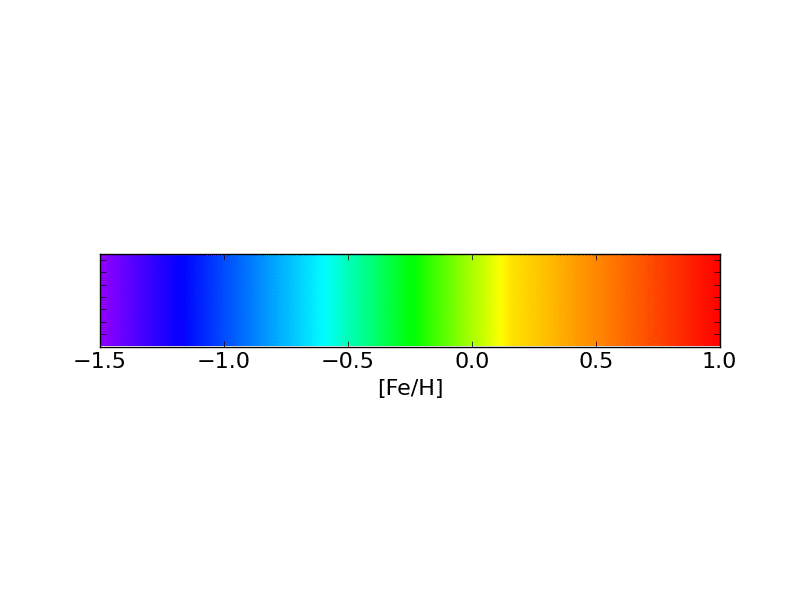} &
         \includegraphics[scale=.3,trim={0.cm 5cm 1.5cm 5cm},clip]{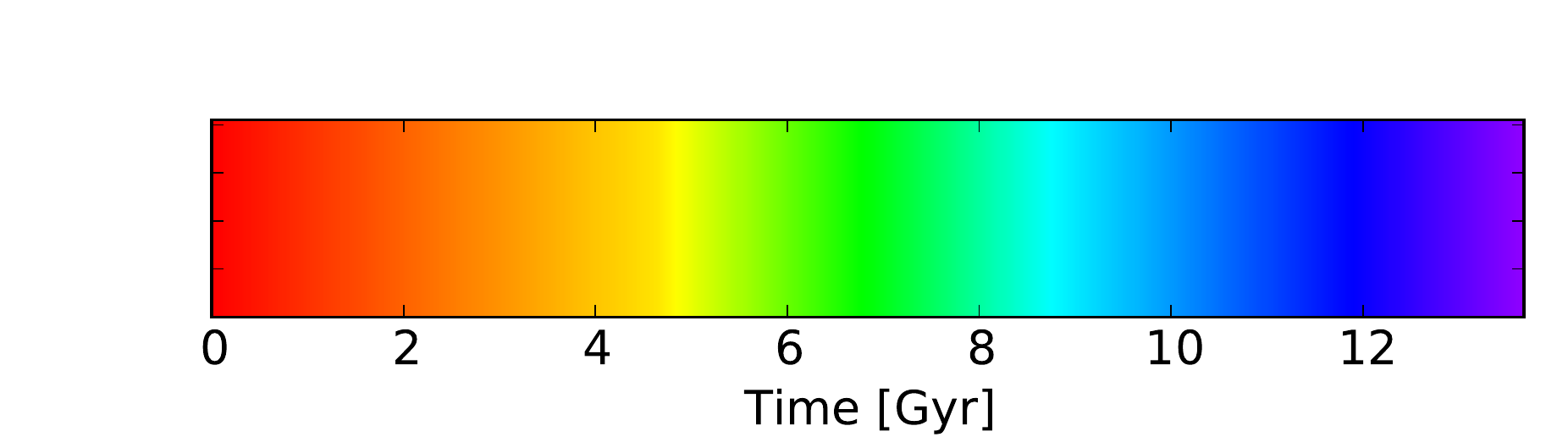} \\         
               
    \end{tabular}
\caption{The MUGS AMR, time-[O/Fe] and metallicity-[O/Fe] distributions of stars in the strictly-defined disc at z=0 which were formed insitu, within different radial bins. The black solid line is the mean metallicity/[O/Fe] value (left column/middle column) of stars in that radial bin, while the dotted line is the mean metallicity/[O/Fe] value of stars formed {\it after 6 Gyr} in that radial bin. Each column is coloured according to the corresponding colour bar and the darkness is  a function to the number of stars, as in Figure \ref{Fig:allgalallplots}. Each panel is individually scaled for brightness via histogram equalization.  Radial bins are defined by the star's position at z=0. Despite removing stars which did not form insitu we see structures in the AMR (at 5 and 7 Gyr) because of the starbursts triggered by the interaction.}
\label{Fig:decomposerad}
\end{figure*}   

\begin{figure*}
\centering
     \begin{tabular}{ccc}
         \includegraphics[scale=.3,trim={0.cm 0 1.5cm 1cm},clip]{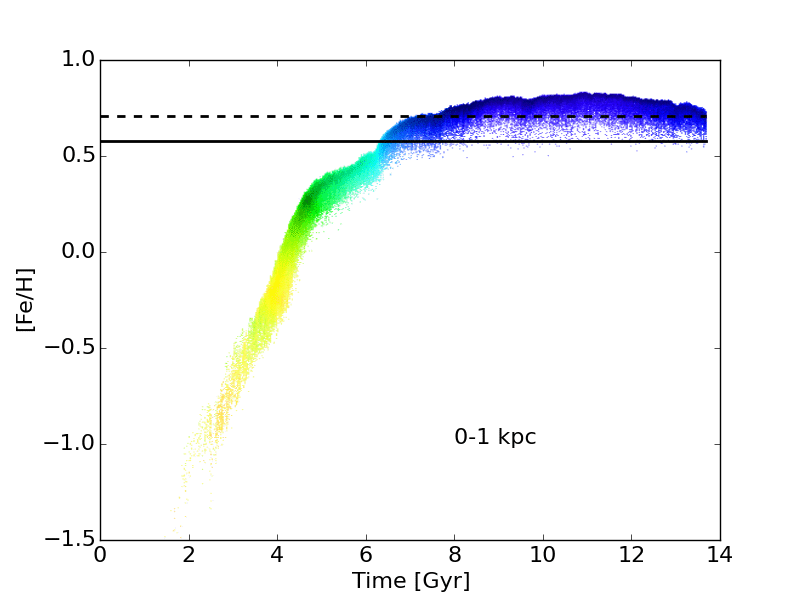} &
         \includegraphics[scale=.3,trim={0.cm 0 1.5cm 1cm},clip]{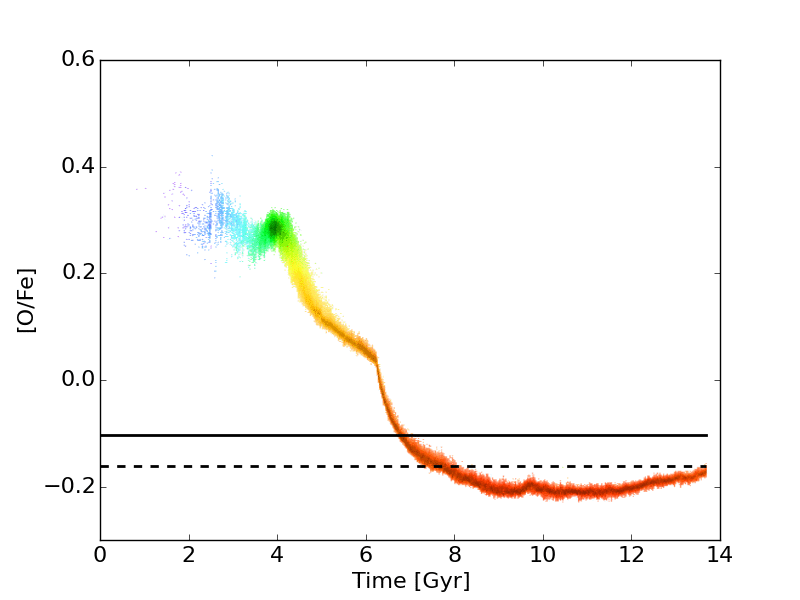}&
         \includegraphics[scale=.3,trim={0.cm 0 1.5cm 1cm},clip]{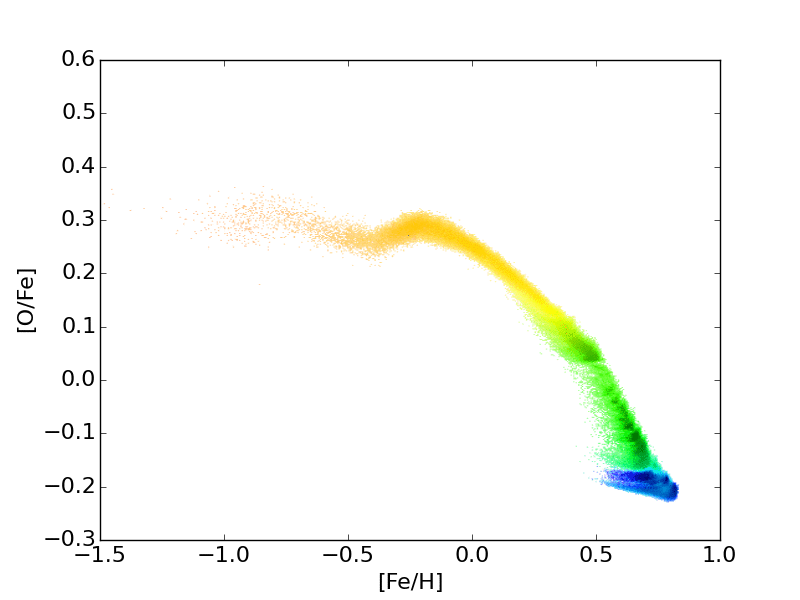} \\

         \includegraphics[scale=.3,trim={0.cm 0 1.5cm 1cm},clip]{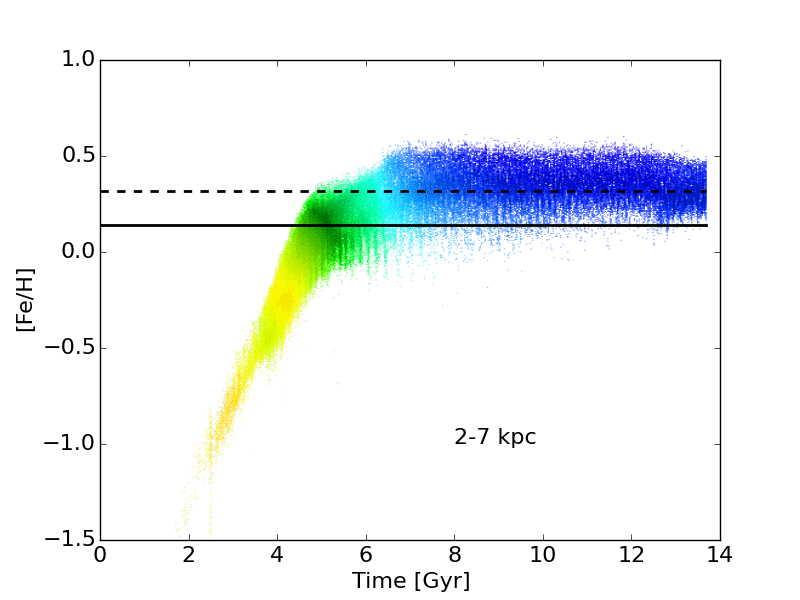} &
         \includegraphics[scale=.3,trim={0.cm 0 1.5cm 1cm},clip]{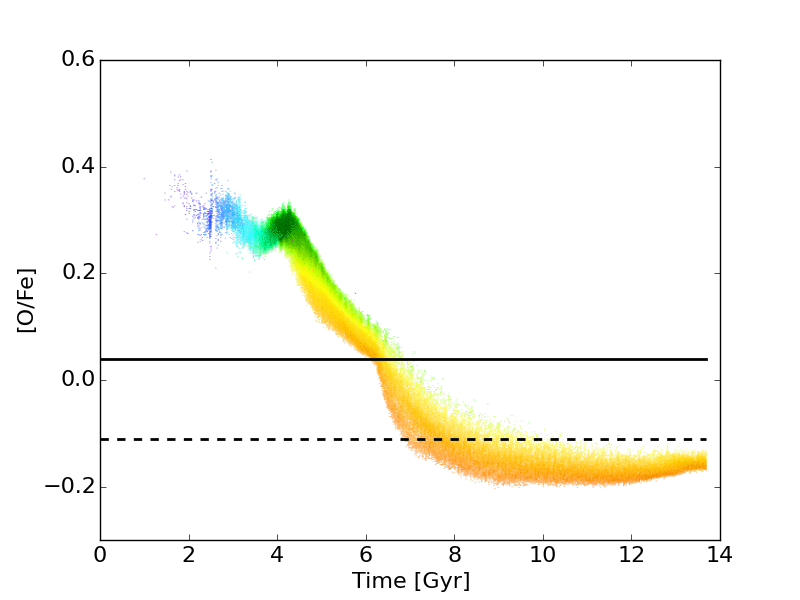}&
         \includegraphics[scale=.3,trim={0.cm 0 1.5cm 1cm},clip]{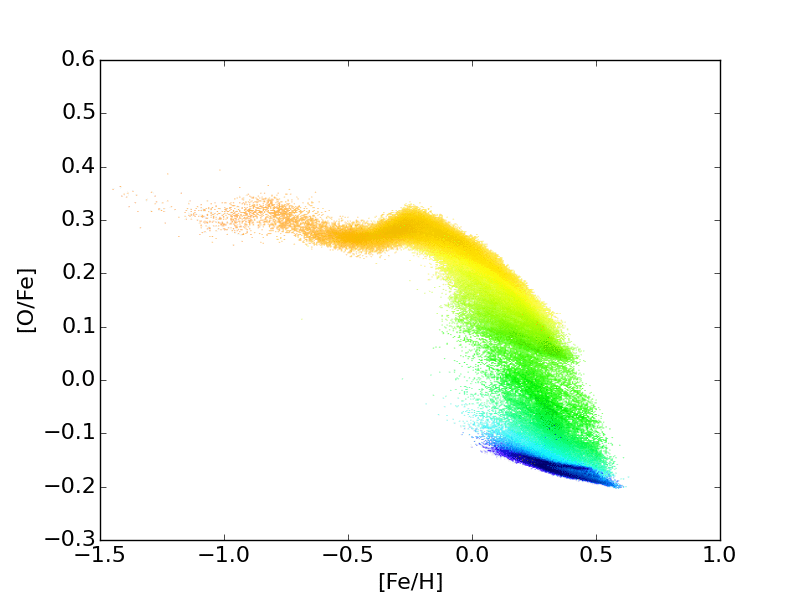} \\
         
         \includegraphics[scale=.3,trim={0.cm 0 1.5cm 1cm},clip]{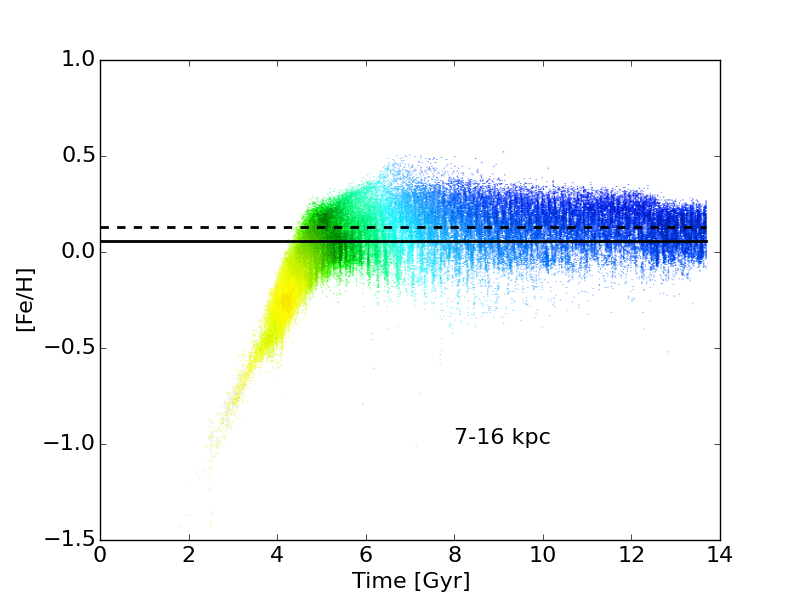} &
         \includegraphics[scale=.3,trim={0.cm 0 1.5cm 1cm},clip]{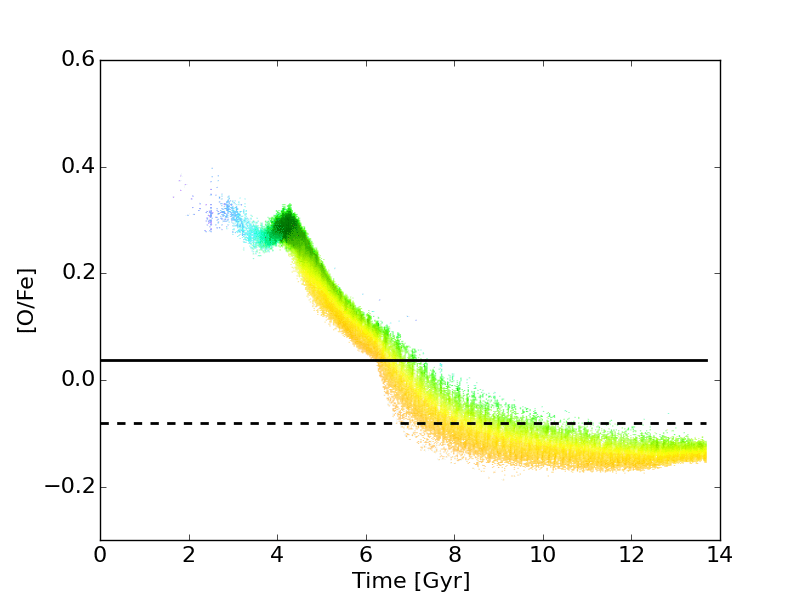}&
         \includegraphics[scale=.3,trim={0.cm 0 1.5cm 1cm},clip]{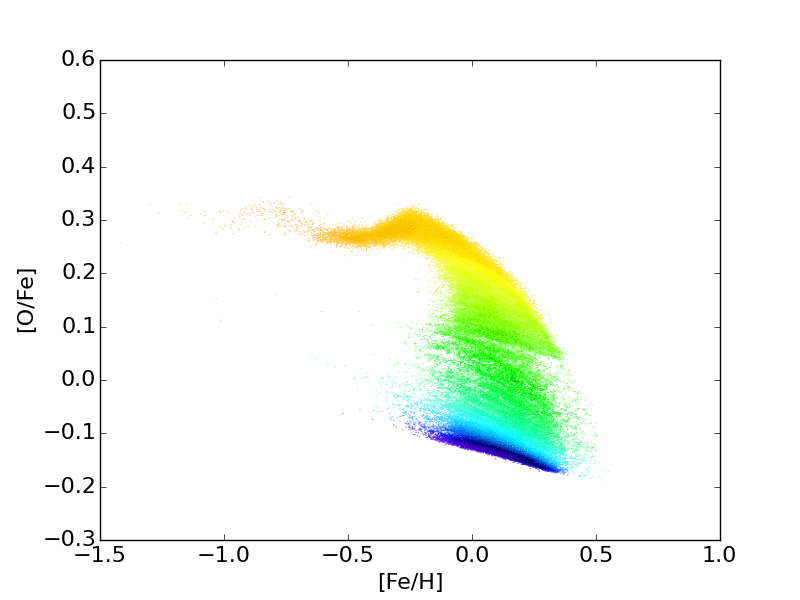} \\
         
         \includegraphics[scale=.3,trim={0.cm 5cm 1.5cm 5cm},clip]{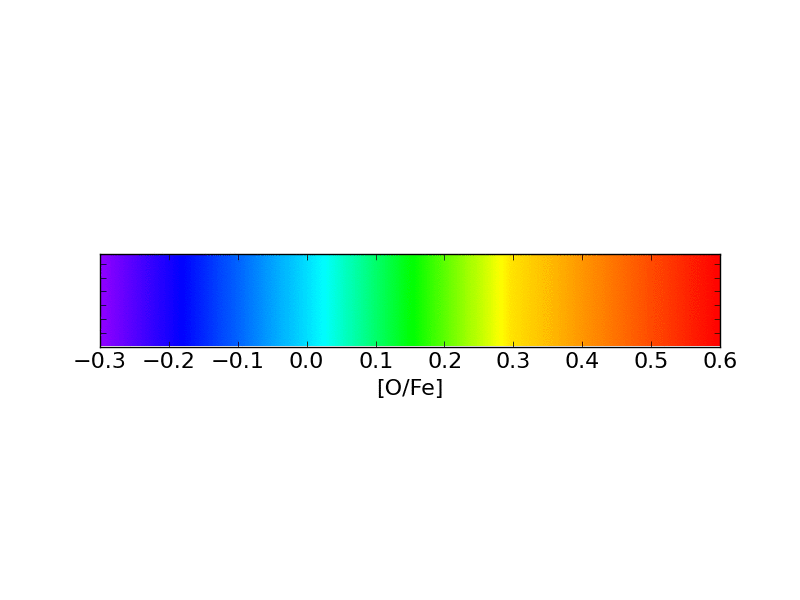} &
         \includegraphics[scale=.3,trim={0.cm 5cm 1.5cm 5cm},clip]{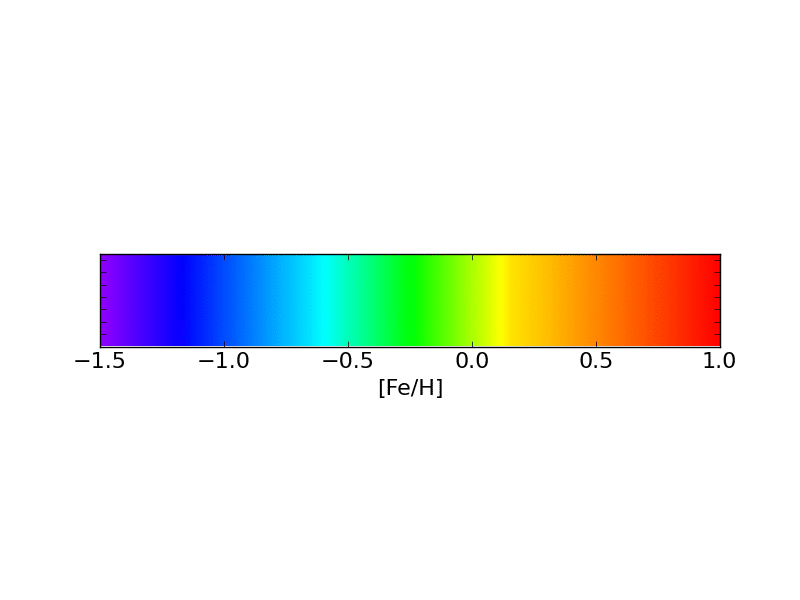} &
         \includegraphics[scale=.3,trim={0.cm 5cm 1.5cm 5cm},clip]{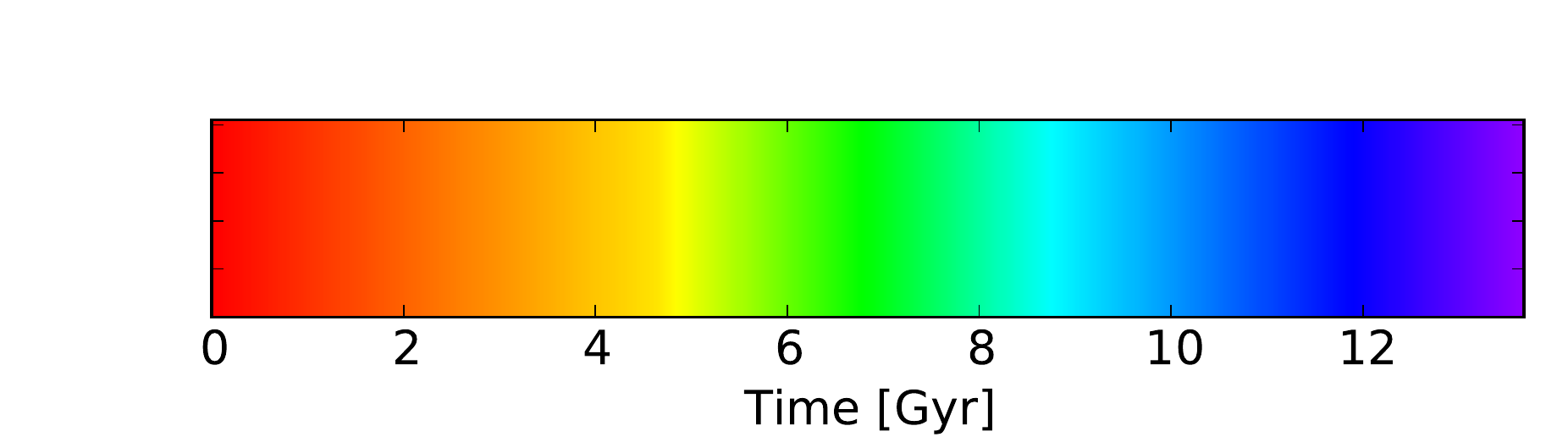} \\   
               
    \end{tabular}
\caption{As in Fig. \ref{Fig:decomposerad} for the MaGICC simulation.}
\label{Fig:decomposeradMaGICC}
\end{figure*}   

In Fig. \ref{Fig:decomposerad}  (for MUGS) and Fig. \ref{Fig:decomposeradMaGICC} (for MaGICC) the shape of the AMR changes from small radii to the edge of the disc. It is well known \citep[e.g.][etc.]{Pilkington2012,Rupke2010} that disc galaxies exhibit a gradient in metallicity with radius, which is often discussed in the context of inside out galaxy formation \citep{Larson1976}, with stars (and gas) in the inner regions of galaxies being more metal rich than stars at larger radii. 

 We see that the general distribution of stars in the left hand column of Fig. \ref{Fig:decomposerad} and  Fig. \ref{Fig:decomposeradMaGICC}  show a general decline in [Fe/H], and broadening of the distribution with increasing radius. If we take the 8 Gyr mark as a base line, the stars have a distribution of approximately 0.1 dex, 0.54 dex and 0.76 dex for the three radial bins in MUGS and 0.1, 0.4 and 0.45 dex in MaGICC. 
 
 At small radii the impact of old stars is clearly evident in both MUGS and MaGICC from the lines of $<$[Fe/H]$>$. The effect of old stars is shown by the difference between solid and dotted  lines in Fig. \ref{Fig:decomposerad} and Fig. \ref{Fig:decomposeradMaGICC}. These lines illustrate the mean [Fe/H] value for all stars at the radius shown, with and without including the stars formed before 6 Gyr.  The difference between the two mean values decreases at larger radii, showing  that early star formation becomes less important further out. This fits well with the standard galaxy formation paradigm  of inside-out formation \citep[as discussed in][ for the MUGS version of g15784 using metallicity gradient evolution through time]{Pilkington2012}.
  
 In Fig. \ref{Fig:decomposerad} (the MUGS galaxy) we see features in the disc, around 5 and 7 Gyr, for example, that are the result of interactions. Although stars which did not form in the disc of the galaxy have been removed, the interactions did cause the gas in the galaxy to be stirred up. This resulted in rapid star formation, which produces the sharp enrichment processes that  are especially prominent in the top row of the figure. This interaction was in the plane of the disc, and retrograde to the rotation of the disc. The interaction pulled gas into the centre of the galaxy, diluting the gas in the bulge with lower metallicity gas from the disc. This triggered a starburst, a brief rise in [O/Fe] and re-enrichment. It also caused a brief hiatus in star formation in the disc, evident in the lower panels.  

Figure \ref{Fig:decomposerad} shows:
\begin{itemize}
\item The {\it lower} sequence becomes increasingly dominant at larger radii, which can also be seen in APOGEE by \citet{Hayden2015}, where the upper sequence is focused in the centre (Fig. \ref{Fig:decomposerad}, right hand column). 
\item The central region of the disc shows a distinctly different distribution to the rest of the disc as it overlaps with the bulge.
\item  The stars in the middle and lower sequences of the [Fe/H]-[O/Fe] distribution can be seen at all radii at $z=0$, but stars from the lower sequence tend to be found further out. 

\item  Outside the inner 2 kpc {\it two} sequences are clearly visible in both simulations: a young inner sequence, which grows increasingly dominant with radius, and an older upper sequence. This corresponds well with the model presented in \citet{Brook2004} and confirmed by \citet{Haywood2013} and \citet{Hayden2015} for Milky Way data. In \citet{Haywood2013} the upper sequence is assumed to belong to the  thick disc of the Milky Way, formed during the high star formation rate phase of the star formation history \citep{Snaith2014b, Snaith2014}, while the lower sequence belongs to the outer disc. We see signs of this here, although both sequences are visible at all radii the middle sequence is significantly more centrally concentrated. 
\end{itemize}

\begin{figure*}
\centering
     \begin{tabular}{cc}    
     MUGS & MaGICC \\      
      \includegraphics[scale=.45,trim={0cm 0 1.5cm 1cm},clip]{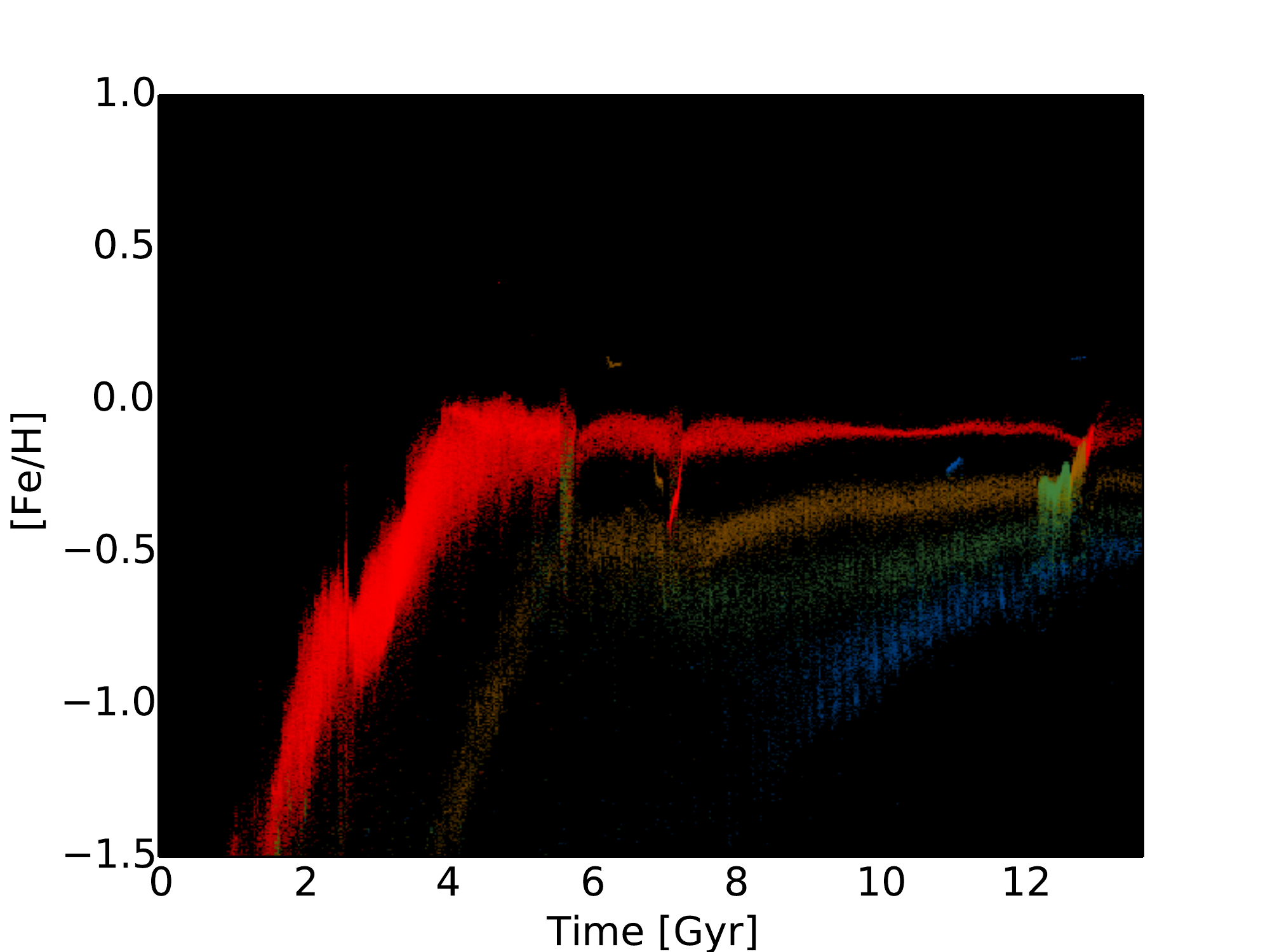}&
      \includegraphics[scale=.45,trim={0cm 0 1.5cm 1cm},clip]{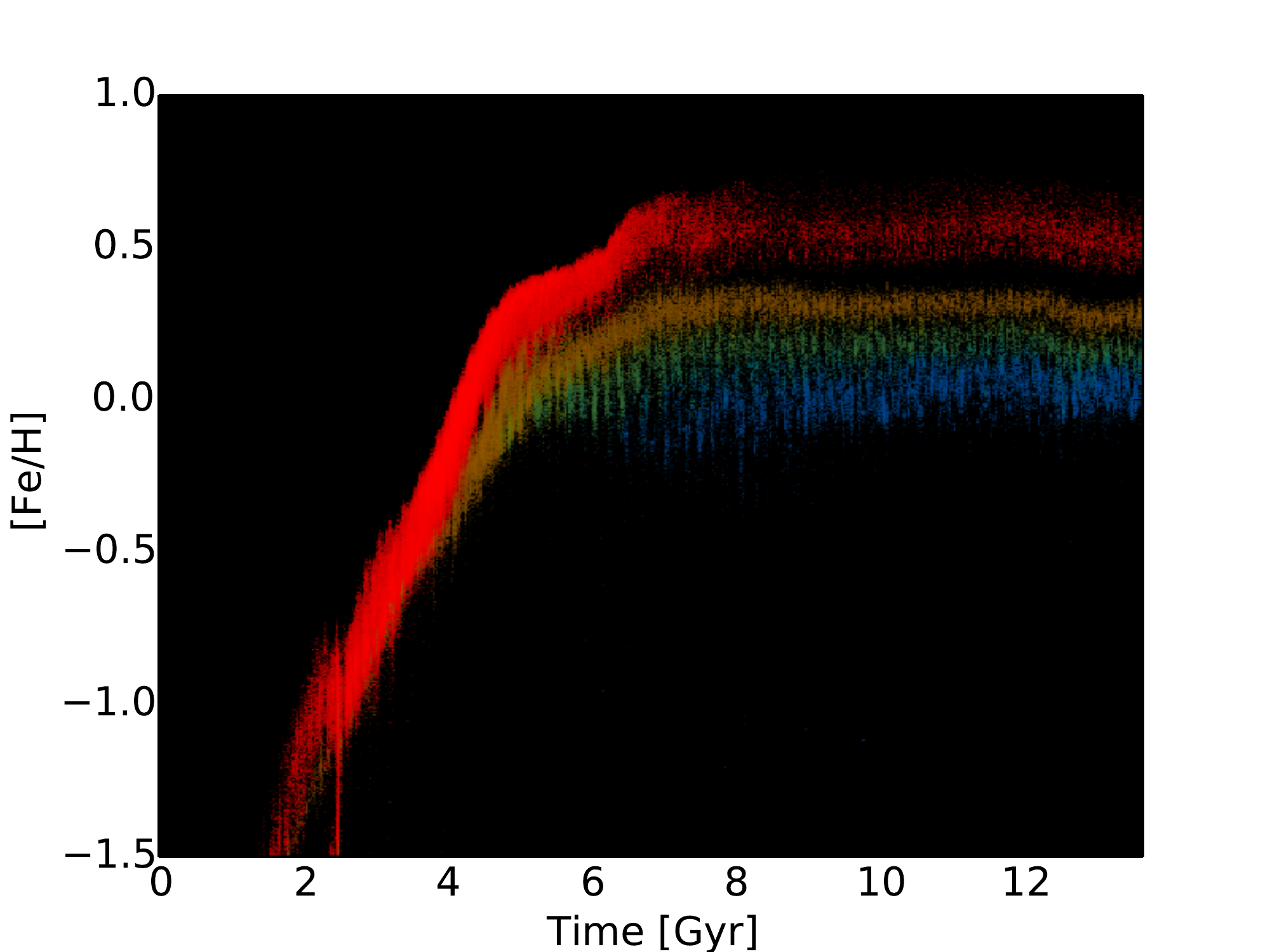}\\  
      (a) MUGS galaxy AMR, stars binned by formation radius & (b) MaGICC galaxy AMR, stars binned by formation radius \\
      \includegraphics[scale=.45,trim={0cm 0 1.5cm 1cm},clip]{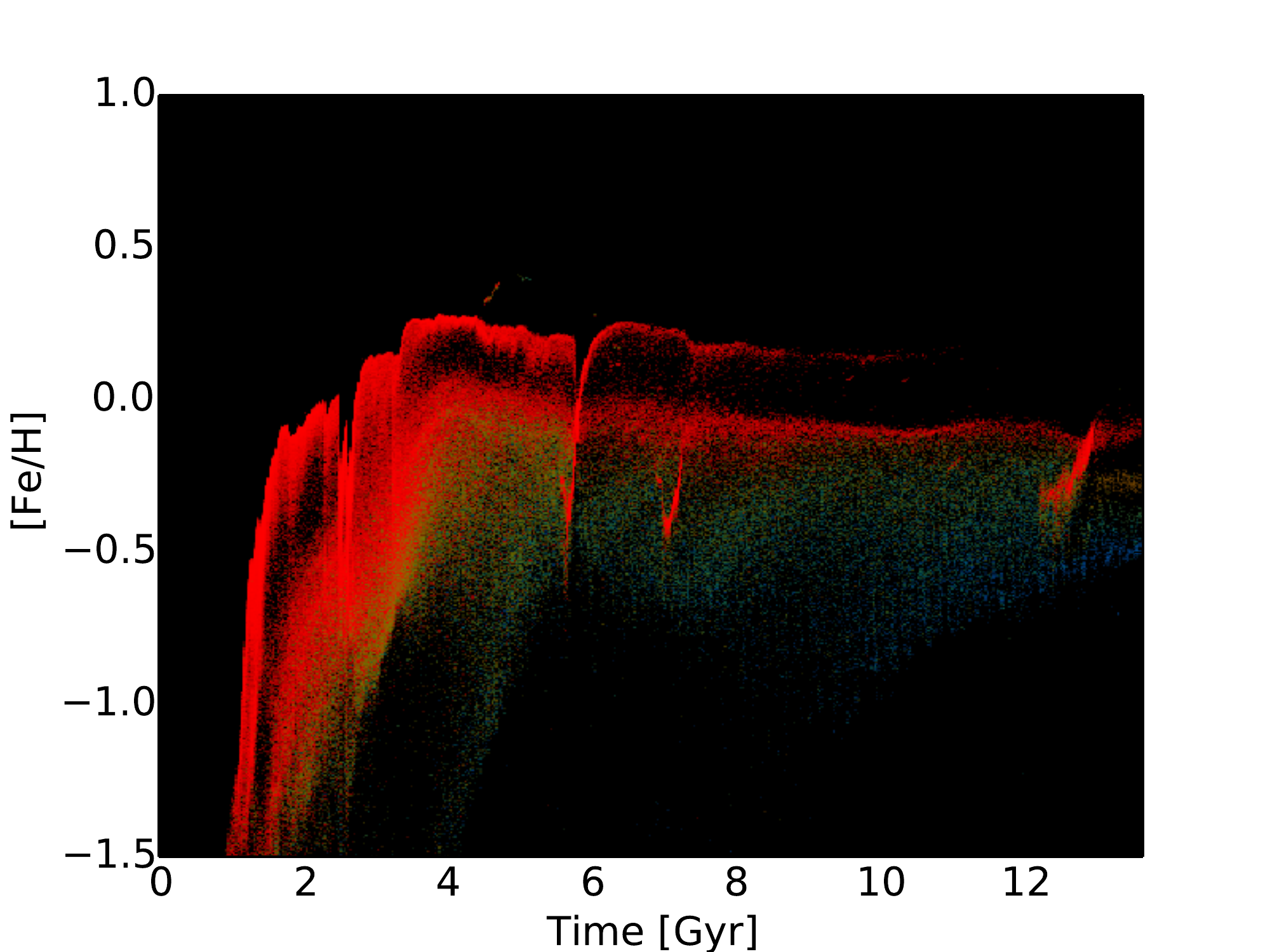}&
      \includegraphics[scale=.45,trim={0cm 0 1.5cm 1cm},clip]{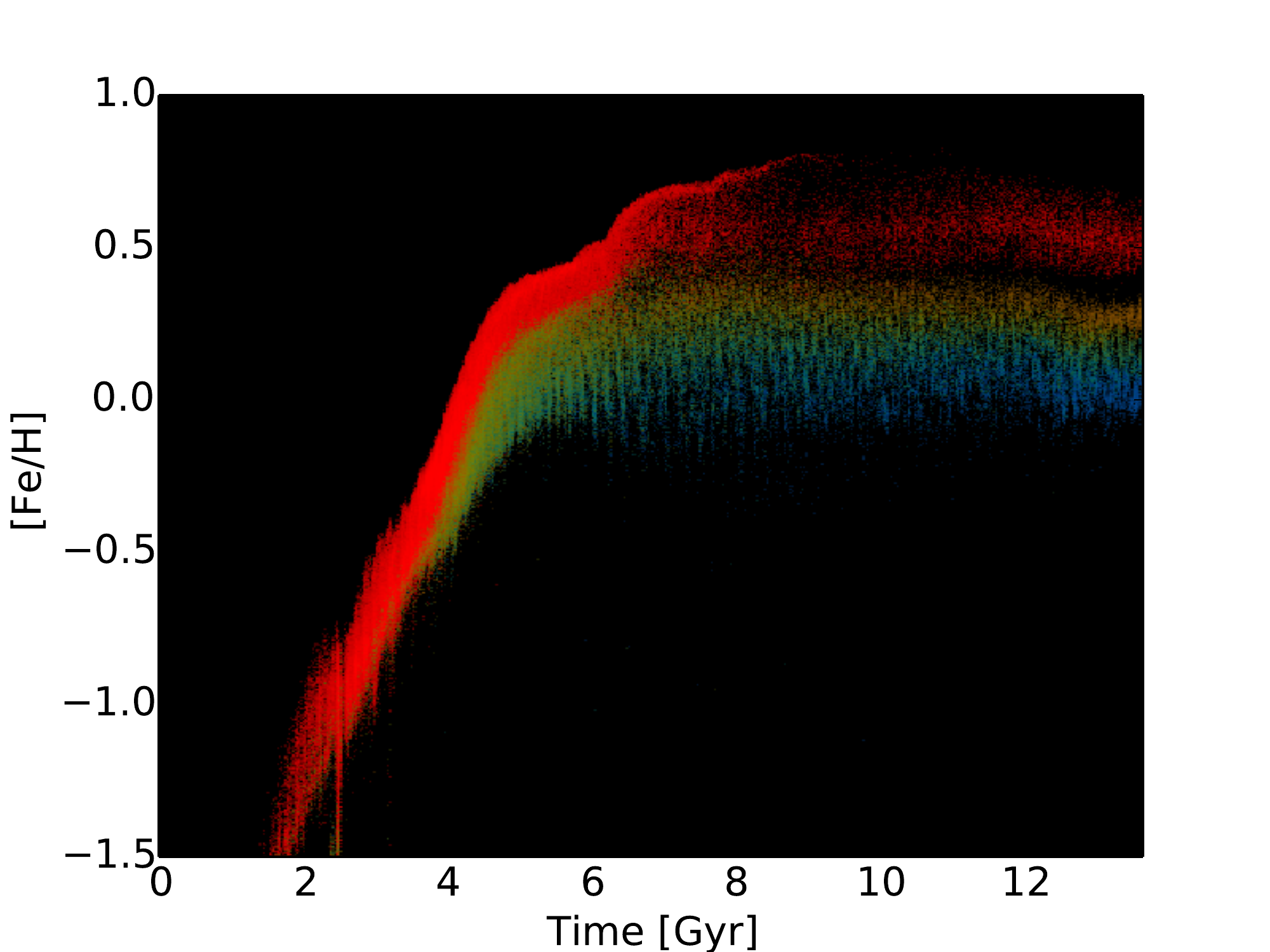}\\
      (c) MUGS galaxy AMR, stars binned by final radius &(d)  MaGICC galaxy AMR, stars binned by final radius \\
            
   \end{tabular}
\caption{The AMRs for MUGS (left) and MaGICC (right) of insitu stars selected according to their radial positions. The top row shows the stars selected according to their radius at the time of formation, while the bottom row shows stars selected according to their z=0 radius. The red sequence is for stars with 1$<$r$<$2 kpc, the orange sequence is for stars 5$<$r$<$6 kpc, the green sequence is for stars with  8$<$r$<$9 kpc and covers the `solar radius'. while blue shows stars 11$<$r$<$15 kpc. }
\label{Fig:insiturad}
\end{figure*}

Using a simulation with early radiative feedback, similar to MaGICC,  \citet{Brook2012} studied a dwarf galaxy (with a stellar mass of $\sim$8$\times$10$^{9}$M$_\odot$) in terms of thin disc, thick disc, bulge etc.  There are two sequences in the [O/Fe]-[Fe/H] distribution for stars at the `solar radius' (r=7-8 kpc) of this galaxy. The distribution of the MaGICC galaxy is considerably different. The [O/Fe]-[Fe/H] distribution shows a `sail'-like distribution rather than the broken line evident in Figure 2 of \citet{Brook2012}, although this may be related to their more strict definition of the solar neighbourhood.

In order to quantify the effect of the processes that change the position of the star's radius over time, we also split the stars by their formation radius (radius at the first simulation output at which they appear, which has an uncertainty of $\pm200$~Myr due to the output cadence of the simulation). In Fig. \ref{Fig:insiturad}, we extract stars from four non-contiguous annulae, 1$<$r$<$2 kpc,  5$<$r$<$6 kpc, 8$<$r$<$9 kpc, and 11$<$r$<$15 kpc, and plot the AMR of those stars using either the formation radius (left) or $z=0$ radius (right). Where stars are subdivided by formation radius, the MUGS stars have tightly correlated, distinct AMRs. The inner sequence is very narrow, and is essentially flat after the first 4 Gyr. The larger radius sequences have increasingly positive slopes with time, and the outermost sequence traces the outer edge of the late time skirt of the Galaxy.  {\it This implies that a given radius in the disc is a particular environment with fairly homogeneous properties, while the disc itself is an ensemble of these  environments.} The MaGICC galaxy does not show an increasing gradient at larger radius: each sequence is essentially lower metallicity, but has a flat AMR in each case. After the first 6 Gyr, MaGICC shows only a small evolution in radial scatter with time, indicating that radial motions are more important in MUGS. If MaGICC is more similar to reality than MUGS then this is good news for Gaia and other astroarcheology missions because mixing is small.  

The MUGS galaxy suggests that metallicity is a poor proxy for age, even based on the 2-7 kpc panels in Figs. \ref{Fig:decomposerad}, \ref{Fig:decomposeradMaGICC} and \ref{Fig:insiturad}, with wide initial scatter followed by a broad, constant metallicity phase. However, the age-[O/Fe] distribution is much more tightly correlated and the early evolution phase shows a significant negative gradient in [O/Fe] with time before 5 Gyr. This corresponds to the high star formation phase. There is considerable overlap of stars in the `solar' vicinity (7-9 kpc) from different populations (at z=0)  due to radial motions of stars. This is less evident in MaGICC, however, where the AMR flattens after 4-4.5 Gyr for each galaxy.  

 In terms of the `cosmic timer' concept discussed in the introduction, it should be possible, for a galaxy which evolves like a MaGICC galaxy, to calculate the age of a star from its metallicity at early times, when the trend is narrow. After z=1 the profile thickens due to the gradient, making it difficult to recover the age from the metallicity without knowing the birth radius.  The birth radius is difficult to know with any certainty because of radial motions.

\subsection{Satellites}
\label{Sec:Sat}
The filamentary substructure in the MUGS chemical evolution distribution (Fig. \ref{Fig:allgalallplots}) is a direct result of the deposition of satellite stars in the galaxies. These stars are either still in satellites or are accreted/commuter stars within the host galaxy. In this section we discuss the influence of these satellites on the chemical distribution of stars using two example satellites. 

Satellite galaxies lie inside their own dark matter subhalos within the host. During the evolution of the universe, dark matter halos form and merge into larger halos. These smaller subhalos form structures within galaxies, each with their own distinct potential well. Some of these subhalos contain satellite galaxies, because gas can remain in the depths of the potential  for a several Gyrs. Satellites can be expected to have different evolutionary paths to the host galaxy due to their different masses and positions.  When the satellites approach the host tidal effects can be expected to disturb the gas and affect the star formation. 

\citet{Nickerson2013} analysed the luminosities of satellites in MUGS, and found that although they produce too large stellar mass, their properties are not greatly different to observations. Subhalos exist in MaGICC, although they are very poor at forming stars. The {\it subhalo} population of MaGICC is not dissimilar to the subhalo population in MUGS, as is expected. However, subhalos in MaGICC are extremely gas rich, and this gas is largely prevented from forming stars. For example, at z=0, the galaxy in MUGS and MaGICC contain 197 and 114 subhalos respectively, with a total dark matter mass of 8.8$\times$10$^{10}$M$_\odot$ and 3.8$\times$10$^{10}$M$_\odot$. Subhalos in MaGICC tend to contain fewer baryons, but have a much higher gas-to-star ratio, where they form stars at all. The average gas-mass-to-stellar-mass in subhalos with stars (satellite galaxies) for MUGS and MaGICC is 0.03 and 14.6 at z=0. This demonstrates the huge impact of feedback on the star formation efficiency of low mass objects in the MaGICC simulations. Indeed, only 7 of the 114 MaGICC satellites contain any stars, while in MUGS 26 contain stars. While MUGS galaxies are not that different to observed satellite populations {\it the MaGICC simulations are comparatively poor at reproducing the low mass end of the stellar mass distribution.} 

Satellites in MaGICC do not have a significant impact on the star formation history, as previously discussed. In the following section we discuss satellites in MUGS because they are more important to our understanding of the detailed chemical evolution in  this simulation. We expect to see a contribution from satellites in  Milky Way observations, and so we need to use MUGS to explore this impact.

\begin{figure*}
\centering
     \begin{tabular}{ccc}
        All satellites & Satellite A & Satellite B \\
         \includegraphics[scale=.3,trim={0.cm 0 1.5cm 1cm},clip]{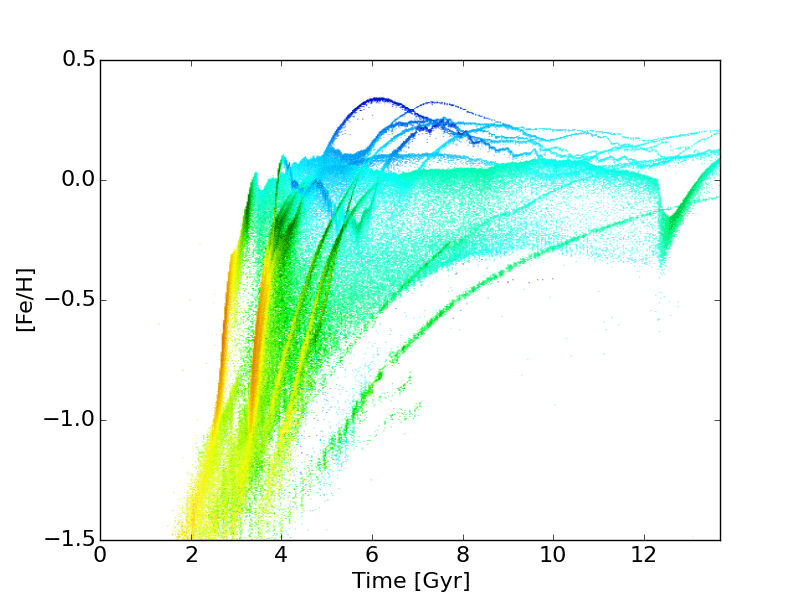} &
         \includegraphics[scale=.3,trim={0.cm 0 1.5cm 1cm},clip]{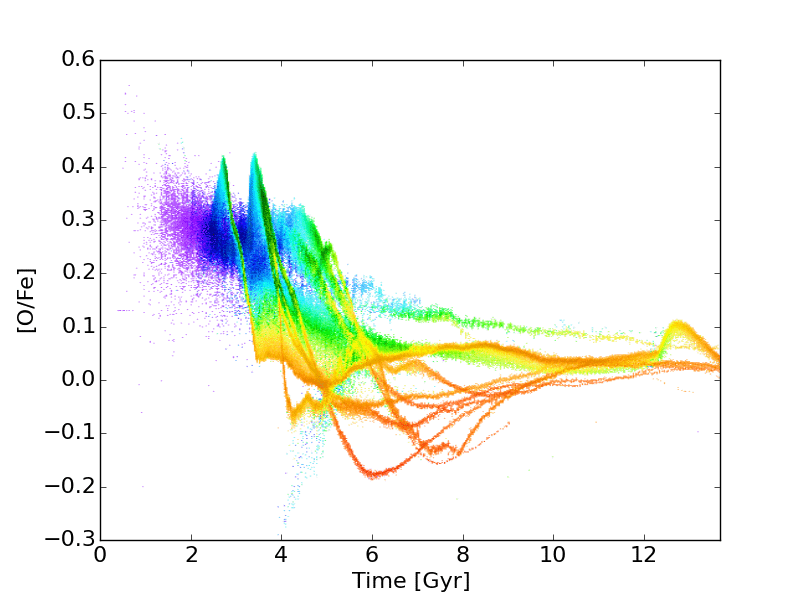} &         
         \includegraphics[scale=.3,trim={0.cm 0 1.5cm 1cm},clip]{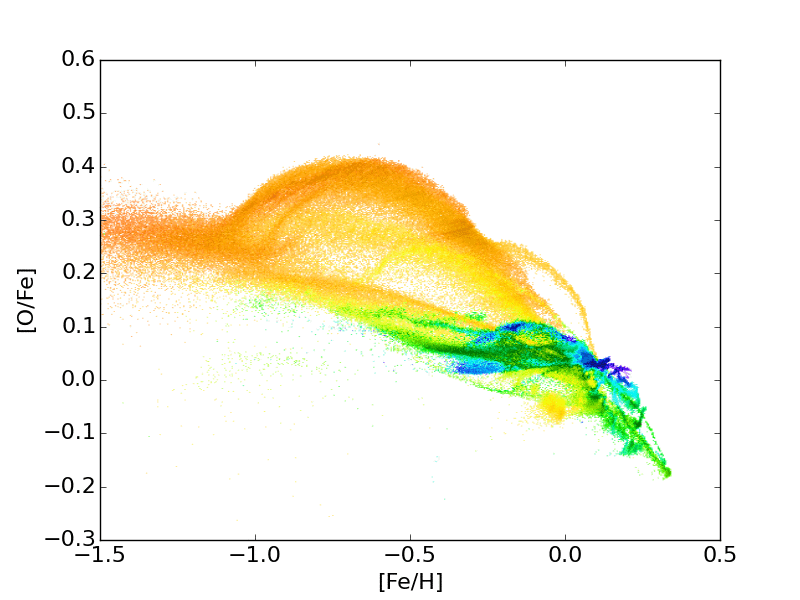} \\            
         (a) & (b) & (c) \\  
         \includegraphics[scale=.3,trim={0.cm 0 1.5cm 1cm},clip]{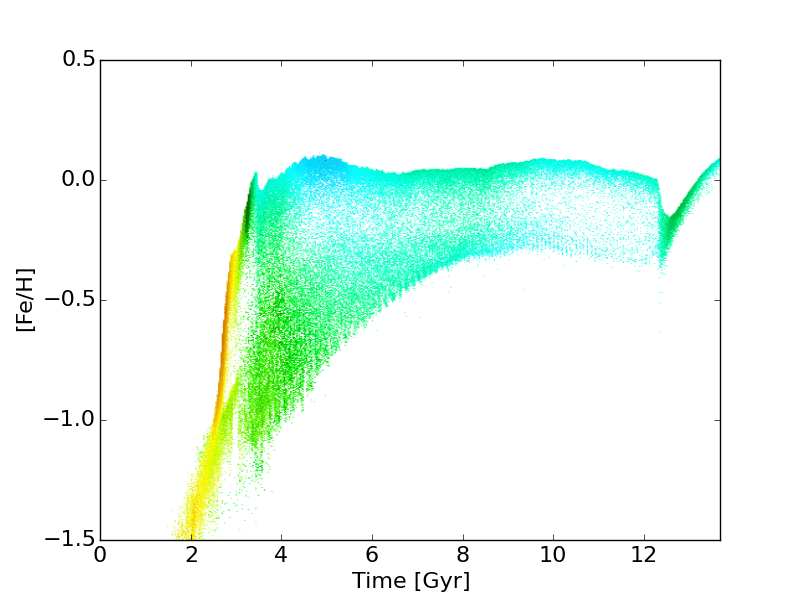} &
         \includegraphics[scale=.3,trim={0.cm 0 1.5cm 1cm},clip]{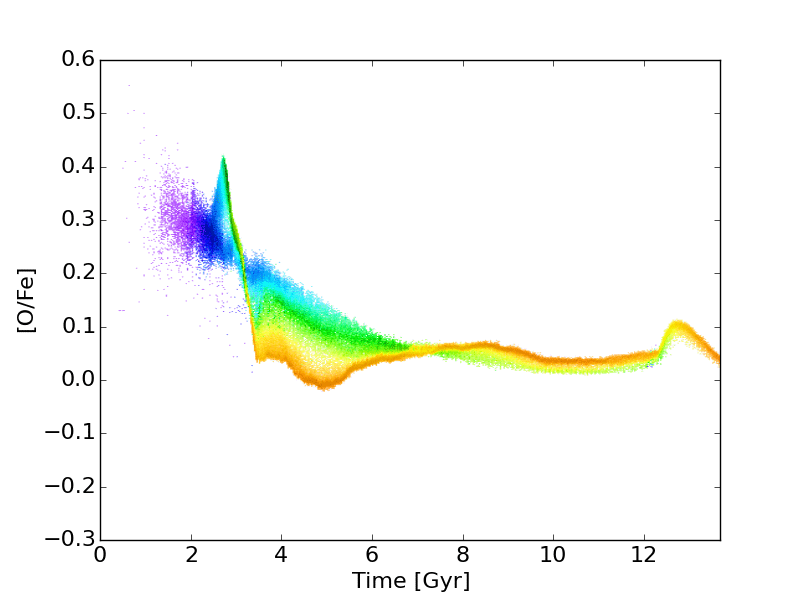} &         
         \includegraphics[scale=.3,trim={0.cm 0 1.5cm 1cm},clip]{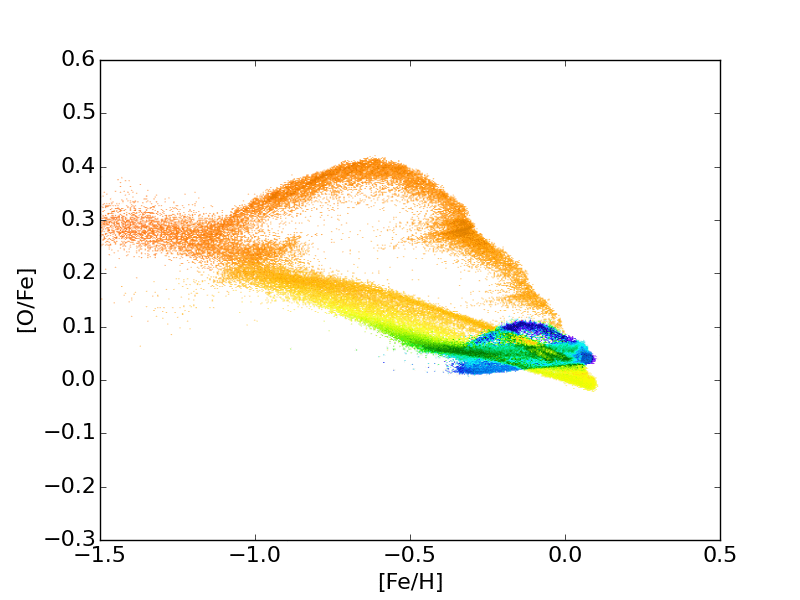} \\              
           (d) & (e) & (f) \\       
         \includegraphics[scale=.3,trim={0.cm 0 1.5cm 1cm},clip]{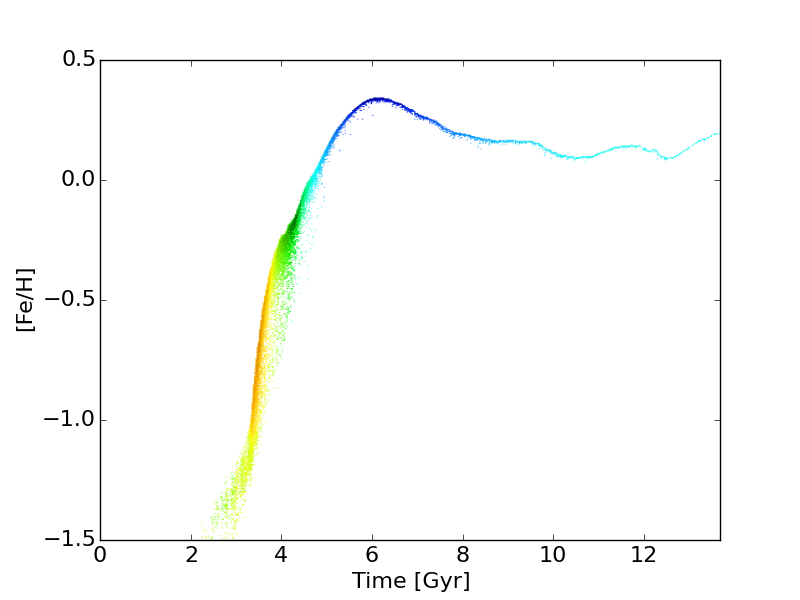} &
         \includegraphics[scale=.3,trim={0.cm 0 1.5cm 1cm},clip]{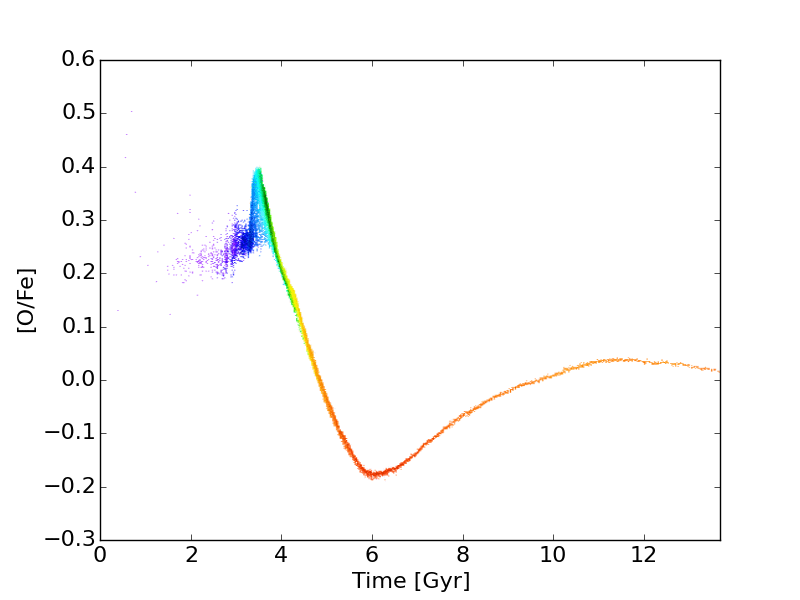} &         
         \includegraphics[scale=.3,trim={0.cm 0 1.5cm 1cm},clip]{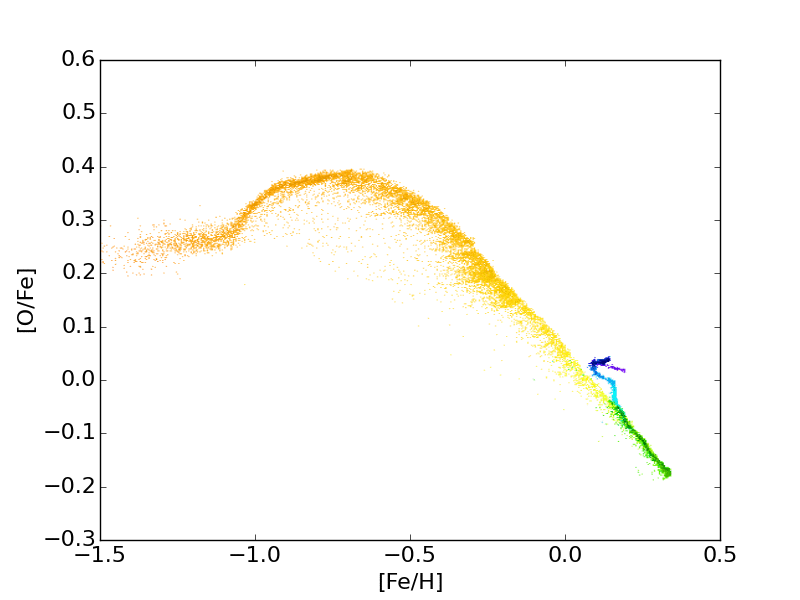} \\              
         (g) & (h) & (i) \\
   \end{tabular}
\caption{The AMR (left), time-[O/Fe] (centre) and [Fe/H]-[O/Fe] (right column) distributions for all stars in MUGS satellites at z=0 (top) and two specific satellites (middle and bottom rows). Plots are coloured as before.    }
\label{Fig:satllitesonly}
\end{figure*}

In the left column of Fig. \ref{Fig:satllitesonly} we extract the satellites from the galaxy and show the total amount of substructure they impart\footnote{An earlier analysis of the AMR, SFH, MDF, and [O/Fe]-[Fe/H] of Satellite A \& B can be found in Section 3.7 of \citet{Pilkington2013} but we will expand on this here}. The largest satellite, Satellite A, (shown in the centre column of Fig. \ref{Fig:satllitesonly})  contains 218000 star particles, 30000 dark matter particles and 2038 gas particles, and shows a spread in metallicity at any given time similar to the host galaxy. Satellite A  is a  large satellite, with a stellar mass of 9.66$\times$10$^9$ M$_\odot$. However, in a less massive satellite, Satellite B (right hand column),  all the gas is very close to the centre of the galaxy, either due to interactions or the resolution of the simulation. This satellite consists of $\sim$42000 star particles and has a stellar mass of 1.83 $\times$10$^9$ M$_\odot$. This narrows the spread of the metallicities and produces tight chemical path. 

Satellite A falls into the host galaxy at around 11 Gyr. During the initial infall there is no significant change in the AMR of the satellite. It is only as the galaxy passes pericentre (at approx 12.5 Gyrs) that the shape of the AMR changes. The narrowing of the spread in the stellar metallicities  (from 0.2 to 0.03 dex)  corresponds to passage close to the centre of the host, and is due to the interaction between the two objects, which is strongest when the separation is small. We see that the metallicity also dips significantly as the satellite comes close to the host. This close approach forces the gas into the centre of the satellite (and strips gas out). Thus, the new star formation all occurs at the centre of the satellite galaxy. The entire satellite galaxy becomes similar to the bulge of the host, although without any noticeable substructure. The pericentric passage of the larger satellite results in an increase in the oxygen to iron ratio (panel (e)), corresponding to a jump in the star formation rate.

Satellite B always has its gas near the centre, and thus, always has a bulge-like evolution. The satellite metallicity peaks at around 6 Gyr, to 0.4 dex, then falls with time to 0.2-0.3 dex by z=0. This is significantly different to the naive expectation that metallicity increases monotonically. This is true for many satellites in the MUGS sample. The origin of this drop in [Fe/H] corresponds to a rise in [O/Fe] (from -0.3 to -0.1 dex) after 6 Gyr in panel (f). After 4 Gyr, the satellite has consumed most of its gas, and has a very low star formation rate (0.03 solar masses per year, after peaking at 1.8 solar masses per year at 3.5 Gyr). It does, however, collect a few particles from the hot halo during its orbit inside the host. The metallicity will fall if the rate at which low metallicity material is picked up is larger than the enrichment rate. However,  the age-[O/Fe] distribution increases after 6 Gyr, so the oxygen is not diluted to the same extent as the iron. After 6 Gyr, the age-[O/H] distribution (not shown) is flat. We suggest the following scenario. The satellite has a fairly shallow potential, so material is easily lost from the halo, if bound only by gravity. However, if the supernova goes off in a high density region hydrodynamical interactions can contain the ejecta.  CCSNae occur in star-forming regions, so their ejecta are contained by the surrounding high-density gas, but SNIa can occur in low gas density regions so the iron is lost. Slow infall of low metallicity gas, therefore, dilutes the iron, but not the oxygen.

The other panels in the figure show the other projections of the metallicity. Satellite A has the same form of multi-sequence [Fe/H]-[O/Fe] distribution as the main galaxy (it differs in the details but not the overall shape). This is because the galaxy is large and its gas distributed over a considerable volume, while the smaller satellite (panel i) shows only a single sequence, because of the concentration of gas at the centre of its potential well.

Satellite galaxies with strong central concentrations of gas also have much lower [O/Fe] minima than either the host, or satellites with more a distributed ISM. This implies that the resolution (mass and spatial) of the simulation is insufficient to accurately model smaller satellites, and that the very iron rich objects are so because they are poorly resolved.

\subsubsection{Sawtooth Galaxy}
\begin{figure}
\centering
     \begin{tabular}{cc}
              \includegraphics[scale=.29,trim={0.cm 1.3cm 1.5cm 1.5cm},clip]{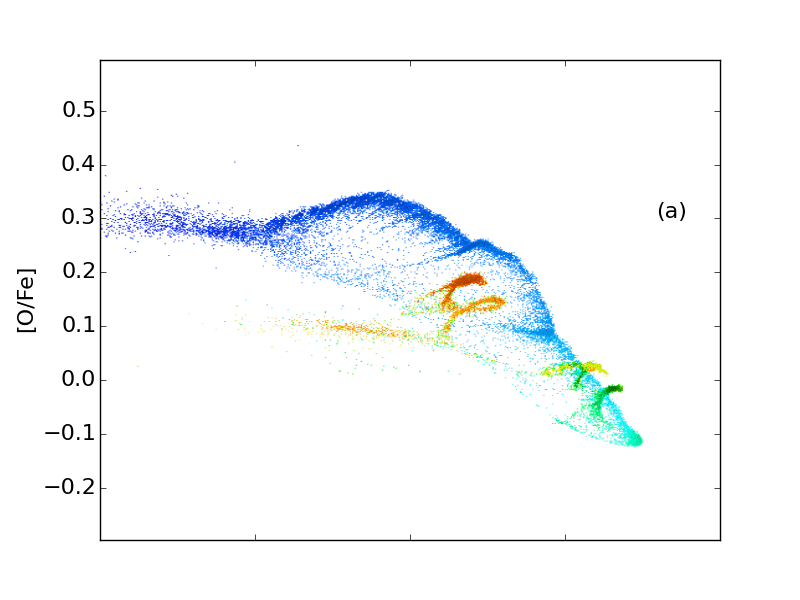} &
         \includegraphics[scale=.29,trim={7.cm 1.3cm 1.5cm 0.55cm},clip]{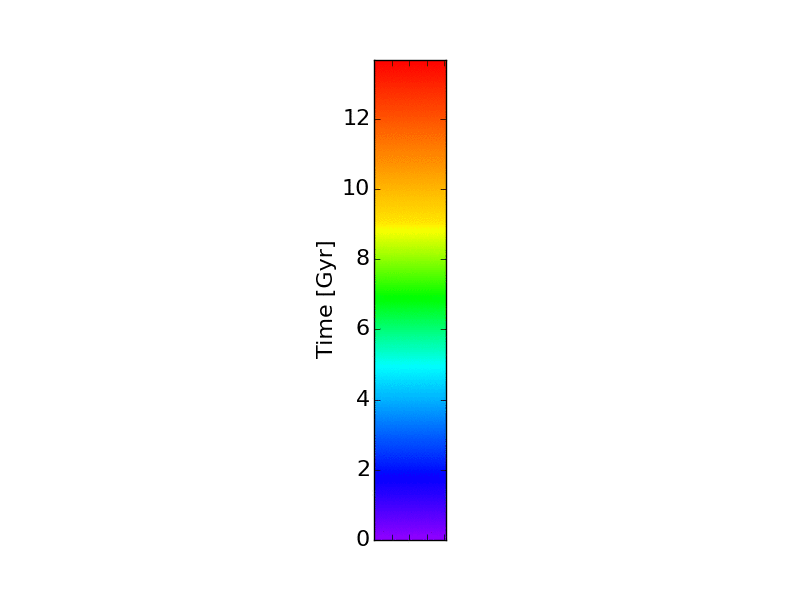}  \\  
               \includegraphics[scale=.29,trim={0.cm 0cm 1.5cm 1.5cm},clip]{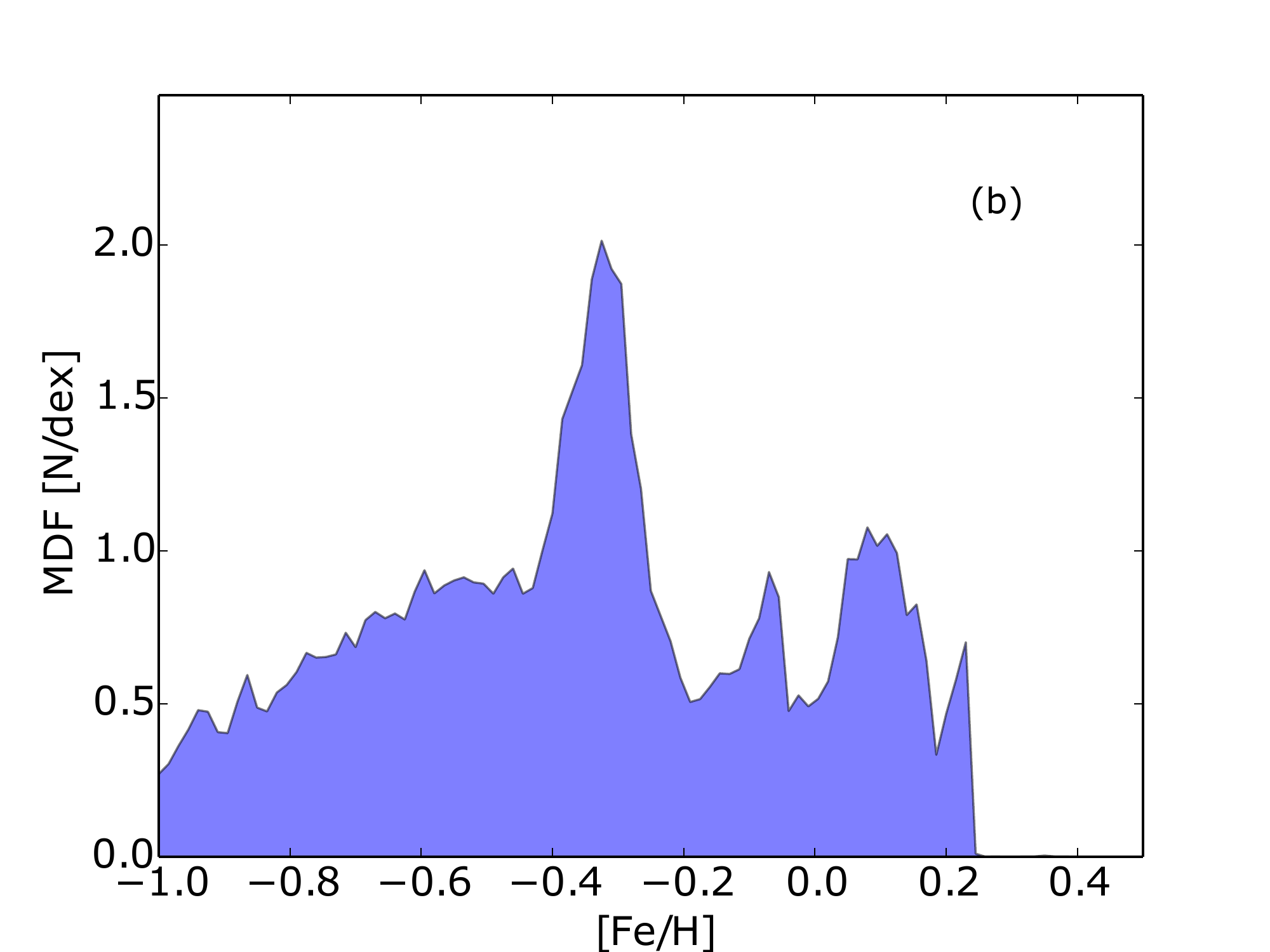} \\
          \includegraphics[scale=.29,trim={0.cm 1.4cm 1.5cm 1cm},clip]{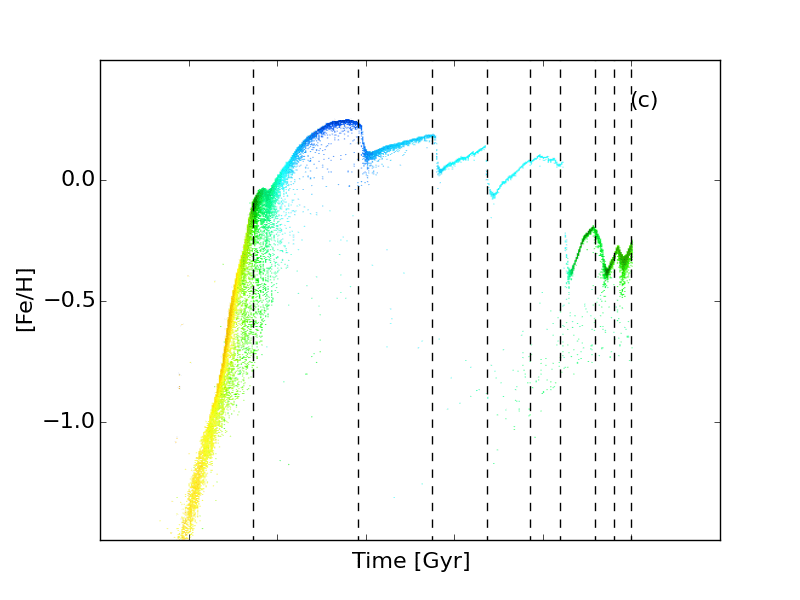} &
          \includegraphics[scale=.29,trim={7.cm 1.4cm 1.5cm 1cm},clip]{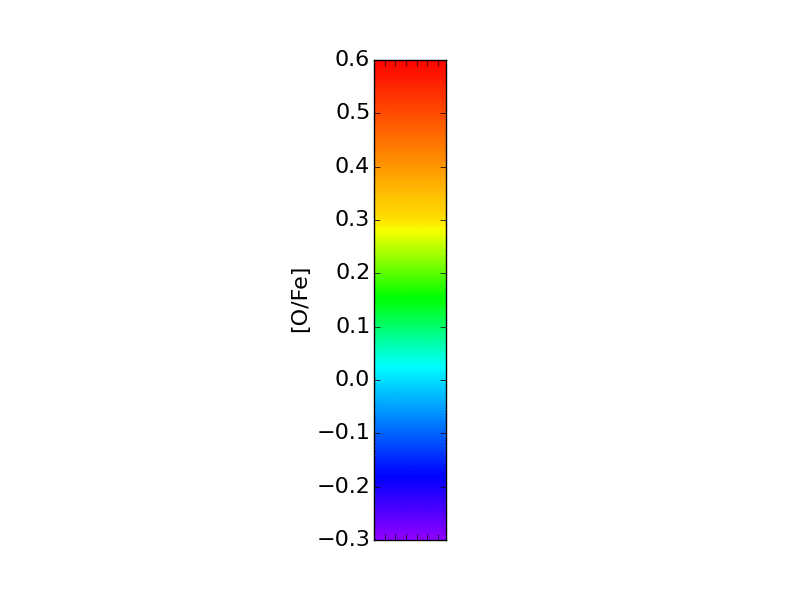} \\              
   
         \includegraphics[scale=.29,trim={0.cm 1.4cm 1.5cm 1.54cm},clip]{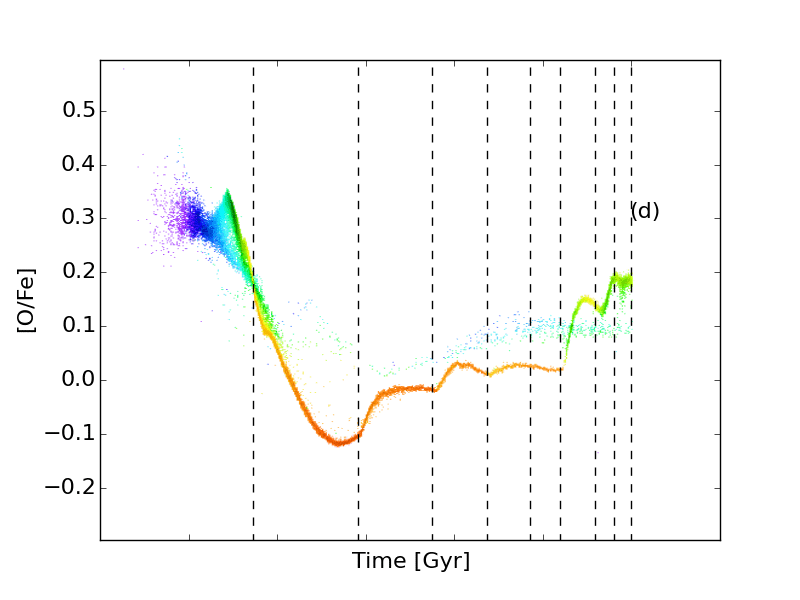} &
         \includegraphics[scale=.29,trim={7.cm 1.4cm 1.5cm 0.55cm},clip]{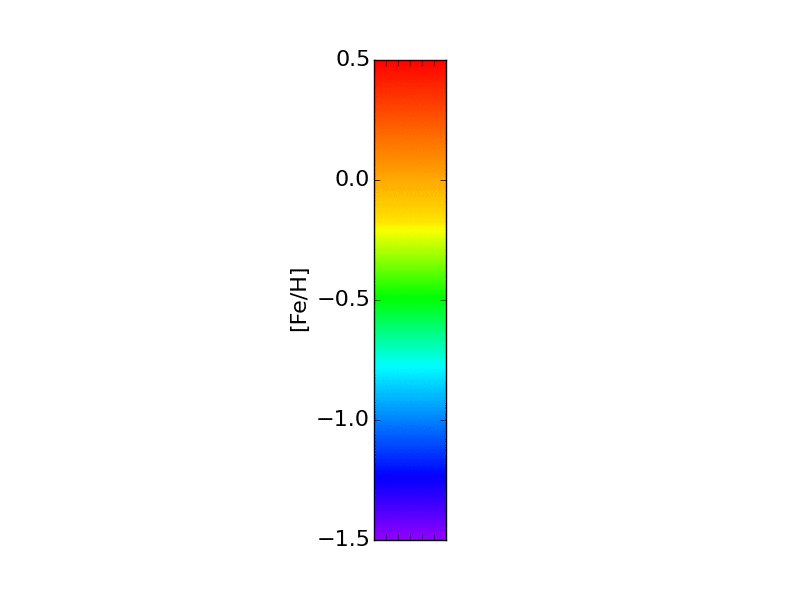}  \\        
               
         \includegraphics[scale=.29,trim={0.cm 0.0cm 1.5cm 1.54cm},clip]{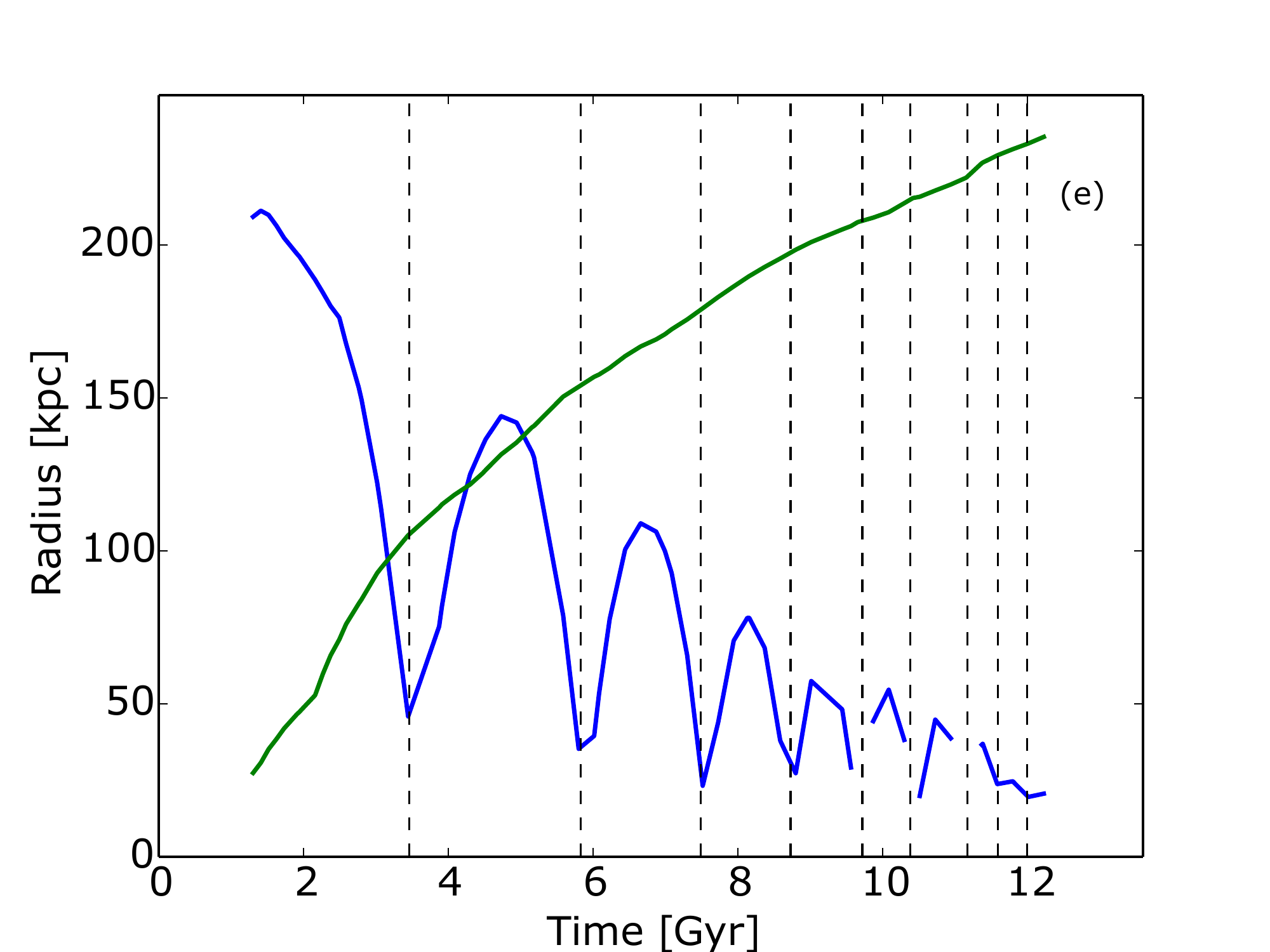} \\

          \\

   \end{tabular}
\caption{Top to bottom:  [Fe/H]-[O/Fe] with time (panel a), metallicity distribution function (panel b) and AMR (panel c), the time-[O/Fe] evolution (panel d). Plots are coloured as in Fig. \ref{Fig:allgalallplots}.   Panel (e) shows the radius of the orbit of the sawtooth galaxy (blue line) and the virial radius of the host galaxy (green). The missing points in panel (e) are where the halo finder cannot distinguish the satellite from the host. In each panel the vertical dotted lines indicate the minimum radius of the various orbits. It is difficult to trace the pericentre at times beyond the last vertical line shown because the output cadence of the snapshots is insufficient. After 10 Gyr, the peak of [Fe/H] corresponds to the pericentre. 
}
\label{Fig:sawtooth}
\end{figure}

An exceptional illustration of  non-monotonic evolution is seen in Fig. \ref{Fig:sawtooth}, which we will call the ``Sawtooth Galaxy'' because of the shape of its AMR. This satellite has a stellar mass of 3.36$\times$10$^{9}$M$_{\odot}$ and a total mass of  6.6$\times$10$^{9}$M$_{\odot}$ at $z=0.12$. It contains $\sim$59000 star particles.  The AMR has a saw-tooth appearance  because the metallicity rises, falls, rises again, etc., with a decreasing period. For much of its lifetime, this satellite galaxy has a metallicity higher than the disc. The lower part of the `sawtooth' feature, however, is of similar metallicity to the disc. This particular satellite has completely merged by $z=0$, and so we identify it at $z=0.12$ and follow it back in time. 

Figure \ref{Fig:sawtooth} shows that the regular drops in the metallicity co-coincide with the pericentre of the orbit of the satellite. The radius of closest approach means that the satellite passes through the outer edge of the galaxy disc. The cold gas in the disc of the galaxy has a metallicity gradient, and thus the gas at the edge of the disc has a comparatively low metallicity. This low $Z$ is close to the metallicity of the stars formed in the satellite at each local minima, at least before the profile changes from `sawtooth' to `m'-shaped  (which occurs when the satellite merges with the bulge). Although the minimum orbital radii of the galaxy shown in Fig. \ref{Fig:sawtooth} implies that the galaxy stays out at approximately 30 kpc, this is due to the output cadence of the simulation. An orbital integration with high time resolution shows that the satellites does in fact pass through the cold gas disc (Fig. \ref{Fig:satorbits}).

\begin{figure}
\centering
     \begin{tabular}{c}
      \includegraphics[scale=.4,trim={0.5cm 0 0cm 1cm},clip]{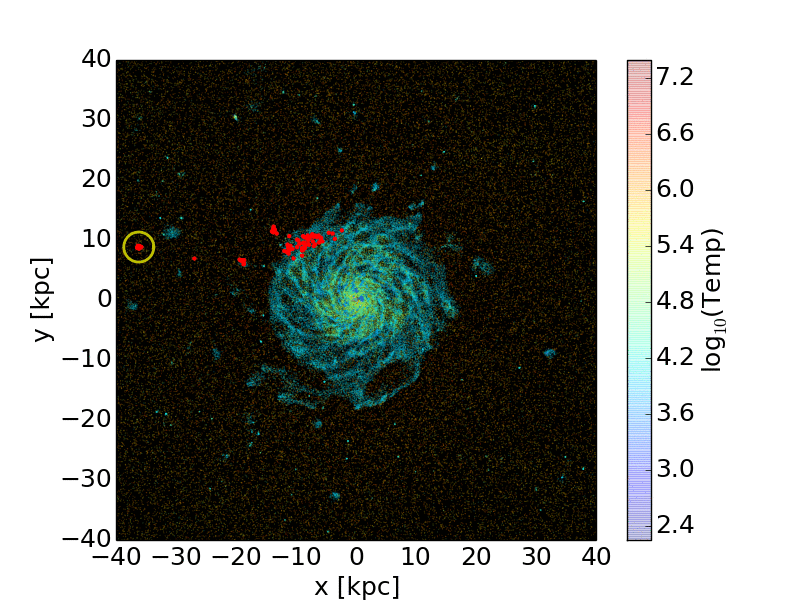}\\
      \includegraphics[scale=.4,trim={0.cm 0 0cm 1cm},clip]{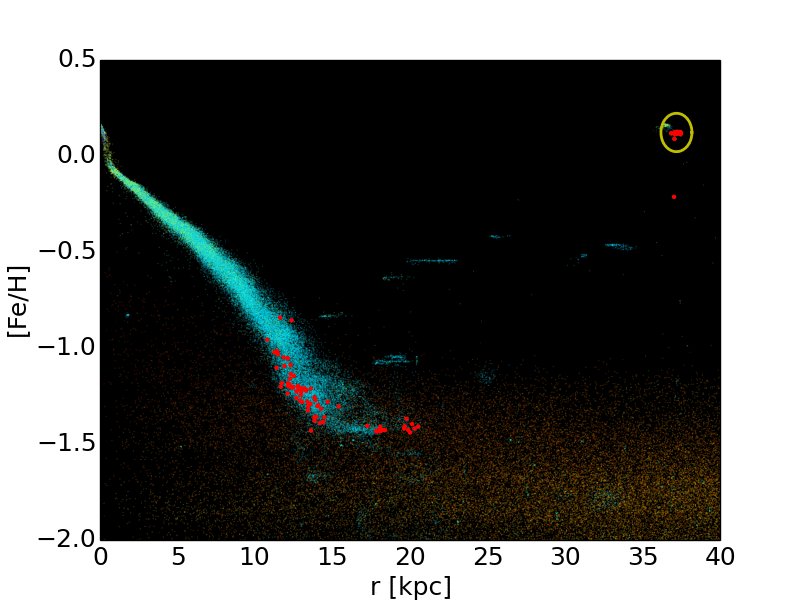}\\  
      \includegraphics[scale=.4,trim={0.cm 0 0cm 1cm},clip]{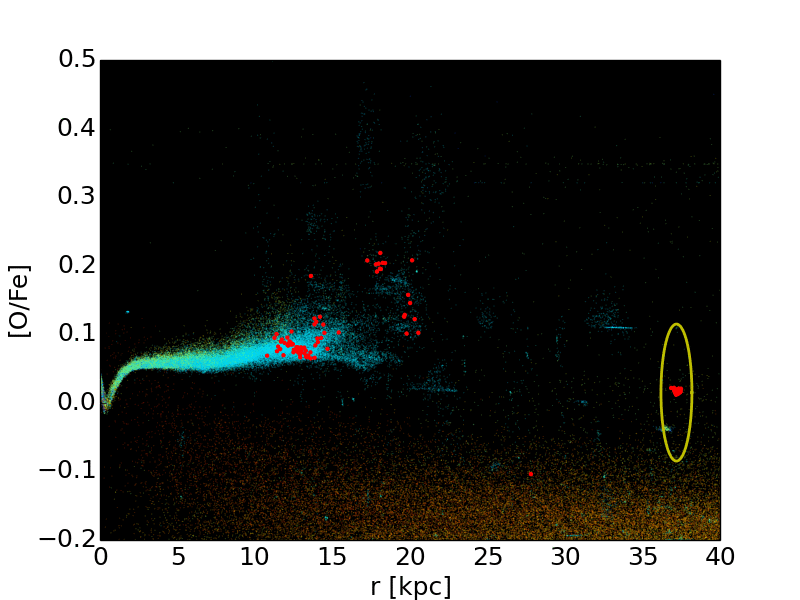}\\            
   \end{tabular}
\caption{Top: The distribution of gas in the host galaxy coloured by the log temperature. The red points are the gas particles at 8.58 Gyrs which form stars between 8.79 and 9.01 Gyrs. These correspond to the 4th dip in the sawtooth shown in Fig. \ref{Fig:sawtooth}. The yellow circle shows the current position of the sawtooth satellite. Middle: The metallicity (y-axis) and radius (x-axis) of gas in the host galaxy and the distribution of gas which form stars between 8.79 and 9.01 Gyrs (red points). The yellow circle shows the mean [Fe/H] and radius of the sawtooth satellite. Bottom: Coloured as before, showing the radial [O/Fe] gas distribution of the host and satellite. }
\label{Fig:wheregasfrom}
\end{figure}

\begin{figure}
\centering
     \begin{tabular}{c}
      \includegraphics[scale=.4,trim={0.5cm 0 0cm 1cm},clip]{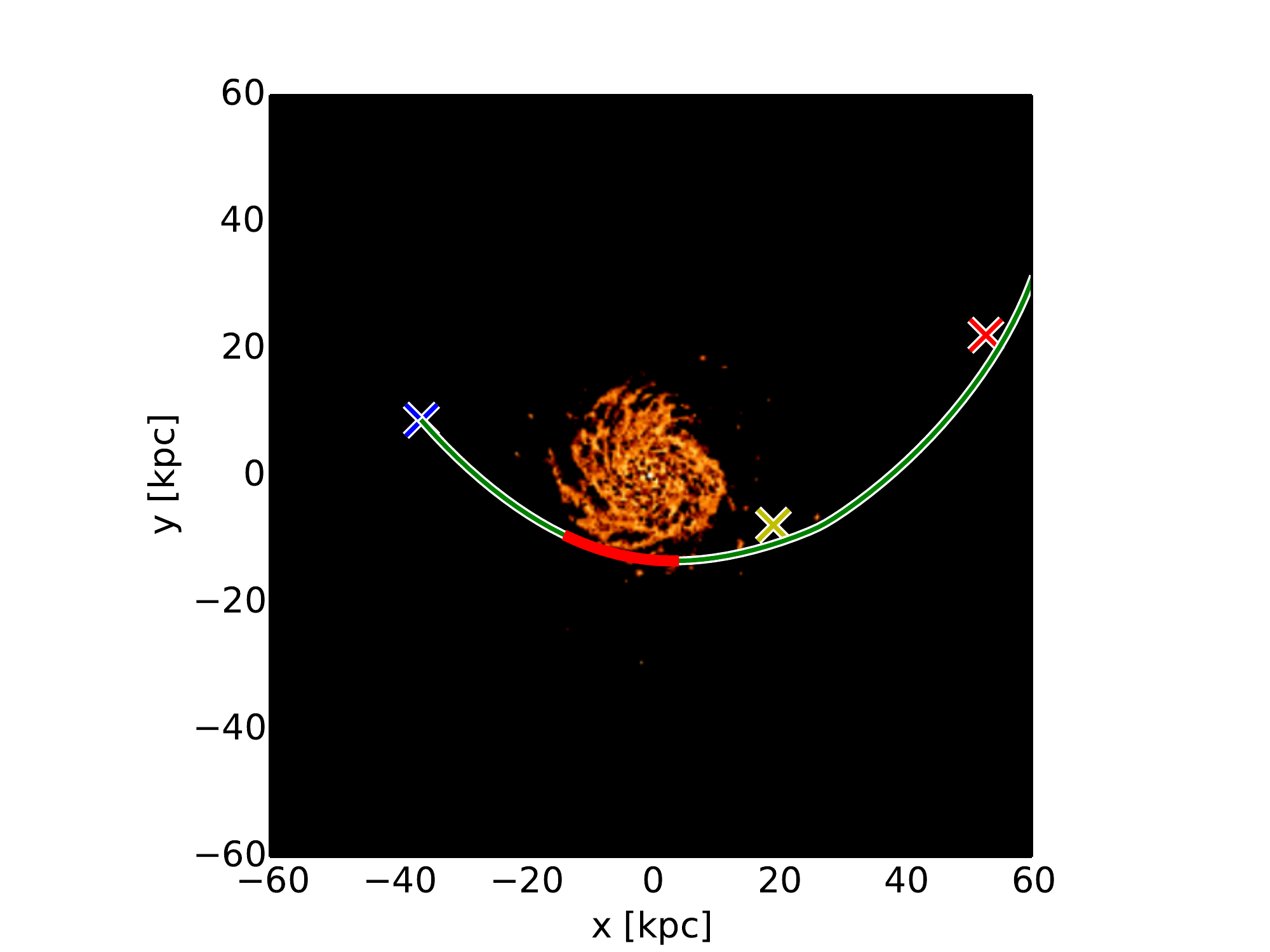}\\
      \includegraphics[scale=.4,trim={0.cm 0 0cm 1cm},clip]{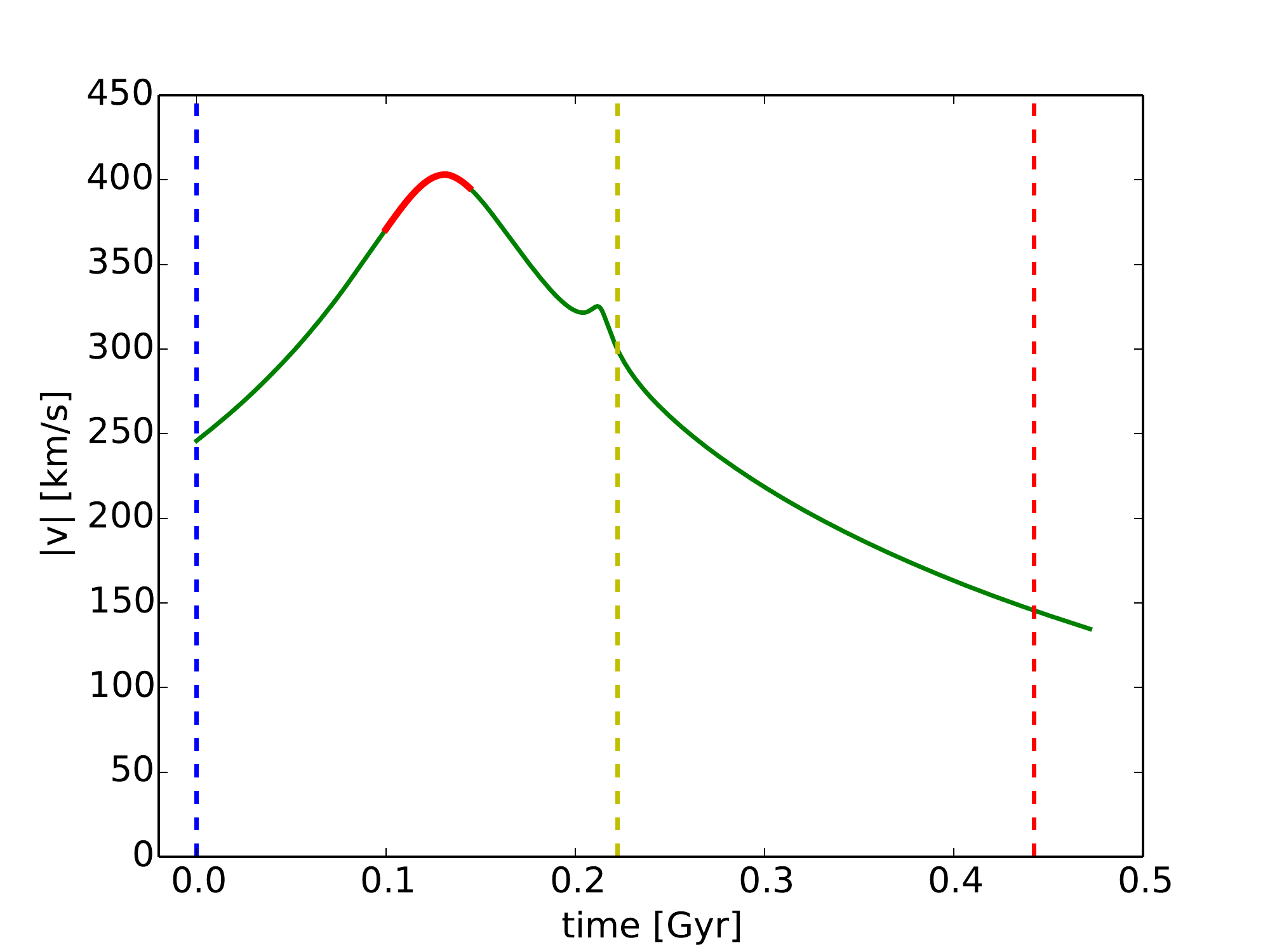}\\        
   \end{tabular}
   \caption{Top: The distribution of gas in the host galaxy coloured by the number density of points at the snapshot corresponding to the yellow cross. The green line follows the orbit of the 'sawtooth' galaxy as calculated in a frozen gravitational potential, while the red line shows the part of the orbit when the satellite is passing through the gas disc of the host. The red line crosses the disc between z= $\pm$1.5 kpc.  The blue, yellow and red crosses show the actual positions of the sawtooth galaxy at three timesteps surrounding one of the sharp drops in metallicity. Bottom: The speed of the galaxy over the orbit shown above. The red line is the part of the orbit the satellite is passing through the disc of the host. The vertical lines correspond to the crosses in the upper panel. }
\label{Fig:satorbits}
\end{figure}

Moreover, careful examination of the stars formed after one of the pericentric passages finds that a sizeable fraction of the stars birthed at that time formed from gas from the cold gas disc of the host in previous outputs (Fig. \ref{Fig:wheregasfrom}). These low metallicity (high oxygen) particles from the outer disc of the host reduced the metallicity of all gas particles in the satellite due to diffusion. When the gas is converted into stars the low $Z$ (high oxygen) population is produced. The metallicity of the satellite gas then rises due to the gas and metals released back into the  ISM from stars via CCSNe and SNIa. Figure \ref{Fig:wheregasfrom} demonstrates that a substantial fraction of the particles which end up as stars in the satellite within 2 snapshots came from the host disc (77 out of 158). At the time shown in the figure the satellite is on a retrograde orbit in the disc plane. It contains 180 gas particles initially and receives another 77 from the disc during the interaction. 

There is one orbit of this satellite, with a pericentric passage at 10 Gyr, which does not correspond to a fall in the  metallicity of the stars formed at that epoch. Although the  cadence of the outputs, and the uncertainty of the evolution of the potential, does not allow us to directly observe why this occurs, we suggest that the satellite misses the edge of the dense cold gas disc during this orbit, and so does not interact as strongly with the host galaxy as on the other orbits.

In order to test whether the accretion of disc gas that causes the `sawtooth' is physically realistic, rather than an artefact of the SPH method, we calculate the approximate orbit of the galaxy between the two simulation outputs that span its passage through the disc plane\footnote{The snapshots are spaced 200 Myr apart, too sparse for us to determine the orbit directly from the simulation outputs. Integrating an orbit does not yield a final state that matches the position of the satellite in the next snapshot. This form of calculation, therefore, is insufficient to precisely discover whether the satellite passes through the disc or not.}. Fig. \ref{Fig:satorbits} shows the orbit calculated in a frozen potential, and demonstrates that the path of the galaxy's orbit passes close to the edge of the gas disc, where it picks up gas. At this point the galaxy is travelling with a speed of $\sim$420 km/s.

Using the classic \citet{Gunn1972} criterion for ram pressure stripping, a medium of density $\rho$ can strip the ISM of a galaxy with gas surface density $\Sigma_{\mathrm{ISM}}$ and total dynamical surface density $\Sigma_*$ if the relative velocity satisfies

\begin{equation}
\label{eq:gunngott}
  V^2 > \frac{2 \pi G \Sigma_* \Sigma_{\mathrm{ISM}}}{\rho}.
\end{equation}

At a radius of 15~kpc, the surface density of the gas and stars in the disc are $\Sigma_{\mathrm{ISM}} \approx 3 \times 10^6~\mathrm{M_{\sun}~kpc^{-2}}$ and $\Sigma_* \approx 3 \times 10^6~\mathrm{M_{\sun}~kpc^{-2}}$ respectively. The stripping medium comes from the satellite, which has a gas mass of $4.5\times 10^7~\mathrm{M_{\sun}}$. In the simulation, this gas all congregates within the central 2-3 smoothing lengths. However, even if we spread it out over a radius of 2~kpc Eqn. \ref{eq:gunngott} implies stripping would occur for  an interaction velocity of $V > 17~\mathrm{km/s}$, which is easily satisfied. It remains a possibility that this process is an artefact of SPH, and the resolution of MUGS. The degree to which the low metallicity gas particles can be incorporated successfully into the satellite, in order to produce the required dilution, is unknown. Much of the satellite gas lies in the very centre of the satellite, partly due to resolution issues with MUGS on these scales. However, whether a higher resolution simulation would destroy the `sawtooth' effect completely is unknown,  and would require further investigation beyond the scope of this paper. If the surface density of the gas in the satellite were sufficiently low, gas ought to be stripped from both the disc and satellite, rather than just being incorporated into the satellite.

The transformation of the gas particles stripped from the disc is rapid, see \citet[][]{Brook2014}. However, the gas particles are drawn quickly into the bottom of the potential well by interactions. In this very dense region star formation takes place especially  rapidly.

Satellite galaxies of the Milky Way, and other nearby galaxies, might exhibit similar features in the metallicities of stars. However, accurate ages are important if we are to unambiguously see this `sawtooth' feature in observations. Using the more easily observed [Fe/H]-[O/Fe] distribution, the `sawtooth' cannot easily be identified. Arcs on the [Fe/H]-[O/Fe] distribution panel of Fig. \ref{Fig:sawtooth} lie directly on top of the main relation, and are only really evident because they have different ages. However, if we plot the MDF of the stars (Fig. \ref{Fig:sawtooth}) we can see significant secondary peaks in the metallicity distribution for [Fe/H]$>-0.4$. This observational clue would be significantly easier in observations than deriving sufficiently good ages. 

We only expect to see this `sawtooth' feature if the satellite galaxy is more metal rich than the outer edge of the cold gas disc. This occurs in the MUGS simulations because the most massive satellites have high stellar masses compared to expectations, due to inefficient feedback. This means that the satellites are also more metal rich than satellites in the real universe, as they still lie on the mass-metallicity relation. Thus, the contrast between the top and bottom of the sawtooth may not be as extreme in reality.  

 However, another place we might expect to see a similar sharp drop in the metallicity of a galaxy is in close pairs. As the two galaxies interact, gas can be stripped from the outer edge of the ISM of one galaxy and be incorporated into the other, so newly-formed stars have lower metallicity \citep[e.g.][]{Kewley2010}. This may also explain the drop in the metallicity of stars formed during the pericentric passage of Satellite A.

\subsection{Comparison to observations}
\label{Sec:Obs}

The chances of observing the rich substructure identified in the previous sections depends on how readily observable a given feature is when the data are convolved with  observational errors. A robust feature is one that would be visible even with a moderate degree of error. In a follow up paper we will go into greater detail comparing the results of simulations and observations using synthetic color-magnitude diagrams and mock observational reconstructions of the AMR \citep[c.f.][where the importance of employing synthetic CMDs in the analysis of simulations and their comparison with empirical data is demonstrated]{Miranda2015a}. In this future paper we will attempt to reconstruct as much information encoded in the simulations using observational techniques as we can. In the mean time we can gain an initial, at least qualitative, estimate by adding a small Gaussian error to each data point in an effort to model observational errors. We first use errors on recently published works, for example, in the Milky Way, and then we decreased those errors to identify the point at which the substructure begins to reveal itself.

For ages, we initially took errors to be 1 Gyr \citep{Haywood2013} which is based on very good Milky Way data \citep{Adibekyan2012}. This error is partly due to  our understanding of stellar physics which relates stellar ages to observations of stars, thus, we require improvements to our theoretical understanding before we can reduce the errors on observed ages in forthcoming surveys such as Gaia and its spectroscopic follow ups. We assume errors in the abundances to be 0.1 dex for metallicity and 0.1 dex for [O/Fe]. We then applied a normal distribution with a standard deviation matching that error. Errors in, for example, Gaia, for M3 stars out to 10 kpc, can be expected to be around 13\% or 1.5 Gyr for the oldest stars \citep{Cacciari2009}.

\begin{figure*}
\centering
     \begin{tabular}{ccc}
          
      \includegraphics[scale=.3,trim={0cm 0 1.5cm 1cm},clip]{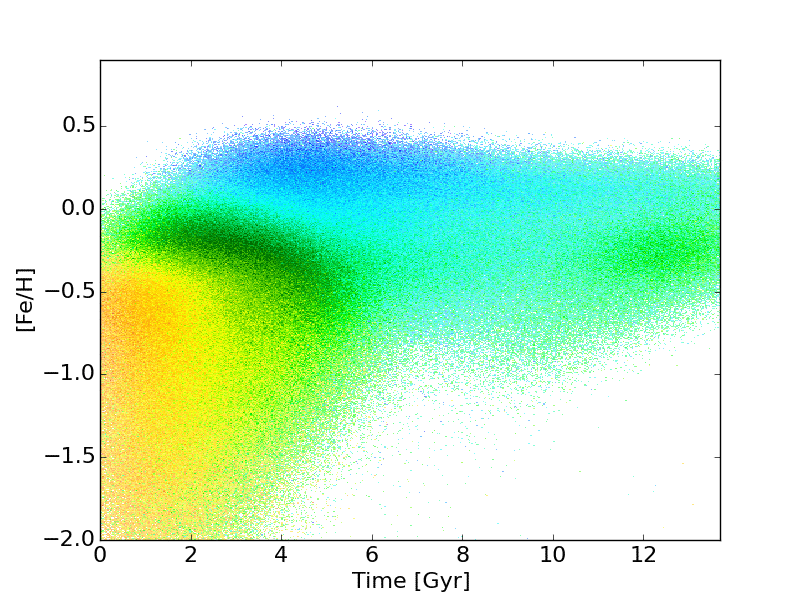}&
      \includegraphics[scale=.3,trim={0cm 0 1.5cm 1cm},clip]{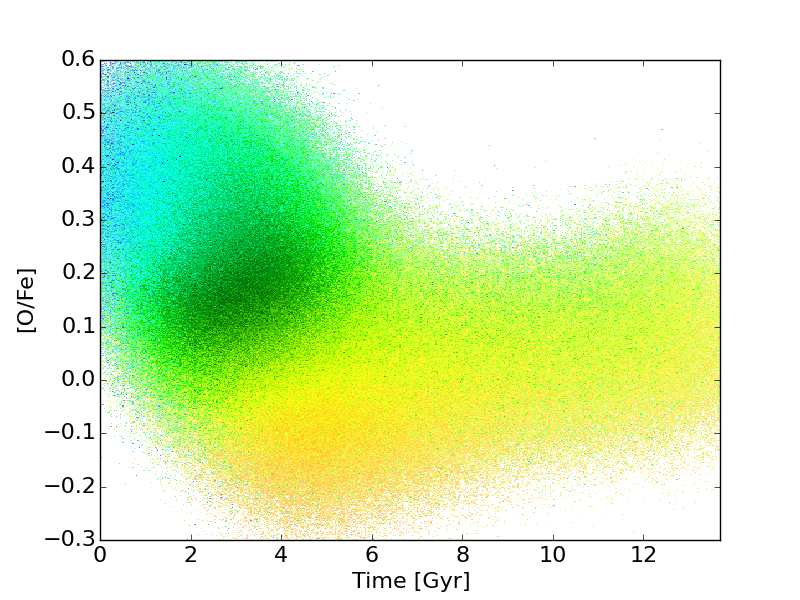}& 
      \includegraphics[scale=.3,trim={0cm 0 1.5cm 1cm},clip]{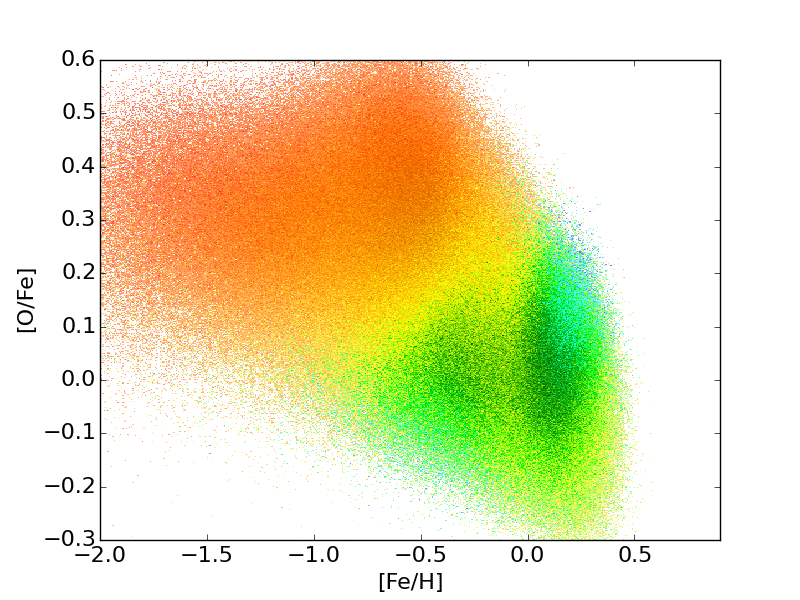}\\   
               
       \includegraphics[scale=.3,trim={0cm 0 1.5cm 1cm},clip]{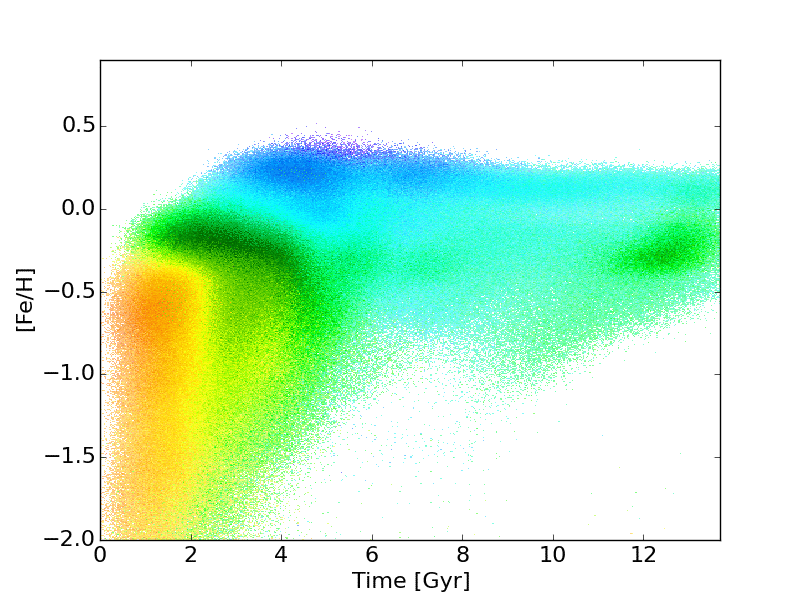}&
       \includegraphics[scale=.3,trim={0cm 0 1.5cm 1cm},clip]{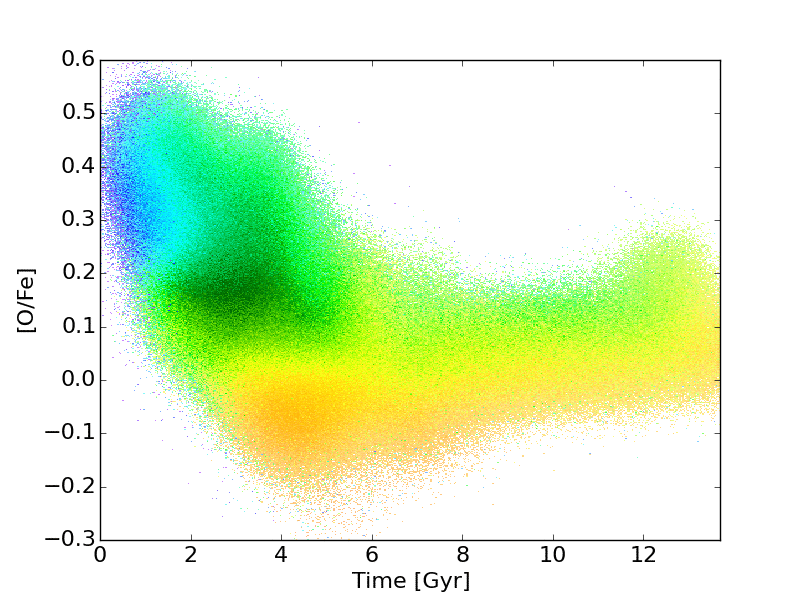}& 
      \includegraphics[scale=.3,trim={0cm 0 1.5cm 1cm},clip]{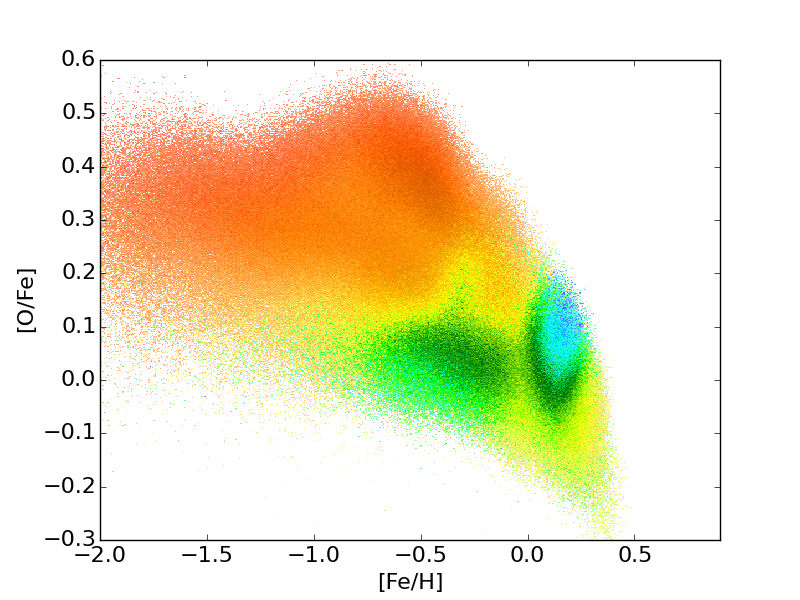}\\ 
           
      \includegraphics[scale=.3,trim={0cm 0 1.5cm 1cm},clip]{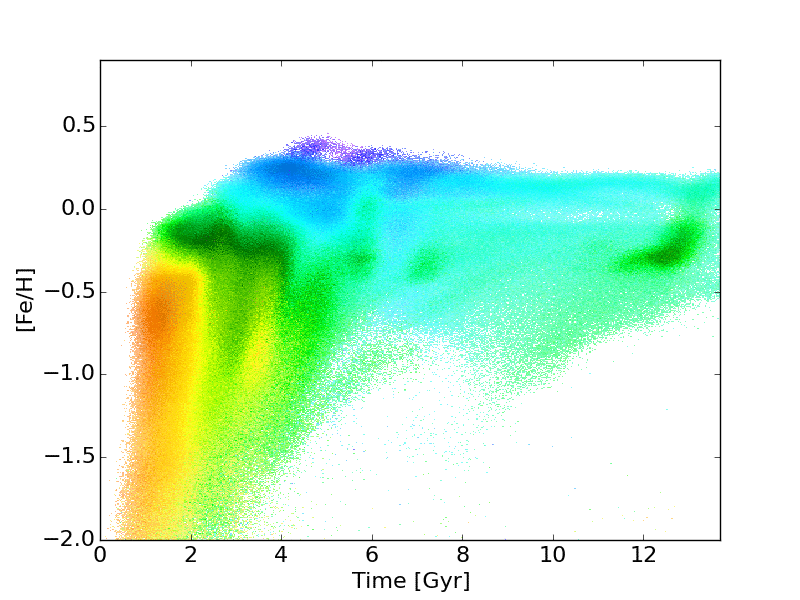}&
      \includegraphics[scale=.3,trim={0cm 0 1.5cm 1cm},clip]{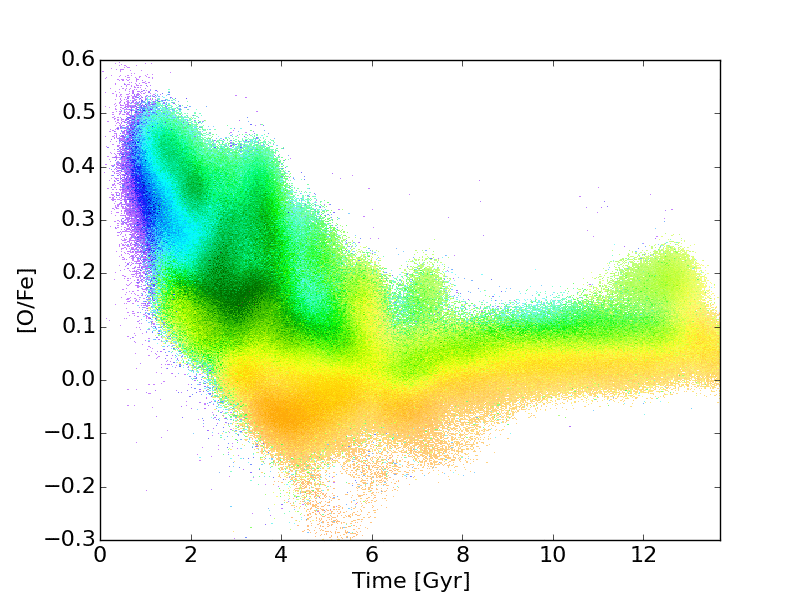}&  
      \includegraphics[scale=.3,trim={0cm 0 1.5cm 1cm},clip]{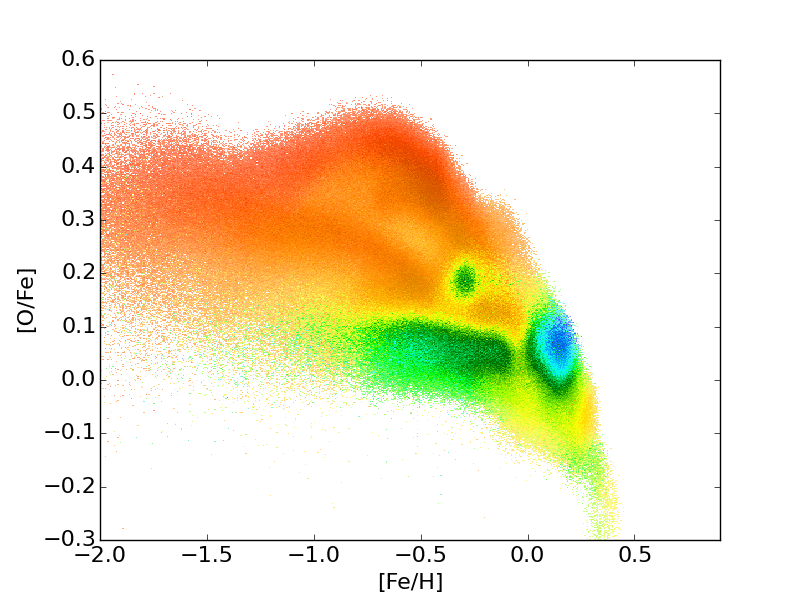}\\    
      
       \includegraphics[scale=.3,trim={0cm 0 1.5cm 1cm},clip]{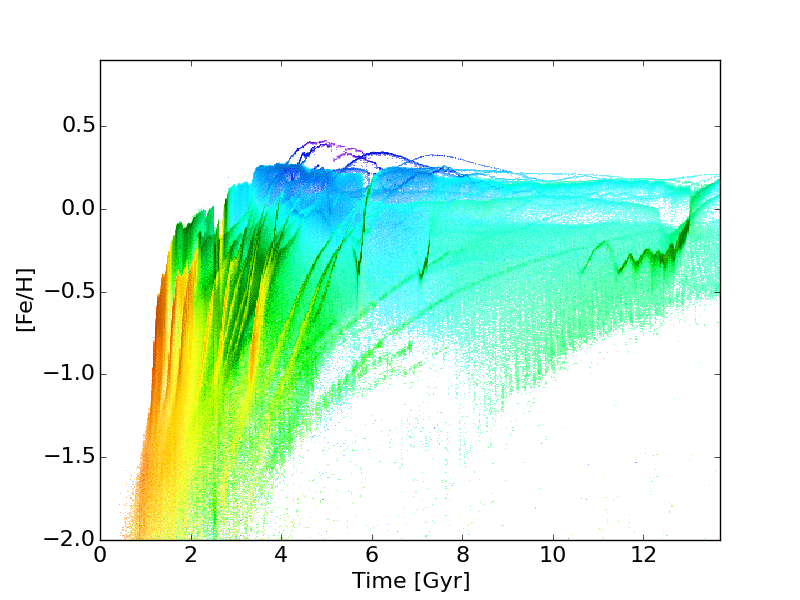}&
      \includegraphics[scale=.3,trim={0cm 0 1.5cm 1cm},clip]{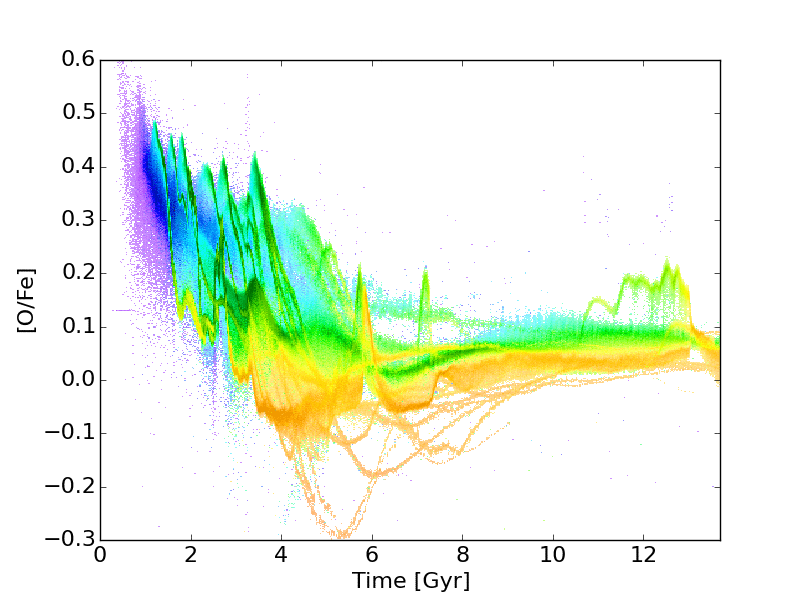} &
      \includegraphics[scale=.3,trim={0cm 0 1.5cm 1cm},clip]{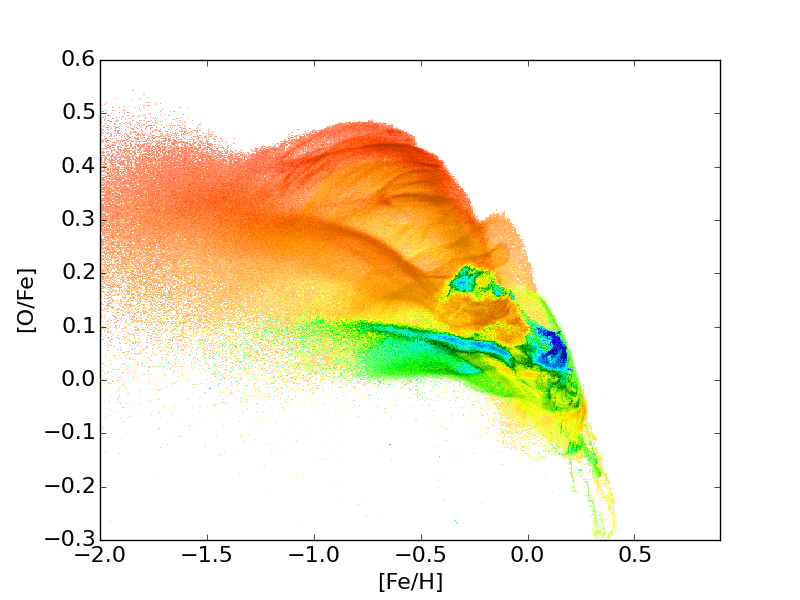}\\         
   \end{tabular}
\caption{ The age-metallicity, age-[O/Fe] and metallicity-[O/Fe] distributions for MUGS convolved with random errors. We add a random number chosen from a normal distribution of width $\sigma$ for age, [Fe/H] and [O/Fe]. From the top row to the bottom row these errors are ($\sigma_{age}$,$\sigma_{[Fe/H]}$,$\sigma_{[O/Fe]}$) = (1 Gyr, 0.1 dex, 0.1 dex), (0.5 Gyr, 0.05 dex, 0.05 dex), (0.25 Gyr, 0.025 dex, 0.025 dex), (0 Gyr, 0.0 dex, 0.0 dex).  }
\label{Fig:convolvederror}
\end{figure*}

\begin{figure*}
\centering
     \begin{tabular}{ccc}
          
      \includegraphics[scale=.3,trim={0cm 0 1.5cm 1cm},clip]{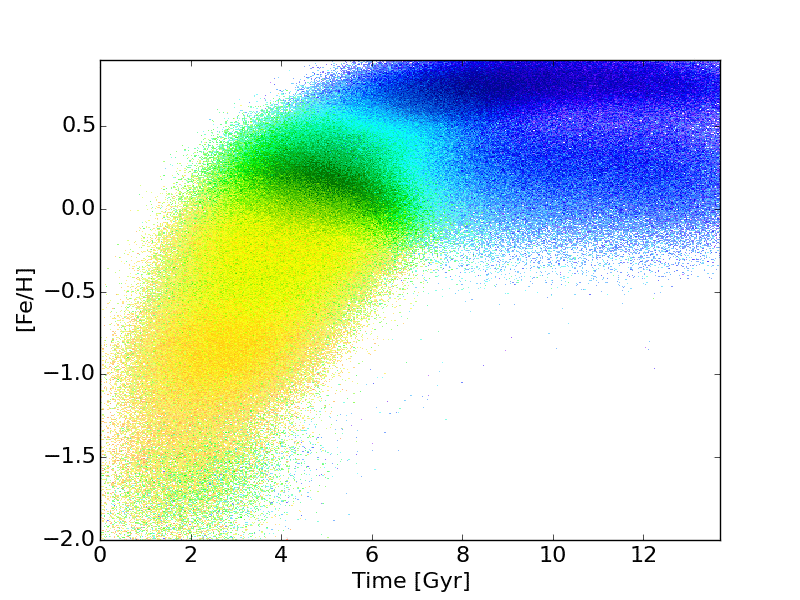}&
      \includegraphics[scale=.3,trim={0cm 0 1.5cm 1cm},clip]{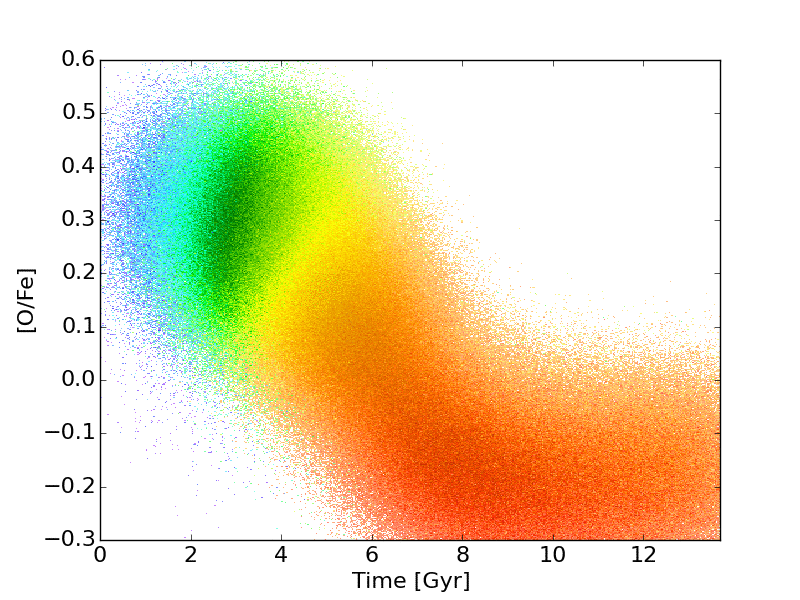}& 
      \includegraphics[scale=.3,trim={0cm 0 1.5cm 1cm},clip]{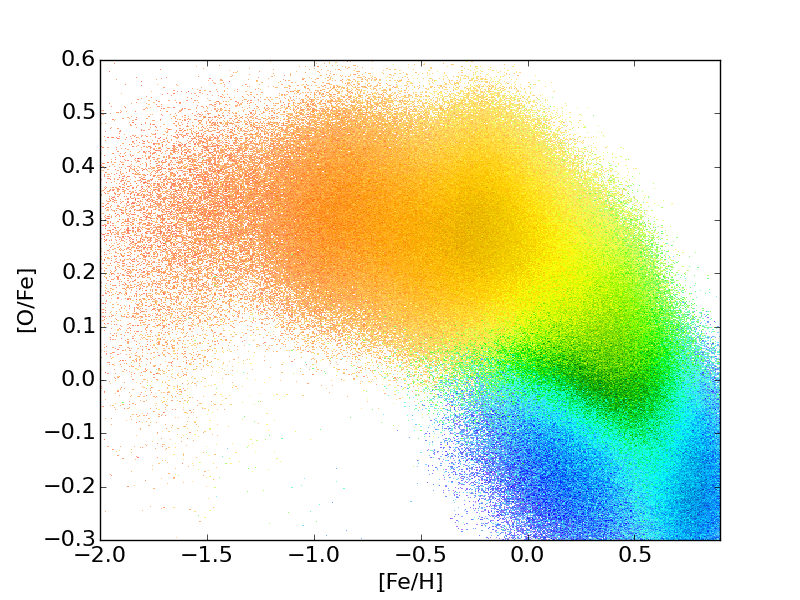}\\   
               
       \includegraphics[scale=.3,trim={0cm 0 1.5cm 1cm},clip]{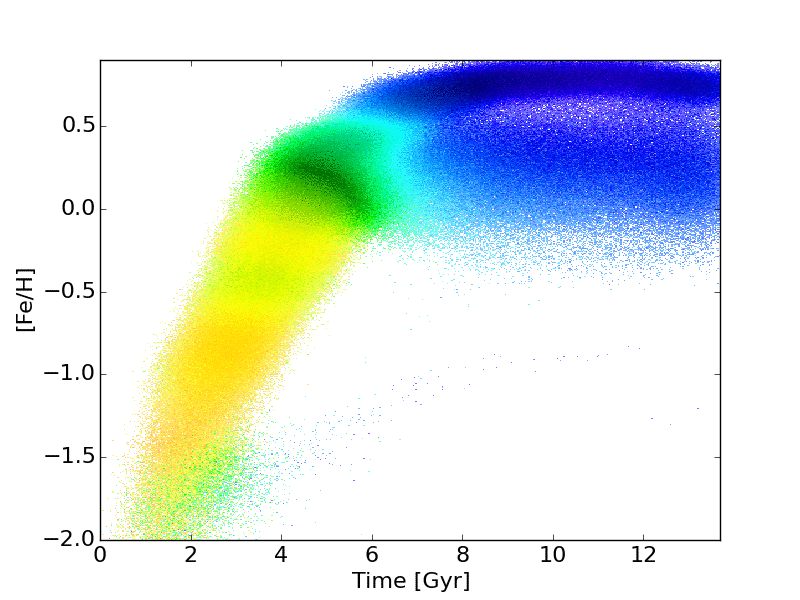}&
       \includegraphics[scale=.3,trim={0cm 0 1.5cm 1cm},clip]{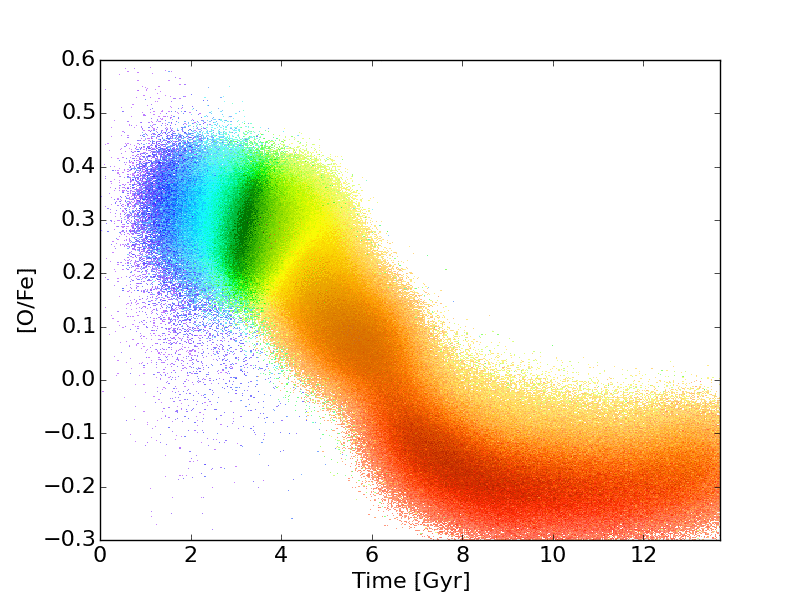}& 
      \includegraphics[scale=.3,trim={0cm 0 1.5cm 1cm},clip]{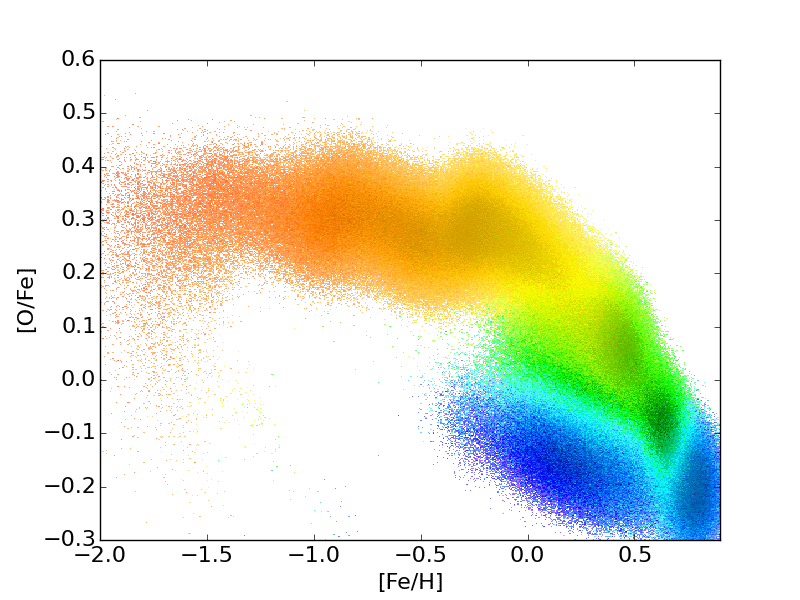}\\ 
           
      \includegraphics[scale=.3,trim={0cm 0 1.5cm 1cm},clip]{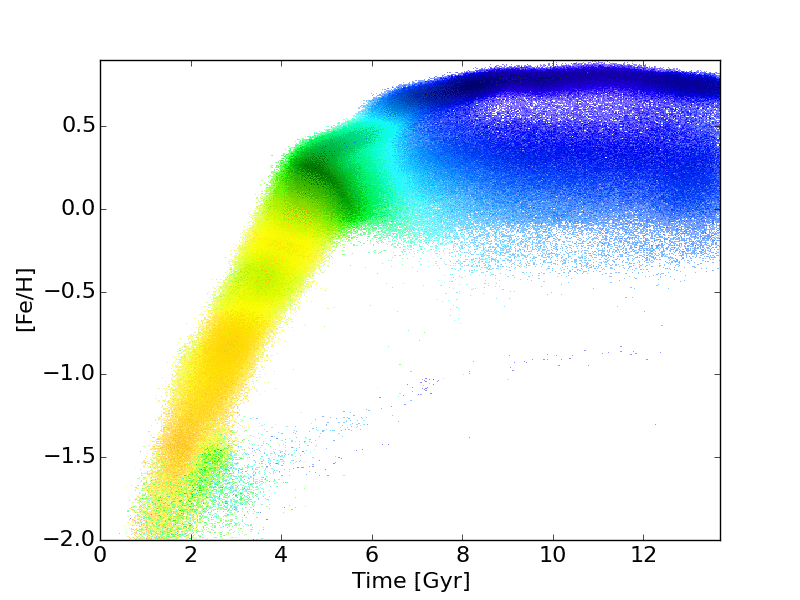}&
      \includegraphics[scale=.3,trim={0cm 0 1.5cm 1cm},clip]{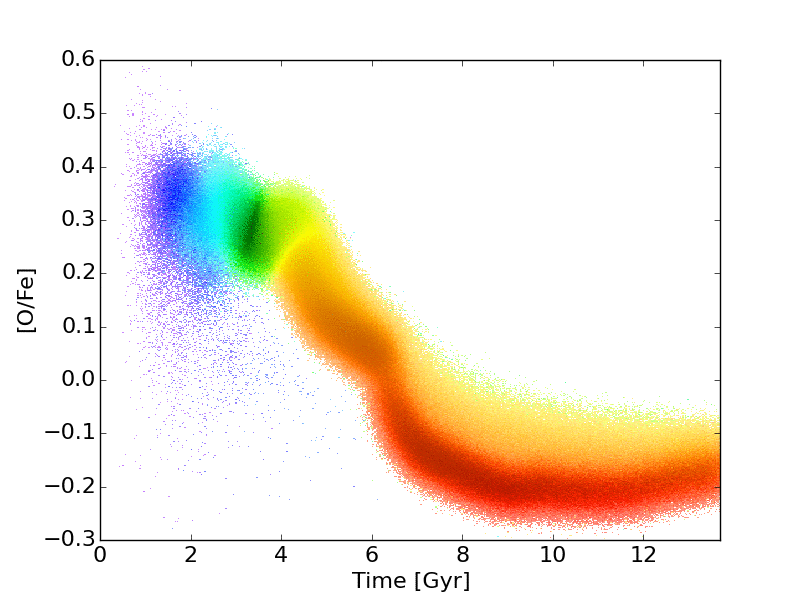}&  
      \includegraphics[scale=.3,trim={0cm 0 1.5cm 1cm},clip]{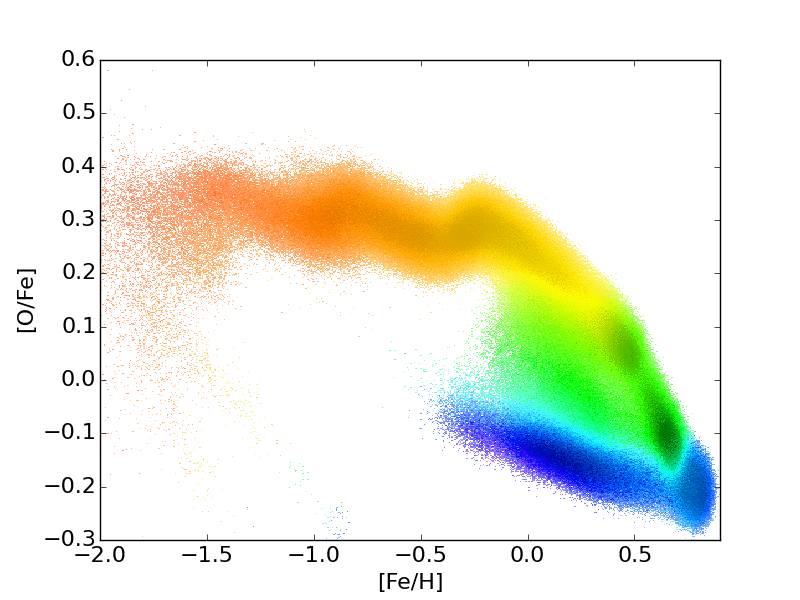}\\    
      
       \includegraphics[scale=.3,trim={0cm 0 1.5cm 1cm},clip]{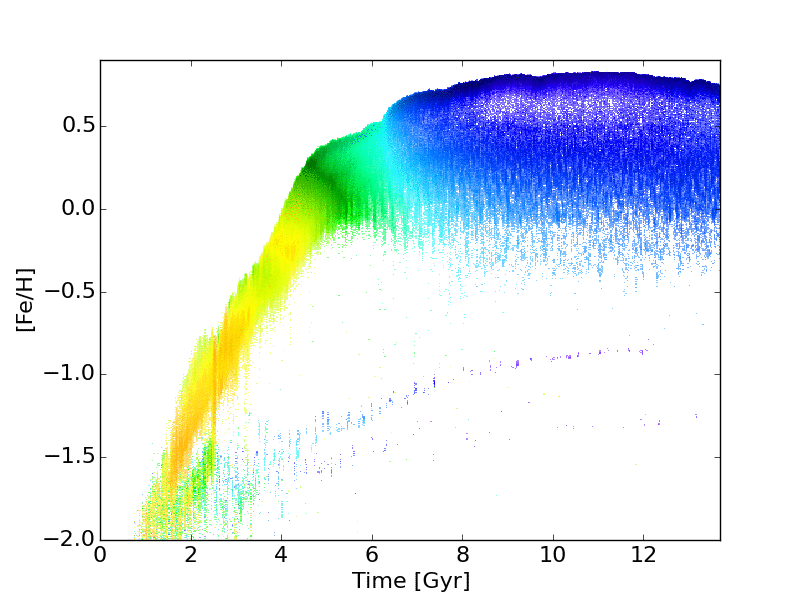}&
      \includegraphics[scale=.3,trim={0cm 0 1.5cm 1cm},clip]{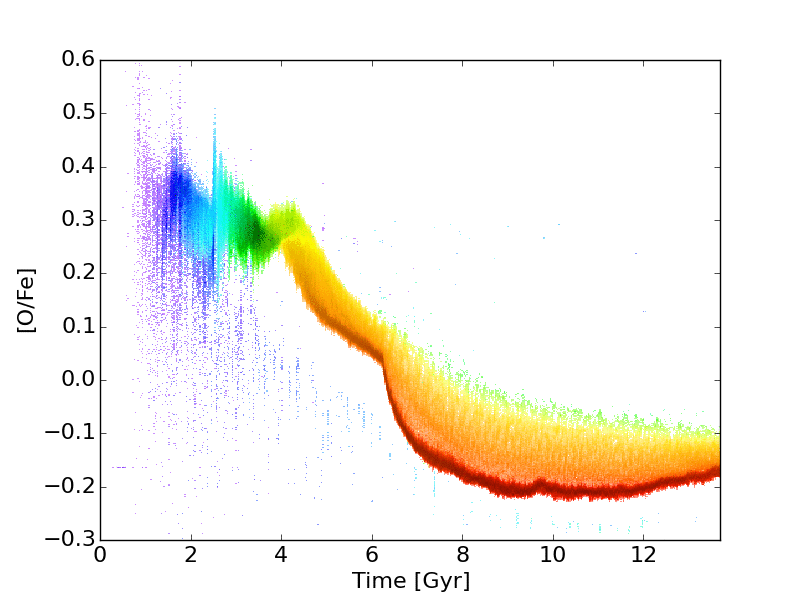} &
      \includegraphics[scale=.3,trim={0cm 0 1.5cm 1cm},clip]{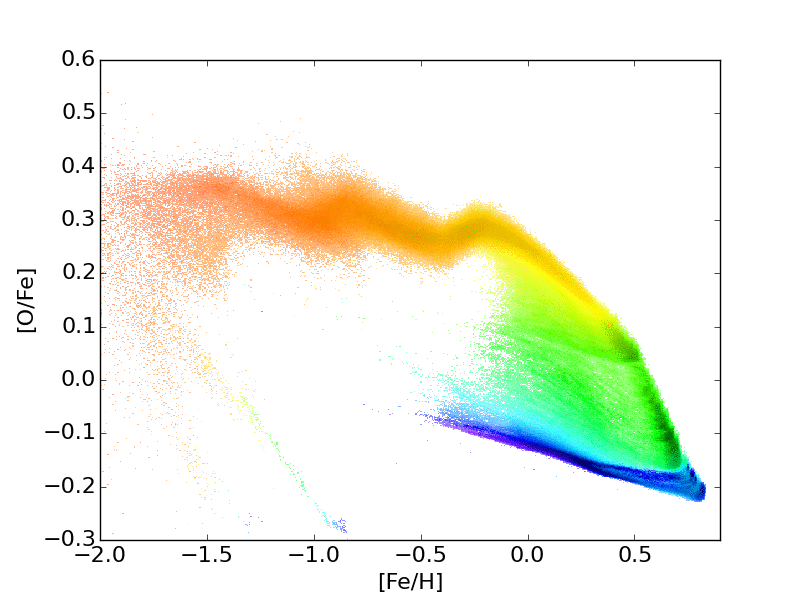}\\       
   \end{tabular}
\caption{ As in Fig. \ref{Fig:convolvederror} for MaGICC.  }
\label{Fig:convolvederrorMAGICC}
\end{figure*}

Figure \ref{Fig:convolvederror} and Fig. \ref{Fig:convolvederrorMAGICC} illustrate how increasingly refined uncertainties in the metallicities and ages of stars allow us to recover more and more of the inherent substructure. For the AMR, the bifurcation between the bulge and the disc becomes clear only where the age, metallicity and [O/Fe] errors are reduced to 0.5 Gyr, 0.05 dex and 0.05 dex. Further substructure becomes clear when the errors are halved again. The global shape of the distribution is visible even with fairly large errors (1 Gyr, 0.1 and 0.1 dex) but all the details are obscured. At this level of error the distributions only recover basic trends, although the `handgun' shape of the MUGS AMR is clear. At the intermediate error (0.5 Gyr, 0.05 and 0.05 dex) some of the substructure is visible, particularly the `m' shaped set of stars formed from the end of the life of the `sawtooth' satellite discussed in the previous section. Although APOGEE does not provide ages, the errors in [O/Fe] and [Fe/H] are around 0.03 dex \citep{Hayden2015}, which is between the second and third rows of the right hand column in Figs. \ref{Fig:convolvederror} and \ref{Fig:convolvederrorMAGICC}. At this level, various features in the distribution are apparent.

As the time-[O/Fe] evolution is more tightly correlated than the AMR, or the [Fe/H]-[O/Fe] distributions, we see that the convolved stars chart the actual distribution fairly well. However, in MUGS, the early slope has a wider spread than the later time-[O/Fe] evolution. This is the opposite to what is observed in the Milky Way using solar vicinity stars \citep{Haywood2013}, and is due to strong merger events during the first 6 Gyr of the formation of g15784. In MaGICC, the distribution is tight throughout its evolution and does not greatly change spread.

Some of the peaks, caused by infalling satellites etc, are evident even in the intermediate error distribution, and the profile has almost recovered most of its detail in the low error profiles. As in \citet{Snaith2014}, it is evident that using the age-[O/Fe] distribution to disentangle the star formation history of a galaxy using observations is probably the best approach. It is also evident that with intermediate errors, the time-[O/Fe] gradient, at early times and with a high star formation rate, is our best chance of recovering a `timer' for stellar ages.

The [Fe/H]-[O/Fe] distributions are also obscured by large errors, although certain broad trends can be seen. However, the splits between the upper, middle and lower sequences have been completely lost. Reducing the error sharpens the distribution, and we can see the emergence of the two sequences. This is not as clear as for observations of the Milky Way \citep[e.g.][]{Haywood2013,Zasowski2013}. Compared to the Milky Way, however, this galaxy does not seem to have as notable separation between sequences \citep{Fuhrmann2008,Adibekyan2012}. This is not surprising, as this galaxy makes no attempt to model the Milky Way evolution, and is its own unique object. Technically, the errors on [Fe/H] and [O/Fe] should be correlated, but that degree of precision is beyond the scope of Figs. \ref{Fig:convolvederror} and \ref{Fig:convolvederrorMAGICC}. which are used simply to illustrate the effect of errors rather than make a precise prediction.

\begin{figure}
\centering
     \begin{tabular}{c}      
      \raisebox{-\height}{\includegraphics[scale=.6,trim={.0cm 0cm 0cm 0cm},clip]{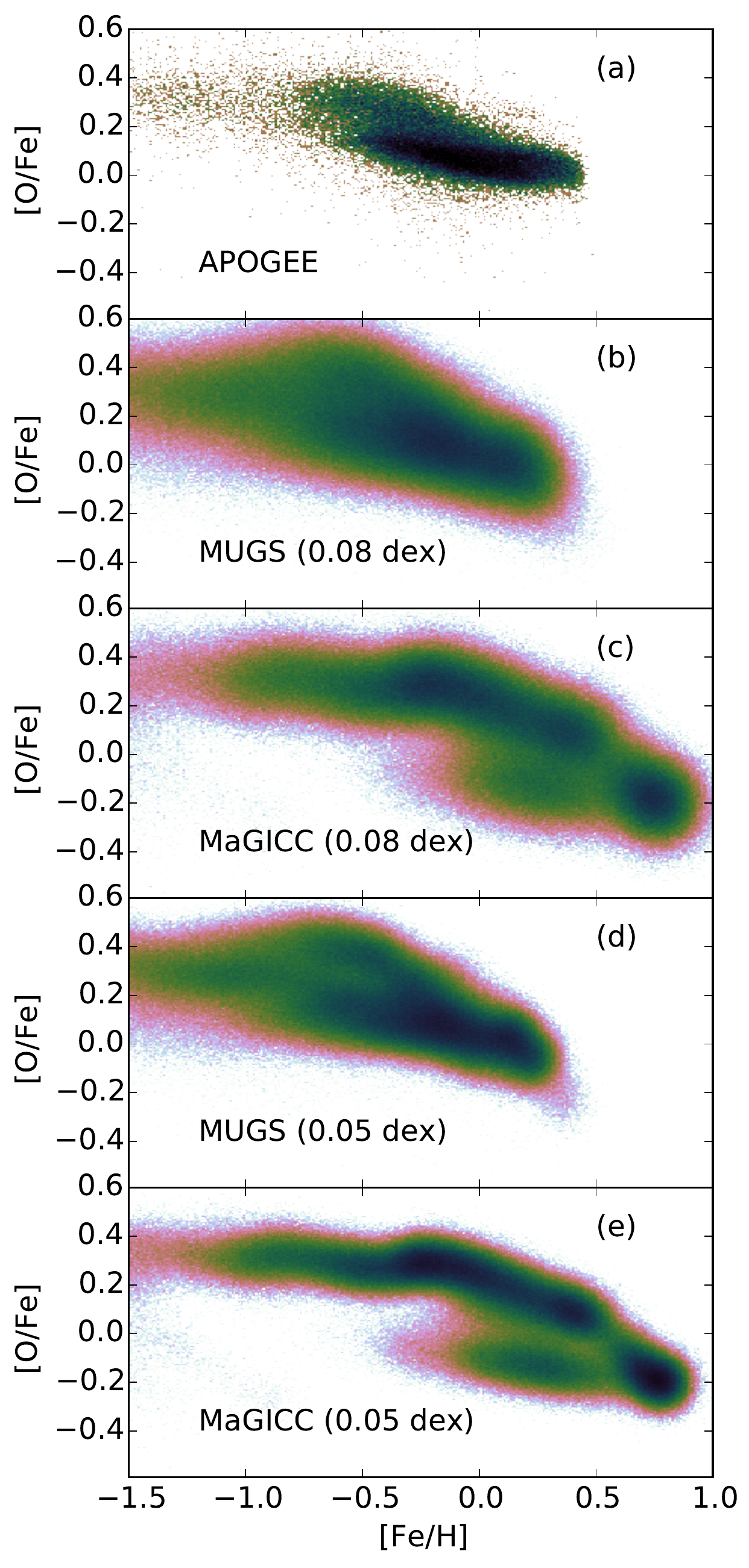}}\\
      \raisebox{-\height}{\includegraphics[scale=.3,trim={.0cm 0cm 0cm 0cm},clip]{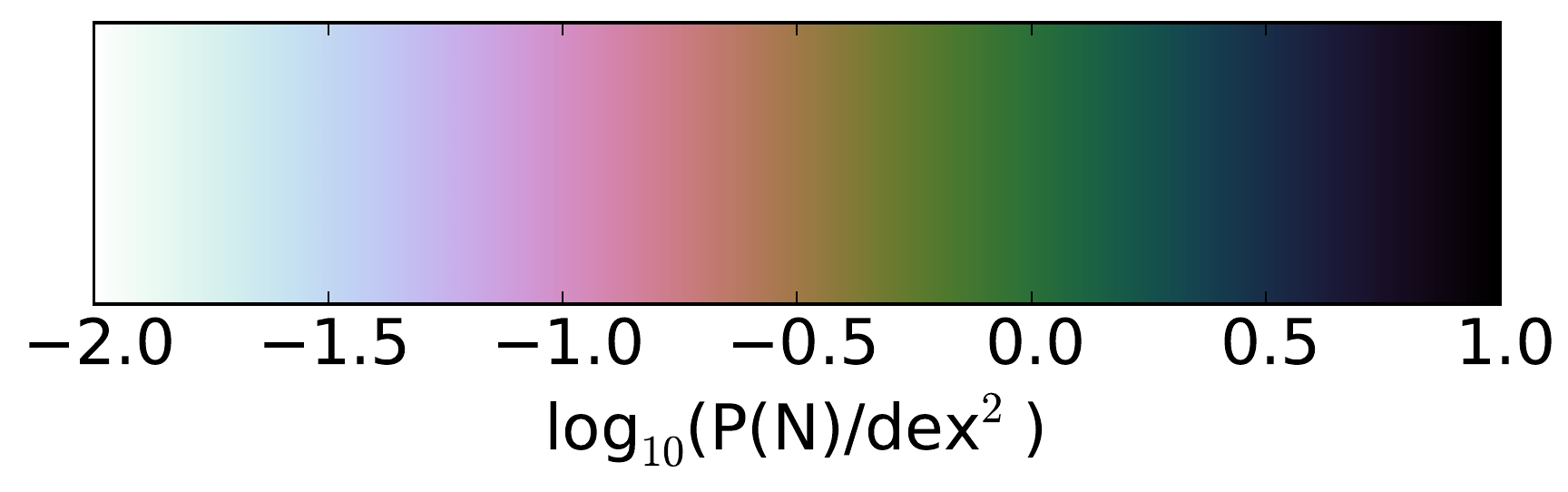}}\\
   \end{tabular}
\caption{The density distribution of stars in APOGEE (panel a), and all stars inside the virial radius  of the MUGS (panels b and d) and MaGICC (panels c and e) galaxies convolved with a random error selected from a Gaussian distribution. The width of the Gaussian is given in each panel.  }
\label{Fig:convolvederrortest1}
\end{figure}

An example of how the results in this paper can be compared with data is shown in Fig. \ref{Fig:convolvederrortest1}. The results of APOGEE show a bimodal distribution \citep[e.g.][see panel (a)]{Hayden2015}, with a high [O/Fe] region and a lower [O/Fe] arm at low to intermediate metallicity. As APOGEE data has a global error between 0.08 and 0.05 dex \citep{Holtzman2015} (higher for low S/N, low metallicity and hotter stars), the distribution of stars in APOGEE is compared with star particles in MUGS and MaGICC that have been convolved by errors of the same size (panels (b) and (c) show data with errors of 0.08 dex, and panels (d) and (e) show the data with errors of 0.05 dex). Even though the MaGICC data is offset to higher metallicities we see a similar bimodal distribution in both MUGS and MaGICC, particularly in panels (d) and (e). However, the highly populated regions of the  [O/Fe]-[Fe/H] distribution are more similar to the Milky Way distribution in MUGS, than in MaGICC. 

Much of the fine structure in MUGS and MaGICC is hidden by the errors, but the global trends in [O/Fe]-metallicity start to become clear between 0.08 and 0.05 dex. The global structure in MaGICC is perhaps clear even at 0.08 dex, but with any larger errors the bifurcation in MUGS  cannot be identified.

MaGICC produces a much stronger high [O/Fe] track than MUGS. Even discarding the inner 2 kpc of the galaxy, containing the dense knot centered on 0.75 dex, does not change this distribution greatly, except to remove the high metallicity region.  However, we must be careful in our comparison with APOGEE because we are using all stars within the virial radius of MUGS and MaGICC, while APOGEE is not as comprehensive a census of Milky Way stars.

This illustrates an inherent difficulty that without distance, spatial, and/or kinematic information, extremely high precision
will be required to recover the features of the galaxy. More informed analysis, which separates out features using spatial and kinematic information, may help to disentangle the various features. This, when combined with metallicity and age data, might assist in the identification of features with even large errors. 

\section{Discussion \& Conclusions}\label{sec:conclusions}
In this era of precision astronomy the metallicity and ages of populations of stars in the Galaxy, and other nearby galaxies, is within reach of observations. However, observations provide only snapshots in the history of any given galaxy. The chemical history of the galaxy does, however, leave a record in the abundances of stars. As the galaxy evolves, the chemical abundances of the ISM change, due to the release of metals from supernovae and AGB stars. We can trace this evolution by observing the chemical makeup of stars. This is because a stellar population retains the chemical abundance of the ISM in which it formed, so as the chemistry of the ISM evolved so did the chemistry of the newly-formed stars. Once stars are produced, however, they act as a permanent record of the conditions in the ISM from which they formed. 

The simulations use the same initial conditions and the same simulation code  \citep[GASOLINE][]{Wadsley2004} but differ primarily by the implementation of early radiative feedback in one simulation. We explore the chemical evolution of the different components of the galaxy g15784, which has been simulated using traditional supernova stellar feedback (MUGS) and traditional+early radiative feedback (MaGICC). We have decomposed the galaxy into its main components (bulge, disc, halo), and by whether stars formed insitu or not. We find that the different components show considerably divergent chemical evolution histories, and vary distinctly between the simulations  --- in particular, the inclusion of early radiative feedback removes substructure and suppresses early star formation.

We find that, contrary to the naive picture of metallicity evolution, the metallicity of the ISM does not rise monotonically, as would be expected in a `closed box' model of Galactic chemical evolution. Infalling gas can dilute the ISM, as can the movement of gas within a galaxy, or between a galaxy and its satellites.   However, even a simple closed box galactic chemical evolution model with a high star formation efficiency  can produce a declining AMR \citep[e.g.][]{Gibson1997}. A declining AMR is not inconsistent with the observations of \citet[][Fig. 18]{SB2009} and the simulations presented in \citet[][Fig. 6]{Few2014} etc. 

The metallicity of the stars reaches a threshold of about [Fe/H]$\sim 0.2$ dex for the MUGS galaxy, and  [Fe/H] $\sim 0.8$ dex for the MaGICC galaxy,  and then does not rise further. For long periods in the latter half of the evolution the upper limit of the metallicity of stars does not change. This means the metallicity of stars is a poor guide to the age of a star. This issue is further enhanced by the broadness of the metallicity distribution in the discs of galaxies, and the substructure introduced by infalling satellite galaxies. The substructure is almost entirely erased in MaGICC, as the enhanced early feedback suppresses star formation in satellites. 

The evolution of [O/Fe] with time is much more tightly correlated than the AMR. During high star formation episodes at early times,there is a rapid evolution of [O/Fe], but this flattens off. Thus, age-[O/Fe] is a better `clock' for stellar ages, but only at early times. This matches recent results in \citet{Haywood2013}.  

When we decompose the galaxy into bulge, disc and halo, and label stars as having either formed insitu, in satellites, or in other dark matter halos, we find that the bulge is more metal rich than the disc, and forms stars over the whole of cosmic time. The halo, however, forms early, during the rising arm of the AMR.  This is true for both MUGS and MaGICC. 

Satellite galaxies show various exotic behaviours. Some satellites have wide AMRs, like the host, while others have very little scatter. When satellites are tidally disturbed all their gas is drawn into the centre, and the AMR scatter drops.
We also see a `sawtooth' shape in the AMR of one satellite galaxy, which results from orbital interactions which allow the satellite to accrete low metallicity gas from the edge of host's disc.

In order to recover substructure in galaxies using abundances alone, small errors are required. Signatures of substructure disappear
when stellar properties are convolved with uncertainties of $\sigma$=1 Gyr, 0.1, 0.1 dex for ages, [Fe/H], and [O/Fe] respectively,
although the general features of the AMR are apparent.
Only when the precision is improved by a factor of four can the substructure be recovered.

We also note that our results cast doubt on the validity of assuming a monotonic AMR, as has regularly been done in algorithms that reconstruct star formation and enrichment histories from resolved stellar populations \citep[e.g.][]{Weisz2014}. However, the [O/Fe] evolution is much closer to monotonic, and so it may be possible to fit it with a curve and recover ages, albeit with large uncertainties. This is not universally the case with all such codes, but a subsample do assume monotonic metallicity, and therefore should be used with care. 

The chemistry of stars is the most effective way of gaining insight into the history of a galaxy. However, due to a number of effects, such as radial motions, decoding observations of galaxy chemistry is non-trivial. Extensive modelling is required, which can look at the time evolution of a galaxy as well as the end-time chemical distribution, in order to effectively interpret the results of forthcoming observations. However, the effect of the physics implemented into simulations must also be taken into account, as we have discussed in this paper. 

\section{acknowledgements}
ONS and JB acknowledge support for program HST-AR-12837 was provided by NASA through a grant from the Space Telescope Science Institute, which is operated by the Association of Universities for Research in Astronomy, Inc., under NASA contract NAS 5-26555.
We gratefully acknowledge the computer resources provided by STFCs DiRAC Facility (through the COSMOS: Galactic Archaeology programme), the DEISA consortium, co-funded through EU FP6 project RI-031513 and the FP7 project
RI-222919 (through the DEISA Extreme Computing Initiative), the PRACE-2IP
Project (FP7 RI-283493), and the University of Central Lancashire’s High
Performance Computing Facility.  This work has made use of the Shared Hierarchical Academic Research Computing Network (SHARCNET) Dedicated Resource Project: “MUGS: The McMaster Unbiased Galaxy Simulations Project” (DR316, DR401 and DR437). We thank the referee for comments which improved the presentation of the paper. 

\bibliographystyle{mn2e}
\bibliography{Mugsbib2}

\begin{thebibliography}{}

\bibitem[\protect\citeauthoryear{{Abadi}, {Navarro}, {Steinmetz} \&
  {Eke}}{{Abadi} et~al.}{2003}]{Abadi2003}
{Abadi} M.~G.,  {Navarro} J.~F.,  {Steinmetz} M.,    {Eke} V.~R.,  2003, \apj,
  597, 21

\bibitem[\protect\citeauthoryear{{Adibekyan}, {Sousa}, {Santos}, {Delgado
  Mena}, {Gonz{\'a}lez Hern{\'a}ndez}, {Israelian}, {Mayor} \&
  {Khachatryan}}{{Adibekyan} et~al.}{2012}]{Adibekyan2012}
{Adibekyan} V.~Z.,  {Sousa} S.~G.,  {Santos} N.~C.,  {Delgado Mena} E.,
  {Gonz{\'a}lez Hern{\'a}ndez} J.~I.,  {Israelian} G.,  {Mayor} M.,
  {Khachatryan} G.,  2012, \aap, 545, A32

\bibitem[\protect\citeauthoryear{{Bird}, {Kazantzidis}, {Weinberg}, {Guedes},
  {Callegari}, {Mayer} \& {Madau}}{{Bird} et~al.}{2013}]{Bird2013}
{Bird} J.~C.,  {Kazantzidis} S.,  {Weinberg} D.~H.,  {Guedes} J.,  {Callegari}
  S.,  {Mayer} L.,    {Madau} P.,  2013, \apj, 773, 43

\bibitem[\protect\citeauthoryear{{Brook}, {Kawata}, {Gibson} \&
  {Freeman}}{{Brook} et~al.}{2004}]{Brook2004}
{Brook} C.~B.,  {Kawata} D.,  {Gibson} B.~K.,    {Freeman} K.~C.,  2004, \apj,
  612, 894

\bibitem[\protect\citeauthoryear{{Brook}, {Stinson}, {Gibson}, {Shen},
  {Macci{\`o}}, {Obreja}, {Wadsley} \& {Quinn}}{{Brook}
  et~al.}{2014}]{Brook2014}
{Brook} C.~B.,  {Stinson} G.,  {Gibson} B.~K.,  {Shen} S.,  {Macci{\`o}} A.~V.,
   {Obreja} A.,  {Wadsley} J.,    {Quinn} T.,  2014, \mnras, 443, 3809

\bibitem[\protect\citeauthoryear{{Brook}, {Stinson}, {Gibson}, {Wadsley} \&
  {Quinn}}{{Brook} et~al.}{2012}]{Brook2012}
{Brook} C.~B.,  {Stinson} G.,  {Gibson} B.~K.,  {Wadsley} J.,    {Quinn} T.,
  2012, \mnras, 424, 1275

\bibitem[\protect\citeauthoryear{{Cacciari}}{{Cacciari}}{2009}]{Cacciari2009}
{Cacciari} C.,  2009, in {Mamajek} E.~E.,  {Soderblom} D.~R.,   {Wyse}
  R.~F.~G.,  eds, IAU Symposium Vol.~258 of IAU Symposium, {The promise of Gaia
  and how it will influence stellar ages}.
pp 409--418

\bibitem[\protect\citeauthoryear{{Calura}, {Gibson}, {Michel-Dansac},
  {Stinson}, {Cignoni}, {Dotter}, {Pilkington}, {House}, {Brook}, {Few},
  {Bailin}, {Couchman} \& {Wadsley}}{{Calura} et~al.}{2012}]{Calura2012}
{Calura} F.,  {Gibson} B.~K.,  {Michel-Dansac} L.,  {Stinson} G.~S.,  {Cignoni}
  M.,  {Dotter} A.,  {Pilkington} K.,  {House} E.~L.,  {Brook} C.~B.,  {Few}
  C.~G.,  {Bailin} J.,  {Couchman} H.~M.~P.,    {Wadsley} J.,  2012, \mnras,
  427, 1401

\bibitem[\protect\citeauthoryear{{Chabrier}}{{Chabrier}}{2003}]{Chabrier2003}
{Chabrier} G.,  2003, \pasp, 115, 763

\bibitem[\protect\citeauthoryear{{Chaplin}, {Basu}, {Huber}, {Serenelli} \&
  {Casagrande}}{{Chaplin} et~al.}{2014}]{Chaplin2014}
{Chaplin} W.~J.,  {Basu} S.,  {Huber} D.,  {Serenelli} A.,    {Casagrande} L.
  e.~a.,  2014, \apjs, 210, 1

\bibitem[\protect\citeauthoryear{{Cole}, {Skillman}, {Tolstoy}, {Gallagher}
  III, {Aparicio}, {Dolphin}, {Gallart}, {Hidalgo}, {Saha}, {Stetson} \&
  {Weisz}}{{Cole} et~al.}{2007}]{Cole2007}
{Cole} A.~A.,  {Skillman} E.~D.,  {Tolstoy} E.,  {Gallagher} III J.~S.,
  {Aparicio} A.,  {Dolphin} A.~E.,  {Gallart} C.,  {Hidalgo} S.~L.,  {Saha} A.,
   {Stetson} P.~B.,    {Weisz} D.~R.,  2007, \apjl, 659, L17

\bibitem[\protect\citeauthoryear{{Dalcanton}, {Williams}, {Lang}, {Lauer},
  {Kalirai}, {Seth}, {Dolphin} \& {Rosenfield}}{{Dalcanton}
  et~al.}{2012}]{Dalcanton2012}
{Dalcanton} J.~J.,  {Williams} B.~F.,  {Lang} D.,  {Lauer} T.~R.,  {Kalirai}
  J.~S.,  {Seth} A.~C.,  {Dolphin} A.,    {Rosenfield} P. e.~a.,  2012, \apjs,
  200, 18

\bibitem[\protect\citeauthoryear{{Epstein} \& {Pinsonneault}}{{Epstein} \&
  {Pinsonneault}}{2014}]{Epstein2014}
{Epstein} C.~R.,  {Pinsonneault} M.~H.,  2014, \apj, 780, 159

\bibitem[\protect\citeauthoryear{{Ferland}, {Korista}, {Verner}, {Ferguson},
  {Kingdon} \& {Verner}}{{Ferland} et~al.}{1998}]{Ferland1998}
{Ferland} G.~J.,  {Korista} K.~T.,  {Verner} D.~A.,  {Ferguson} J.~W.,
  {Kingdon} J.~B.,    {Verner} E.~M.,  1998, \pasp, 110, 761

\bibitem[\protect\citeauthoryear{{Few}, {Courty}, {Gibson}, {Michel-Dansac} \&
  {Calura}}{{Few} et~al.}{2014}]{Few2014}
{Few} C.~G.,  {Courty} S.,  {Gibson} B.~K.,  {Michel-Dansac} L.,    {Calura}
  F.,  2014, \mnras, 444, 3845

\bibitem[\protect\citeauthoryear{{Fuhrmann}}{{Fuhrmann}}{2008}]{Fuhrmann2008}
{Fuhrmann} K.,  2008, \mnras, 384, 173

\bibitem[\protect\citeauthoryear{{Gibson}}{{Gibson}}{1997}]{Gibson1997}
{Gibson} B.~K.,  1997, \mnras, 290, 471

\bibitem[\protect\citeauthoryear{{Gibson}, {Pilkington}, {Brook}, {Stinson} \&
  {Bailin}}{{Gibson} et~al.}{2013}]{Gibson2013}
{Gibson} B.~K.,  {Pilkington} K.,  {Brook} C.~B.,  {Stinson} G.~S.,    {Bailin}
  J.,  2013, \aap, 554, A47

\bibitem[\protect\citeauthoryear{{Gill}, {Knebe} \& {Gibson}}{{Gill}
  et~al.}{2004}]{Gill2004}
{Gill} S.~P.~D.,  {Knebe} A.,    {Gibson} B.~K.,  2004, \mnras, 351, 399

\bibitem[\protect\citeauthoryear{{Gogarten}, {Dalcanton}, {Williams}, {Ro{\v
  s}kar}, {Holtzman}, {Seth}, {Dolphin}, {Weisz}, {Cole}, {Debattista},
  {Gilbert}, {Olsen}, {Skillman}, {de Jong}, {Karachentsev} \&
  {Quinn}}{{Gogarten} et~al.}{2010}]{Gogarten2010}
{Gogarten} S.~M.,  {Dalcanton} J.~J.,  {Williams} B.~F.,  {Ro{\v s}kar} R.,
  {Holtzman} J.,  {Seth} A.~C.,  {Dolphin} A.,  {Weisz} D.,  {Cole} A.,
  {Debattista} V.~P.,  {Gilbert} K.~M.,  {Olsen} K.,  {Skillman} E.,  {de Jong}
  R.~S.,  {Karachentsev} I.~D.,    {Quinn} T.~R.,  2010, \apj, 712, 858

\bibitem[\protect\citeauthoryear{{Gunn} \& {Gott} III}{{Gunn} \&
  {Gott}}{1972}]{Gunn1972}
{Gunn} J.~E.,  {Gott} III J.~R.,  1972, \apj, 176, 1

\bibitem[\protect\citeauthoryear{{Hayden}, {Bovy}, {Holtzman}, {Nidever},
  {Bird}, {Weinberg}, {Andrews} \& {Majewski}}{{Hayden}
  et~al.}{2015}]{Hayden2015}
{Hayden} M.~R.,  {Bovy} J.,  {Holtzman} J.~A.,  {Nidever} D.~L.,  {Bird} J.~C.,
   {Weinberg} D.~H.,  {Andrews} B.~H.,    {Majewski} S.~R. e.~a.,  2015, \apj,
  808, 132

\bibitem[\protect\citeauthoryear{{Haywood}, {Di Matteo}, {Lehnert}, {Katz} \&
  {G{\'o}mez}}{{Haywood} et~al.}{2013}]{Haywood2013}
{Haywood} M.,  {Di Matteo} P.,  {Lehnert} M.~D.,  {Katz} D.,    {G{\'o}mez} A.,
   2013, \aap, 560, A109

\bibitem[\protect\citeauthoryear{{Haywood}, {Di Matteo}, {Snaith} \&
  {Lehnert}}{{Haywood} et~al.}{2015}]{Haywood2015}
{Haywood} M.,  {Di Matteo} P.,  {Snaith} O.,    {Lehnert} M.~D.,  2015, \aap,
  579, A5

\bibitem[\protect\citeauthoryear{{Holmberg}, {Nordstr{\"o}m} \&
  {Andersen}}{{Holmberg} et~al.}{2009}]{Holmberg2009}
{Holmberg} J.,  {Nordstr{\"o}m} B.,    {Andersen} J.,  2009, \aap, 501, 941

\bibitem[\protect\citeauthoryear{{Holtzman}, {Shetrone}, {Johnson}, {Allende
  Prieto}, {Anders} \& et al.}{{Holtzman} et~al.}{2015}]{Holtzman2015}
{Holtzman} J.~A.,  {Shetrone} M.,  {Johnson} J.~A.,  {Allende Prieto} C.,
  {Anders} F.,    et al. 2015, \aj, 150, 148

\bibitem[\protect\citeauthoryear{{Hopkins}, {Kere{\v s}}, {O{\~n}orbe},
  {Faucher-Gigu{\`e}re}, {Quataert}, {Murray} \& {Bullock}}{{Hopkins}
  et~al.}{2014}]{Hopkins2014}
{Hopkins} P.~F.,  {Kere{\v s}} D.,  {O{\~n}orbe} J.,  {Faucher-Gigu{\`e}re}
  C.-A.,  {Quataert} E.,  {Murray} N.,    {Bullock} J.~S.,  2014, \mnras, 445,
  581

\bibitem[\protect\citeauthoryear{{Howard}, {Rich}, {Reitzel}, {Koch}, {De
  Propris} \& {Zhao}}{{Howard} et~al.}{2008}]{Howard2008}
{Howard} C.~D.,  {Rich} R.~M.,  {Reitzel} D.~B.,  {Koch} A.,  {De Propris} R.,
    {Zhao} H.,  2008, \apj, 688, 1060

\bibitem[\protect\citeauthoryear{{Iwamoto}, {Brachwitz}, {Nomoto}, {Kishimoto},
  {Umeda}, {Hix} \& {Thielemann}}{{Iwamoto} et~al.}{1999}]{Iwamoto1999}
{Iwamoto} K.,  {Brachwitz} F.,  {Nomoto} K.,  {Kishimoto} N.,  {Umeda} H.,
  {Hix} W.~R.,    {Thielemann} F.-K.,  1999, \apjs, 125, 439

\bibitem[\protect\citeauthoryear{{Kennicutt}
  Jr.}{{Kennicutt}}{1998}]{Kennicutt1998}
{Kennicutt} Jr. R.~C.,  1998, \apj, 498, 541

\bibitem[\protect\citeauthoryear{{Kewley}, {Rupke}, {Zahid}, {Geller} \&
  {Barton}}{{Kewley} et~al.}{2010}]{Kewley2010}
{Kewley} L.~J.,  {Rupke} D.,  {Zahid} H.~J.,  {Geller} M.~J.,    {Barton}
  E.~J.,  2010, \apjl, 721, L48

\bibitem[\protect\citeauthoryear{{Kirby}, {Cohen}, {Smith}, {Majewski}, {Sohn}
  \& {Guhathakurta}}{{Kirby} et~al.}{2011}]{Kirby2011}
{Kirby} E.~N.,  {Cohen} J.~G.,  {Smith} G.~H.,  {Majewski} S.~R.,  {Sohn}
  S.~T.,    {Guhathakurta} P.,  2011, \apj, 727, 79

\bibitem[\protect\citeauthoryear{{Knebe}, {Knollmann}, {Muldrew}, {Pearce},
  {Aragon-Calvo}, {Ascasibar} \& {Behroozi}}{{Knebe} et~al.}{2011}]{Knebe2011}
{Knebe} A.,  {Knollmann} S.~R.,  {Muldrew} S.~I.,  {Pearce} F.~R.,
  {Aragon-Calvo} M.~A.,  {Ascasibar} Y.,    {Behroozi} P.~S. e.~a.,  2011,
  \mnras, 415, 2293

\bibitem[\protect\citeauthoryear{{Knollmann} \& {Knebe}}{{Knollmann} \&
  {Knebe}}{2009}]{Knollmann2009}
{Knollmann} S.~R.,  {Knebe} A.,  2009, \apjs, 182, 608

\bibitem[\protect\citeauthoryear{{Kroupa}, {Tout} \& {Gilmore}}{{Kroupa}
  et~al.}{1993}]{Kroupa1993}
{Kroupa} P.,  {Tout} C.~A.,    {Gilmore} G.,  1993, \mnras, 262, 545

\bibitem[\protect\citeauthoryear{{Larson}}{{Larson}}{1976}]{Larson1976}
{Larson} R.~B.,  1976, \mnras, 176, 31

\bibitem[\protect\citeauthoryear{{Madau} \& {Dickinson}}{{Madau} \&
  {Dickinson}}{2014}]{Madau2014}
{Madau} P.,  {Dickinson} M.,  2014, \araa, 52, 415

\bibitem[\protect\citeauthoryear{{McMillan}}{{McMillan}}{2011}]{McMillan2011}
{McMillan} P.~J.,  2011, \mnras, 414, 2446

\bibitem[\protect\citeauthoryear{{Miranda}, {Macfarlane} \& {Gibson}}{{Miranda}
  et~al.}{2015}]{Miranda2015a}
{Miranda} M.~S.,  {Macfarlane} B.~A.,    {Gibson} B.~K.,  2015, ArXiv
  e-prints:1502.00444

\bibitem[\protect\citeauthoryear{{Miranda}, {Pilkington}, {Gibson}, {Brook} \&
  et al.}{{Miranda} et~al.}{2015b}]{Miranda2015b}
{Miranda} M.~S.,  {Pilkington} K.,  {Gibson} B.~K.,  {Brook} C.~B.,    et al.
  2015b, \mnrass

\bibitem[\protect\citeauthoryear{{Monachesi}, {Trager}, {Lauer}, {Hidalgo},
  {Freedman}, {Dressler}, {Grillmair} \& {Mighell}}{{Monachesi}
  et~al.}{2012}]{Monachesi2012}
{Monachesi} A.,  {Trager} S.~C.,  {Lauer} T.~R.,  {Hidalgo} S.~L.,  {Freedman}
  W.,  {Dressler} A.,  {Grillmair} C.,    {Mighell} K.~J.,  2012, \apj, 745, 97

\bibitem[\protect\citeauthoryear{{Nickerson}, {Stinson}, {Couchman}, {Bailin}
  \& {Wadsley}}{{Nickerson} et~al.}{2011}]{Nickerson2011}
{Nickerson} S.,  {Stinson} G.,  {Couchman} H.~M.~P.,  {Bailin} J.,    {Wadsley}
  J.,  2011, \mnras, 415, 257

\bibitem[\protect\citeauthoryear{{Nickerson}, {Stinson}, {Couchman}, {Bailin}
  \& {Wadsley}}{{Nickerson} et~al.}{2013}]{Nickerson2013}
{Nickerson} S.,  {Stinson} G.,  {Couchman} H.~M.~P.,  {Bailin} J.,    {Wadsley}
  J.,  2013, \mnras, 429, 452

\bibitem[\protect\citeauthoryear{{Obreja}, {Brook}, {Stinson},
  {Dom{\'{\i}}nguez-Tenreiro}, {Gibson}, {Silva} \& {Granato}}{{Obreja}
  et~al.}{2014}]{Obreja2014}
{Obreja} A.,  {Brook} C.~B.,  {Stinson} G.,  {Dom{\'{\i}}nguez-Tenreiro} R.,
  {Gibson} B.~K.,  {Silva} L.,    {Granato} G.~L.,  2014, \mnras, 442, 1794

\bibitem[\protect\citeauthoryear{{Pagel} \& {Edmunds}}{{Pagel} \&
  {Edmunds}}{1981}]{Pagel1981}
{Pagel} B.~E.~J.,  {Edmunds} M.~G.,  1981, \araa, 19, 77

\bibitem[\protect\citeauthoryear{{Pilkington}}{{Pilkington}}{2013}]{Pilkington%
2013}
{Pilkington} K.,  2013, PhD thesis, University of Central Lancashire (United
  Kingdom)

\bibitem[\protect\citeauthoryear{{Pilkington}, {Few}, {Gibson}, {Calura},
  {Michel-Dansac}, {Thacker}, {Moll{\'a}}, {Matteucci}, {Rahimi}, {Kawata},
  {Kobayashi}, {Brook}, {Stinson}, {Couchman}, {Bailin} \&
  {Wadsley}}{{Pilkington} et~al.}{2012}]{Pilkington2012}
{Pilkington} K.,  {Few} C.~G.,  {Gibson} B.~K.,  {Calura} F.,  {Michel-Dansac}
  L.,  {Thacker} R.~J.,  {Moll{\'a}} M.,  {Matteucci} F.,  {Rahimi} A.,
  {Kawata} D.,  {Kobayashi} C.,  {Brook} C.~B.,  {Stinson} G.~S.,  {Couchman}
  H.~M.~P.,  {Bailin} J.,    {Wadsley} J.,  2012, \aap, 540, A56

\bibitem[\protect\citeauthoryear{{Pilkington}, {Gibson}, {Brook}, {Calura},
  {Stinson}, {Thacker}, {Michel-Dansac}, {Bailin}, {Couchman}, {Wadsley},
  {Quinn} \& {Maccio}}{{Pilkington} et~al.}{2012a}]{Pilkington2012b}
{Pilkington} K.,  {Gibson} B.~K.,  {Brook} C.~B.,  {Calura} F.,  {Stinson}
  G.~S.,  {Thacker} R.~J.,  {Michel-Dansac} L.,  {Bailin} J.,  {Couchman}
  H.~M.~P.,  {Wadsley} J.,  {Quinn} T.~R.,    {Maccio} A.,  2012a, \mnras, 425,
  969

\bibitem[\protect\citeauthoryear{{Pilkington}, {Gibson}, {Brook}, {Calura},
  {Stinson}, {Thacker}, {Michel-Dansac}, {Bailin}, {Couchman}, {Wadsley},
  {Quinn} \& {Maccio}}{{Pilkington} et~al.}{2012b}]{aPilkington2012}
{Pilkington} K.,  {Gibson} B.~K.,  {Brook} C.~B.,  {Calura} F.,  {Stinson}
  G.~S.,  {Thacker} R.~J.,  {Michel-Dansac} L.,  {Bailin} J.,  {Couchman}
  H.~M.~P.,  {Wadsley} J.,  {Quinn} T.~R.,    {Maccio} A.,  2012b, \mnras, 425,
  969

\bibitem[\protect\citeauthoryear{{Pillepich}, {Madau} \& {Mayer}}{{Pillepich}
  et~al.}{2014}]{Pillepich2014}
{Pillepich} A.,  {Madau} P.,    {Mayer} L.,  2014, ArXiv e-prints

\bibitem[\protect\citeauthoryear{{Pontzen}, {Ro{\v s}kar}, {Stinson}, {Woods},
  {Reed}, {Coles} \& {Quinn}}{{Pontzen} et~al.}{2013}]{pynbody}
{Pontzen} A.,  {Ro{\v s}kar} R.,  {Stinson} G.~S.,  {Woods} R.,  {Reed} D.~M.,
  {Coles} J.,    {Quinn} T.~R., , 2013, {pynbody: Astrophysics Simulation
  Analysis for Python}

\bibitem[\protect\citeauthoryear{{Putman}, {Peek} \& {Joung}}{{Putman}
  et~al.}{2012}]{Putman2012}
{Putman} M.~E.,  {Peek} J.~E.~G.,    {Joung} M.~R.,  2012, \araa, 50, 491

\bibitem[\protect\citeauthoryear{{Ram{\'{\i}}rez}, {Allende Prieto} \&
  {Lambert}}{{Ram{\'{\i}}rez} et~al.}{2013}]{Ramirez2013}
{Ram{\'{\i}}rez} I.,  {Allende Prieto} C.,    {Lambert} D.~L.,  2013, \apj,
  764, 78

\bibitem[\protect\citeauthoryear{{Rupke}, {Kewley} \& {Barnes}}{{Rupke}
  et~al.}{2010}]{Rupke2010}
{Rupke} D.~S.~N.,  {Kewley} L.~J.,    {Barnes} J.~E.,  2010, \apjl, 710, L156

\bibitem[\protect\citeauthoryear{{S{\'a}nchez-Bl{\'a}zquez}, {Courty}, {Gibson}
  \& {Brook}}{{S{\'a}nchez-Bl{\'a}zquez} et~al.}{2009}]{SB2009}
{S{\'a}nchez-Bl{\'a}zquez} P.,  {Courty} S.,  {Gibson} B.~K.,    {Brook} C.~B.,
   2009, \mnras, 398, 591

\bibitem[\protect\citeauthoryear{{Scannapieco}, {Wadepuhl}, {Parry}, {Navarro},
  {Jenkins}, {Springel}, {Teyssier}, {Carlson}, {Couchman}, {Crain}, {Dalla
  Vecchia}, {Frenk} \& et al.}{{Scannapieco} et~al.}{2012}]{Scannapieco2012}
{Scannapieco} C.,  {Wadepuhl} M.,  {Parry} O.~H.,  {Navarro} J.~F.,  {Jenkins}
  A.,  {Springel} V.,  {Teyssier} R.,  {Carlson} E.,  {Couchman} H.~M.~P.,
  {Crain} R.~A.,  {Dalla Vecchia} C.,  {Frenk} C.~S.,    et al. 2012, \mnras,
  423, 1726

\bibitem[\protect\citeauthoryear{{Schmidt}}{{Schmidt}}{1959}]{Schmidt1959}
{Schmidt} M.,  1959, \apj, 129, 243

\bibitem[\protect\citeauthoryear{{Sellwood} \& {Binney}}{{Sellwood} \&
  {Binney}}{2002}]{Sellwood2002}
{Sellwood} J.~A.,  {Binney} J.~J.,  2002, \mnras, 336, 785

\bibitem[\protect\citeauthoryear{{Shen}, {Wadsley} \& {Stinson}}{{Shen}
  et~al.}{2010}]{Shen2010}
{Shen} S.,  {Wadsley} J.,    {Stinson} G.,  2010, \mnras, 407, 1581

\bibitem[\protect\citeauthoryear{{Skillman}, {Tolstoy}, {Cole}, {Dolphin},
  {Saha}, {Gallagher}, {Dohm-Palmer} \& {Mateo}}{{Skillman}
  et~al.}{2003}]{Skillman2003}
{Skillman} E.~D.,  {Tolstoy} E.,  {Cole} A.~A.,  {Dolphin} A.~E.,  {Saha} A.,
  {Gallagher} J.~S.,  {Dohm-Palmer} R.~C.,    {Mateo} M.,  2003, \apj, 596, 253

\bibitem[\protect\citeauthoryear{{Snaith}, {Haywood}, {Di Matteo}, {Lehnert},
  {Combes}, {Katz} \& {G{\'o}mez}}{{Snaith} et~al.}{2015}]{Snaith2014b}
{Snaith} O.,  {Haywood} M.,  {Di Matteo} P.,  {Lehnert} M.~D.,  {Combes} F.,
  {Katz} D.,    {G{\'o}mez} A.,  2015, \aap, 578, A87

\bibitem[\protect\citeauthoryear{{Snaith}, {Haywood}, {Di Matteo}, {Lehnert},
  {Combes}, {Katz} \& {G{\'o}mez}}{{Snaith} et~al.}{2014}]{Snaith2014}
{Snaith} O.~N.,  {Haywood} M.,  {Di Matteo} P.,  {Lehnert} M.~D.,  {Combes} F.,
   {Katz} D.,    {G{\'o}mez} A.,  2014, \apjl, 781, L31

\bibitem[\protect\citeauthoryear{{Sommer-Larsen}}{{Sommer-Larsen}}{2006}]{Somm%
er2006}
{Sommer-Larsen} J.,  2006, \apjl, 644, L1

\bibitem[\protect\citeauthoryear{{Spergel}, {Bean}, {Dor{\'e}}, {Nolta} \&
  {Bennett}}{{Spergel} et~al.}{2007}]{Spergel2007a}
{Spergel} D.~N.,  {Bean} R.,  {Dor{\'e}} O.,  {Nolta} M.~R.,    {Bennett}
  e.~a.,  2007, \apjs, 170, 377

\bibitem[\protect\citeauthoryear{{Stinson}, {Seth}, {Katz}, {Wadsley},
  {Governato} \& {Quinn}}{{Stinson} et~al.}{2006}]{Stinson2006}
{Stinson} G.,  {Seth} A.,  {Katz} N.,  {Wadsley} J.,  {Governato} F.,
  {Quinn} T.,  2006, \mnras, 373, 1074

\bibitem[\protect\citeauthoryear{{Stinson}, {Bailin}, {Couchman}, {Wadsley},
  {Shen}, {Nickerson}, {Brook} \& {Quinn}}{{Stinson}
  et~al.}{2010}]{Stinson2010}
{Stinson} G.~S.,  {Bailin} J.,  {Couchman} H.,  {Wadsley} J.,  {Shen} S.,
  {Nickerson} S.,  {Brook} C.,    {Quinn} T.,  2010, \mnras, 408, 812

\bibitem[\protect\citeauthoryear{{Stinson}, {Bovy}, {Rix}, {Brook}, {Ro{\v
  s}kar}, {Dalcanton}, {Macci{\`o}}, {Wadsley}, {Couchman} \&
  {Quinn}}{{Stinson} et~al.}{2013}]{Stinson2013b}
{Stinson} G.~S.,  {Bovy} J.,  {Rix} H.-W.,  {Brook} C.,  {Ro{\v s}kar} R.,
  {Dalcanton} J.~J.,  {Macci{\`o}} A.~V.,  {Wadsley} J.,  {Couchman} H.~M.~P.,
    {Quinn} T.~R.,  2013, \mnras, 436, 625

\bibitem[\protect\citeauthoryear{{Stinson}, {Brook}, {Macci{\`o}}, {Wadsley},
  {Quinn} \& {Couchman}}{{Stinson} et~al.}{2013}]{Stinson2013}
{Stinson} G.~S.,  {Brook} C.,  {Macci{\`o}} A.~V.,  {Wadsley} J.,  {Quinn}
  T.~R.,    {Couchman} H.~M.~P.,  2013, \mnras, 428, 129

\bibitem[\protect\citeauthoryear{{Tinsley}}{{Tinsley}}{1972}]{Tinsley1972}
{Tinsley} B.~M.,  1972, \aap, 20, 383

\bibitem[\protect\citeauthoryear{{Tissera}, {Scannapieco}, {Beers} \&
  {Carollo}}{{Tissera} et~al.}{2013}]{Tissera2013}
{Tissera} P.~B.,  {Scannapieco} C.,  {Beers} T.~C.,    {Carollo} D.,  2013,
  \mnras, 432, 3391

\bibitem[\protect\citeauthoryear{{Valluri}, {Debattista}, {Stinson}, {Bailin},
  {Quinn}, {Couchman} \& {Wadsley}}{{Valluri} et~al.}{2013}]{Valluri2013}
{Valluri} M.,  {Debattista} V.~P.,  {Stinson} G.~S.,  {Bailin} J.,  {Quinn}
  T.~R.,  {Couchman} H.~M.~P.,    {Wadsley} J.,  2013, \apj, 767, 93

\bibitem[\protect\citeauthoryear{{Wadsley}, {Stadel} \& {Quinn}}{{Wadsley}
  et~al.}{2004}]{Wadsley2004}
{Wadsley} J.~W.,  {Stadel} J.,    {Quinn} T.,  2004, \na, 9, 137

\bibitem[\protect\citeauthoryear{{Wadsley}, {Veeravalli} \&
  {Couchman}}{{Wadsley} et~al.}{2008}]{Wadsley2008}
{Wadsley} J.~W.,  {Veeravalli} G.,    {Couchman} H.~M.~P.,  2008, \mnras, 387,
  427

\bibitem[\protect\citeauthoryear{{Weisz}, {Dalcanton}, {Williams}, {Gilbert},
  {Skillman}, {Seth}, {Dolphin}, {McQuinn}, {Gogarten}, {Holtzman}, {Rosema},
  {Cole}, {Karachentsev} \& {Zaritsky}}{{Weisz} et~al.}{2011}]{Weisz2011}
{Weisz} D.~R.,  {Dalcanton} J.~J.,  {Williams} B.~F.,  {Gilbert} K.~M.,
  {Skillman} E.~D.,  {Seth} A.~C.,  {Dolphin} A.~E.,  {McQuinn} K.~B.~W.,
  {Gogarten} S.~M.,  {Holtzman} J.,  {Rosema} K.,  {Cole} A.,  {Karachentsev}
  I.~D.,    {Zaritsky} D.,  2011, \apj, 739, 5

\bibitem[\protect\citeauthoryear{{Weisz}, {Dolphin}, {Skillman}, {Holtzman},
  {Gilbert}, {Dalcanton} \& {Williams}}{{Weisz} et~al.}{2014}]{Weisz2014}
{Weisz} D.~R.,  {Dolphin} A.~E.,  {Skillman} E.~D.,  {Holtzman} J.,  {Gilbert}
  K.~M.,  {Dalcanton} J.~J.,    {Williams} B.~F.,  2014, \apj, 789, 147

\bibitem[\protect\citeauthoryear{{Williams}, {Dalcanton}, {Seth}, {Weisz},
  {Dolphin}, {Skillman}, {Harris}, {Holtzman}, {Girardi}, {de Jong}, {Olsen},
  {Cole}, {Gallart}, {Gogarten}, {Hidalgo}, {Mateo}, {Rosema}, {Stetson} \&
  {Quinn}}{{Williams} et~al.}{2009}]{Williams2009}
{Williams} B.~F.,  {Dalcanton} J.~J.,  {Seth} A.~C.,  {Weisz} D.,  {Dolphin}
  A.,  {Skillman} E.,  {Harris} J.,  {Holtzman} J.,  {Girardi} L.,  {de Jong}
  R.~S.,  {Olsen} K.,  {Cole} A.,  {Gallart} C.,  {Gogarten} S.~M.,  {Hidalgo}
  S.~L.,  {Mateo} M.,  {Rosema} K.,  {Stetson} P.~B.,    {Quinn} T.,  2009,
  \aj, 137, 419

\bibitem[\protect\citeauthoryear{{Woods}, {Wadsley}, {Couchman}, {Stinson} \&
  {Shen}}{{Woods} et~al.}{2014}]{Woods2014}
{Woods} R.~M.,  {Wadsley} J.,  {Couchman} H.~M.~P.,  {Stinson} G.,    {Shen}
  S.,  2014, \mnras, 442, 732

\bibitem[\protect\citeauthoryear{{Woosley} \& {Weaver}}{{Woosley} \&
  {Weaver}}{1995}]{WW95}
{Woosley} S.~E.,  {Weaver} T.~A.,  1995, \apjs, 101, 181

\bibitem[\protect\citeauthoryear{{Zasowski}, {Johnson}, {Frinchaboy},
  {Majewski}, {Nidever}, {Rocha Pinto}, {Girardi}, {Andrews}, {Chojnowski},
  {Cudworth}, {Jackson}, {Munn}, {Skrutskie}, {Beaton}, {Blake} \&
  {Covey}}{{Zasowski} et~al.}{2013}]{Zasowski2013}
{Zasowski} G.,  {Johnson} J.~A.,  {Frinchaboy} P.~M.,  {Majewski} S.~R.,
  {Nidever} D.~L.,  {Rocha Pinto} H.~J.,  {Girardi} L.,  {Andrews} B.,
  {Chojnowski} S.~D.,  {Cudworth} K.~M.,  {Jackson} K.,  {Munn} J.,
  {Skrutskie} M.~F.,  {Beaton} R.~L.,  {Blake} C.~H.,    {Covey} K. e.~a.,
  2013, \aj, 146, 81

\end{thebibliography}

\end{document}